\documentclass[useAMS,usenatbib]{mn2e}

\usepackage{graphicx}
\usepackage{natbib}

\catcode`\@=11
\def\gsim{\ifmmode{\mathrel{\mathpalette\@versim>}}
    \else{$\mathrel{\mathpalette\@versim>}$}\fi}
\def\lsim{\ifmmode{\mathrel{\mathpalette\@versim<}}
    \else{$\mathrel{\mathpalette\@versim<}$}\fi}
\def\@versim#1#2{\lower 2.9truept \vbox{\baselineskip 0pt \lineskip
    0.5truept \ialign{$\m@th#1\hfil##\hfil$\crcr#2\crcr\sim\crcr}}}
\catcode`\@=12

\newcommand{\beq}{\begin{equation}}
\newcommand{\eeq}{\end{equation}}

\newcommand{\ant}{\rm {\alpha_{nt}}}
\newcommand{\fth}{f_{\rm th}}
\def\hii{{H{\sc ii}~}}

\title[GMRT 333 MHz observations]{GMRT 333 MHz observations of 6 nearby normal galaxies}

\author[Basu et al.]{A. Basu$^{1}$\thanks{E-mail:
    aritra@ncra.tifr.res.in (AB), dmitra@ncra.tifr.res.in (DM), yogesh@ncra.tifr.res.in (YW),
ishwar@ncra.tifr.res.in (CHI-C)}, D. Mitra$^{1\star}$, Y. Wadadekar$^{1\star}$ and 
C.~H. Ishwara-Chandra$^{1\star}$ \\ 
$^{1}$National Centre for Radio Astrophysics, TIFR, Post Bag 3, Ganeshkhind, Pune - 411007, India.}
\begin{document}

\date{\it Accepted for publication in MNRAS}

\pagerange{\pageref{firstpage}--\pageref{lastpage}} \pubyear{2011}

\maketitle

\label{firstpage}

\begin{abstract}

We report Giant Meterwave Radio Telescope (GMRT) continuum
observations of six nearby normal galaxies at 333 MHz. The galaxies
are observed with angular resolutions better than $\sim 20\arcsec$
(corresponding to a linear scale of about 0.4 -- 1 kpc). 
These observations are sensitive to all the angular scales of interest,
since the resolution of the shortest baseline in GMRT is greater than the angular size 
of the galaxies.
Further, for five of these galaxies we show that at 333 MHz, the mean
thermal fraction is less than 5\%. Using archival data at about 1 GHz,
we estimate the mean thermal fraction to be about 10\% at that
frequency. We also find that the nonthermal spectral index is
generally steeper in regions with low thermal fraction and/or located
in the outer parts of the galaxy. In regions of high thermal fraction,
the nonthermal spectral index is flatter, and has a narrow
distribution peaking at $\sim -0.78$ with a spread of 0.16, putting
stringent constraints on the physical conditions for generation,
diffusion and energy losses of cosmic ray electrons at scales of
$\sim$ 1 kpc.

\end{abstract}


\section{Introduction}
\label{section1}

\begin{table*}
 \centering
  \caption{The sample galaxies observed with the GMRT at 333 MHz.}
   \begin{tabular}{@{}lcccccccc@{}}
  \hline
     Name  & RA           & Dec			       &Morphological	  &Angular	     & $i$& Distance                &Spatial & Other data \\
           &		  &			       &type	  & size (D$_{25}$)($^\prime$) & ($^\circ$) & (Mpc) 			& resolution (pc/$\arcsec$) & available$^\P$\\
      (1)     &	(2)	  &(3)			       &(4)	  & (5) & (6) & (7) 			& (8) & (9)\\
 \hline
 NGC 1097  & 02h46m19.0s  & -30d16$^\prime$30$\arcsec$ & SBbc     & 9.3$\times$6.3  & 45 & 14.5$^\dagger$  	& 70.3  & H$\alpha$, $\lambda$70$\mu$m, $\lambda$160$\mu$m,\\
	       &	      &				   &	      &	&		  &			&	&  $\lambda$20cm\\
 NGC 3034$^*$  &  09h55m52.7s & +69d40$^\prime$46$\arcsec$ & IO       & 11.2$\times$4.3  & 75 & 3.97$^\dagger$ 		& 21    & H$\alpha$, $\lambda$20cm                    \\
(M82)	       &	      &				   &	      &		&	  &			&	&  \\
 NGC 4736  &  12h50m53.0s & +41d07$^\prime$14$\arcsec$ & SAab     & 11.2$\times$9.1  & 41 & 4.66$^1$ 		& 22.3  & H$\alpha$,$\lambda$70$\mu$m, $\lambda$160$\mu$m,\\
(M94)	       &	      &				   &	      &		&	  &			&	&  $\lambda$20cm\\
 NGC 5055  &  13h15m49.3s & +42d01$^\prime$45$\arcsec$ & SAbc     & 12.6$\times$7.2  & 59 & 9.2$^\dagger$ 		& 44.6  & H$\alpha$, $\lambda$70$\mu$m, $\lambda$160$\mu$m,\\
(M63)	       &	      &				   &	      &		&	  &			&	& $\lambda$18cm \\
 NGC 5236  &  13h37m00.9s & -29d51$^\prime$56$\arcsec$ & SABc	  & 11.2$\times$11      & 24 & 4.51$^2$ 		& 22    &  H$\alpha$, $\lambda$70$\mu$m, $\lambda$160$\mu$m\\
(M83)	       &	      &				   &	      &		&	  &			&	&  $\lambda$20cm\\
 NGC 6946  &  20h34m52.3s & +60d09$^\prime$14$\arcsec$ & SABcd    & 11.5$\times$9.8  & 33 & 6.8$^3$  		& 33    & H$\alpha$, $\lambda$70$\mu$m, $\lambda$160$\mu$m,\\
	       &	      &				   &	      &		&	  &			&	&  $\lambda$6cm, $\lambda$20cm\\
\hline 
\end{tabular}
 
In column (5)  D$_{25}$ refers to the optical
diameter measured at the 25 magnitude arcsec$^{-2}$ contour from \cite{vauco91}.
Column (6) gives the inclination angle ($i$) defined such that $0^\circ$ is face-on. Distances in column (7) are taken from: $^1$ \cite{karac03}, $^2$ \cite{karac02}, $^3$ \cite{karac00} 
and the NED $^\dagger$.Column (9) lists available ancillary data that have been used in this paper.
$^\P$ The wavelengths represents the waveband in which other data were available/used (see Section \ref{section2} for exact frequencies).
$^*$ Thermal-nonthermal emission separation was not done for this galaxy (see text for details).
\label{table1}
\end{table*}

The radio continuum emission from normal galaxies originates from two
emission processes: thermal free-free emission from \hii regions
predominantly seen at recent star formation sites, and the nonthermal
synchrotron emission (hereafter referred to as nonthermal emission)
due to acceleration of cosmic ray electrons (CRe) in the ambient
galactic magnetic field. A typical galactic magnetic field strength of
few microgauss (1 -- 15 $\mu$G) and CRe in the energy range 1 to 10
GeV give rise to radio nonthermal emission between 0.1--10 GHz.
However, only at frequencies below 1 GHz, does the thermal fraction of
the emission reduce significantly, and hence low frequency studies are
direct probes  of nonthermal emission in galaxies.

It is now abundantly clear that CRe in normal galaxies are accelerated
in supernova remnant (SNR) shock fronts (with typical linear size $<$ 1 pc)
of Type II and Type Ib supernovae produced by short lived ($< 10^7$ yr) massive
stars. Subsequently, the CRe diffuse away from their sites of origin
and lose their energy primarily via synchrotron and inverse-Compton
radiation in about 10$^8$ yrs (see \citealt{condo92} for a review).
In turbulent magnetic fields, such as those encountered in the spiral
arms, the CRe diffusion speed is the Alfv\'{e}n speed ($\sim$ 100 km/s), and hence the
CRe can expand to a volume of radius $\sim$ 1 kpc without loosing much
energy, thus increasing the observed extent of the nonthermal emitting
source.  Note that the CRe would occupy a larger  volume
when observed at lower frequencies, since lower energy electrons suffer
lesser energy loss and hence can diffuse farther. The CRe generated at
the shock front have steep electron energy spectra which results in
a nonthermal synchrotron spectrum, well represented by a power law,
$S_{\rm \nu,nt} \propto \nu^{\ant}$, where $S_{\nu,\rm nt}$ is the 
nonthermal radio flux density at a
frequency $\nu$ and $\ant$ is the nonthermal spectral index. The
effects of CRe generation and propagation will cause
$\ant$ to vary from point to point in the galaxy.  At the acceleration
sites one expects $\ant \sim -0.5 \rm ~{to}~ -0.7$ \citep{bell78,bogda83,bierm93}, 
getting steeper due to various losses as the CRe diffuse away.

To confirm the aforementioned scenarios a host of studies have been
done in the past (see e.g. \citealt{kruit77, segal77a, segal77b,
  beck82, klein82, klein83, caril92, nikla97a, berkh03, beck05,
  murgi05, palad09}).  However, due to the presence of thermal
free-free emission, the $\ant$ is contaminated, making the measured value 
flatter than the true value. This is because the thermal emission has a spectral 
index of $-0.1$, while the nonthermal emission is significantly steeper 
($\ant< -0.5$)\footnote{Note that we use the
parameter $\alpha$ for the total spectral index, 
where $S_{\nu,\rm tot} \propto \nu^{\alpha}$, and $S_{\nu,\rm tot}$ 
is the total (thermal$+$nonthermal) flux density.}. At
frequencies below 1 GHz, the nonthermal emission is expected to
dominate.  \cite{nikla97} have shown for a sample of 74 galaxies,
about 8\% of the emission at 1 GHz is thermal in origin. At even lower
frequencies, the contribution of thermal fraction reduces
significantly, although there can be considerable local variation
between, e.g, arms, interarms and giant \hii regions etc., so as to
make the separation of the thermal emission important.  Previous
studies at low frequencies probed linear scales larger than 1 kpc, and
hence any small scale structures were lost.  These considerations
emphasize the importance of high resolution, low frequency
observations.

In this paper we report 333 MHz interferometric observation of
six nearby galaxies, NGC 1097, NGC 3034, NGC 4736, NGC 5055, NGC 5236
and NGC 6946, using the Giant Meterwave Radio Telescope (GMRT).  The
galaxies have angular size $\sim10\arcmin$  in the optical and the GMRT
observations have a resolution better than 20$\arcsec$, probing linear
scales of 0.4--1 kpc. The typical noise in the
maps is $\sim 300~\mu$Jy/beam (except for NGC 3034), making them 10
times more sensitive than previous studies at similar frequency of
other nearby galaxies \citep{palad09,heese09,sukum87}.  All our maps
have a dynamic range $>1000$, which allows us to reliably determine
the flux densities in regions near strong emission sites, mainly the
nuclear region.  We also use the technique developed by
\cite{tabat07}, using archival H$\alpha$ maps, for spatially
separating thermal and nonthermal emission as discussed in Section
\ref{section4}. However, we emphasize that in using this robust
technique, the resolution is limited by much coarser ($40\arcsec$) far
infrared maps (discussed in Section 4).  In Section~\ref{section2} we
define the sample and discuss the observation and analysis procedure
in Section 3. Section 4 discusses the procedure to estimate $\ant$.
Results on individual galaxies are presented in
Section~\ref{section5}.  We summarize the results in
Section~\ref{section6}. The thermal nonthermal emission separation
procedure using H$\alpha$ and infrared data is discussed in
Appendix~\ref{appendixa1}.

\begin{figure*}
 \includegraphics[width=7.2cm,height=7.2cm,angle=0]{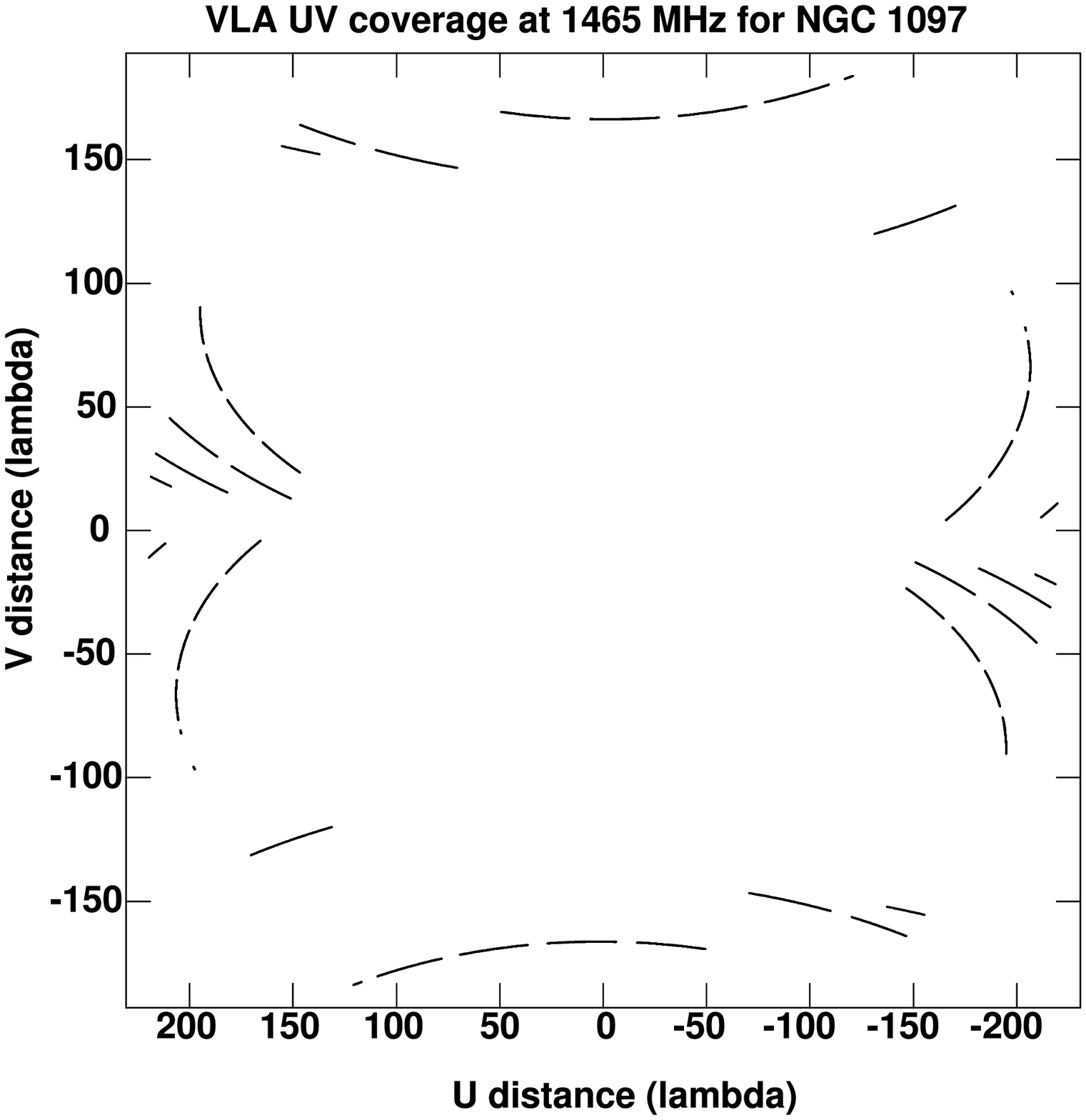}
 \includegraphics[width=7.2cm,height=7.2cm,angle=0]{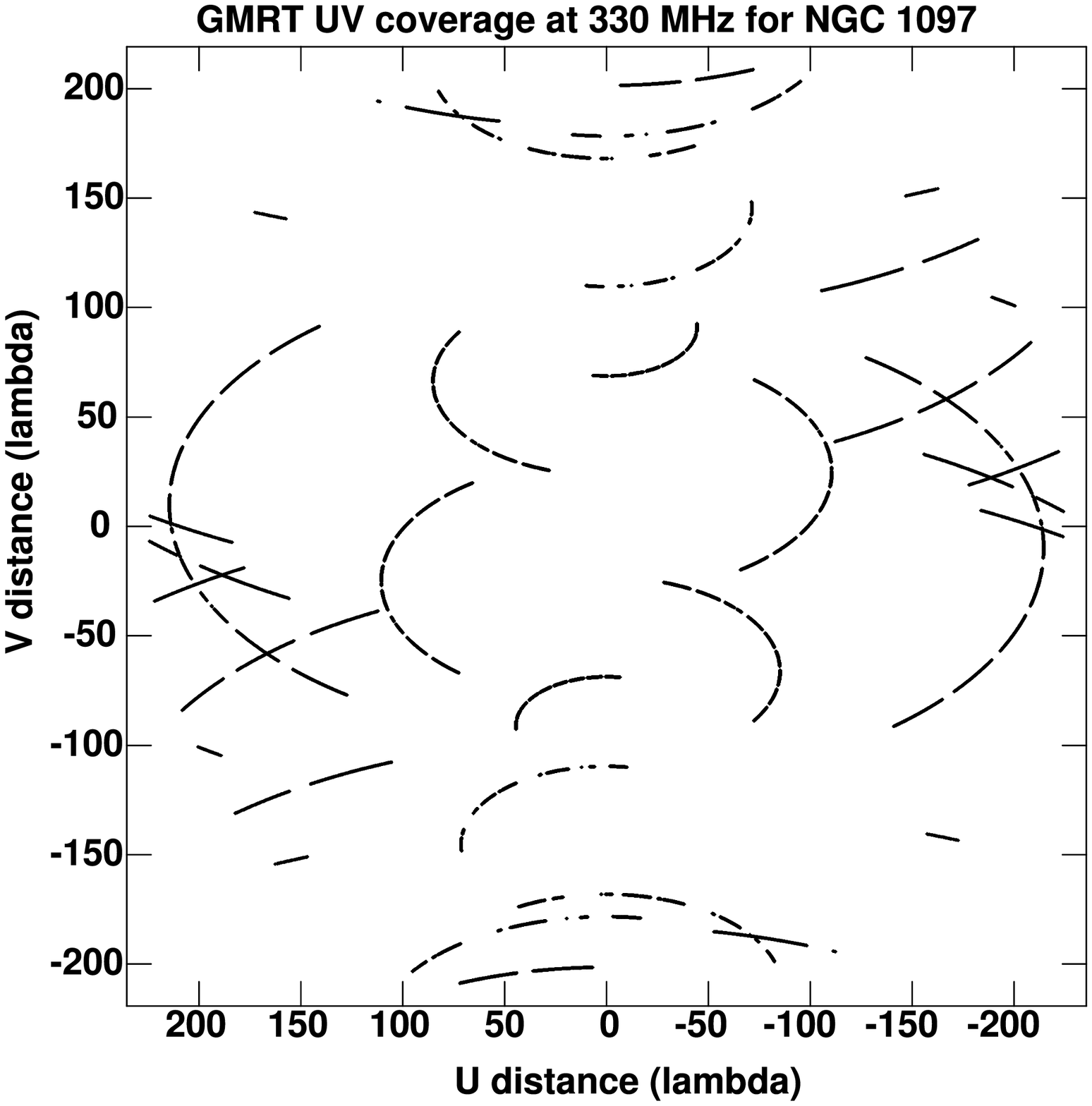}
 \caption{ {\sc uv} coverage in the range of {\sc uv} distance 0--220$\lambda$ for 
NGC 1097 at a declination of $\sim-30.25^\circ$. The left plot is for the VLA in CD array and the right
plot is for the GMRT. Sources of angular size above $\sim9\arcmin$ should be detected in this {\sc uv} range.
Note that the {\sc uv} coverage for the VLA is sparse resulting in incomplete sampling of the 
{\sc uv} plane, which leads to missing flux density  problems above this angular scale. 
However, for the GMRT, several complete {\sc uv} tracks are present.}
\label{uvcov}
\end{figure*}

\section{Sample definition}
\label{section2}

Our sample (see Table~\ref{table1}) includes six nearby spiral
galaxies, with clearly demarcated arm and interarm regions as seen
both in the optical and radio continuum. The optical size of these
galaxies are typically 9$\arcmin$--12$\arcmin$ and the GMRT
synthesized beam (typically $\sim15\arcsec$) at 333 MHz can resolve
the arm interarm regions.

All the galaxies have 1.4 GHz integrated continuum flux
density greater than 200 mJy (as measured in
NVSS, \citealt{condo98}, data), which is necessary to ensure detection
of low surface brightness diffuse emission across the galaxy. High resolution infrared and
H$\alpha$ data (see Table~\ref{table1}) are also used.  The infrared data were all taken from the {\it Spitzer Infrared Nearby Galaxy Survey}
({\sc sings}) \citep{kenni03} obtained using {\it Multiband
Imaging Photometer} ({\sc mips}, \citealt{rieke04}) on the {\it Spitzer} Space Telescope at
$\lambda$70 $\mu$m and $\lambda$160$\mu$m, while the H$\alpha$ images
were obtained from the ancillary datasets of the {\sc sings} and other
public datasets, as described in Appendix~\ref{appendixa1}.  For
five galaxies the data could be used for thermal nonthermal
separation, however in the case of NGC 3034, the $\lambda$160$\mu$m
{\sc mips} image was not usable due to nonlinearity and streaking effects.
Finally, a higher frequency, near 1.4 GHz, archival continuum
interferometric data/maps were used to find the spectral index with
333 MHz (see Table~\ref{table1}).    

An interferometer with the shortest
  baseline $D_{\rm min}$, can detect all the flux from angular scales
  less than $\sim 0.6\lambda/D_{\rm min}$, provided the {\sc uv}-plane
  is densely sampled at the shortest spacings.  The galaxies in our
  sample have optical angular size less than $\sim12\arcmin$.  For
  face-on galaxies, the optical and radio continuum sizes are
  comparable, however, for highly inclined galaxies, the sizes may
  differ slightly.
In GMRT  $D_{\rm min}\sim$100 meters,
at 333 MHz allows us to detect all the flux from angular scales less than
$\sim$18${\arcmin}$. Given that all the galaxies in our sample are
less than $\sim$12$\arcmin$, the {\sc uv}-plane is 
well sampled (see Figure~\ref{uvcov}) and hence we do not expect any
{\it missing flux density}\footnote{The extent of missing flux density with 
our GMRT data  was tested as
follows: we took an existing archival 1.465 GHz map of NGC 6946,
which was made by combining interferometric data from the Very Large
Array (VLA) using C and D array \citep{beck07}. This galaxy is $\sim$11$\arcmin$ in
size.  The map was then sampled to the 333 MHz GMRT UV coverage and
subsequently imaged. All the flux in the source ($>99$\%) could be
recovered.}. This is an issue for the archival higher frequency
data which is discussed below along with a short description of the sample galaxies.

{\bf NGC 1097}, in the optical is a spectacular barred spiral galaxy
with prominent dust lanes in the bar. The central bar extends for
about 20 kpc and then continues into the two optical spiral arms.
The galaxy has an active Seyfert nucleus, and a circumnuclear ring of
about 1.5 kpc diameter. Amongst the major radio continuum studies of
this galaxy, \cite{ondre83} reported radio emission at 1465 MHz
coincident with the narrow dust lanes in the bar. Recently
\cite{beck05} obtained a spectral index map for this galaxy from
observations at 4.8 and 8.4 MHz, which revealed relatively steep
spectral index of $-1$ in the inner ridge of the bar, and becoming
shallower to $-0.7$ as one moves to the outer ridge.  

To find the spectral index with GMRT 333 MHz, we analyzed the archival
1465 MHz data observed using CD array from the VLA\footnote{The Very
Large Array (VLA) is operated by the NRAO. The NRAO is a facility of
the National Science Foundation operated under cooperative agreement
by Associated Universities, Inc.} (project code: AW237).  The map
has an angular resolution of $40\arcsec \times 40\arcsec$, and the
shortest measured baseline is $\sim$150$\lambda$, which can detect
angular scales up to $\sim$14$\arcmin$.  However, the {\sc uv}
coverage is sparse for angular scales above 8$\arcmin$ (see
Figure~\ref{uvcov}). Since the size of the galaxy is about 9$\arcmin$,
we expect missing flux density at large angular scales in these data.

{\bf NGC 3034 (M82)} is a prototypical, starburst, edge-on galaxy.
The galaxy was initially classified as irregular, however recently
\cite{mayya05}, have reported the discovery of two symmetric spiral
arms in the near infra-red. The galaxy is bright in the radio
continuum, and has a synchrotron emitting halo extending up to a
diameter of about 8 kpc. Based on a low frequency spectral index study
of the whole galaxy  from 330 MHz to 4835 MHz, \cite{seaqu91} found a
spectral index of $-0.4$ in the nuclear region, steepening to $-1$ at
a radius of 1 kpc due to transport of CRe from the disk by galactic winds.

We used the archival 1490 MHz VLA map (\citealt{condo87}, obtained
from the NED) at a resolution $60\arcsec \times 60\arcsec$ to obtain the spectral 
index with 333 MHz.
 The 1490 MHz observations were made with the VLA D configuration
with shortest baseline length of $\sim170\lambda$, and are sensitive to angular 
sizes less than $\sim12\arcmin$.  This should be sufficient to detect the galaxy, which
has an optical size ($\rm D_{25}$) of $\sim11\arcmin$.
However, due to very bright emission from the core, the high resolution radio 
continuum map is dynamic range limited.

{\bf NGC 4736 (M94)} is the nearest large spiral galaxy, which has
double ringed morphology. The inner ring is mainly made of recent star forming
\hii regions, showing bremsstrahlung spectra \citep{duric88a} and 
is distinctly visible in the H$\alpha$.
The outer ring however is a low
surface brightness feature in the optical. The rings were suggested to be caused by
inner and outer Lindblad resonances \citep{schom76}. The radio continuum study
by \cite{bruyn77} between  610 MHz and 1415 MHz, reported spectral index maps
with the spectral index being $\sim-0.5$ towards the center of the galaxy
and steepening towards the outer parts of the galaxy.  

 We used the archival WSRT\footnote{The Westerbork
Synthesis Radio Telescope (WSRT) is operated by the Netherlands
Foundation for Research in Astronomy (NFRA) with financial support
from the Netherlands Organization for scientific research (NWO).}
1374.5 MHz map taken from the Westerbork {\sc sings} sample \citep{braun07}
at a resolution of $19\arcsec \times 12.5\arcsec$ after smoothing
to $20\arcsec\times20\arcsec$ to obtain the spectral index.
The observations was made using the ``maxi-short'' array configuration 
of the WSRT where the shortest east-west baseline was 
$\sim170\lambda$, $250\lambda$ and $330\lambda$ (36, 54
and 72 meters respectively), and are sensitive to angular scales less 
than $\sim12\arcmin$. NGC 4736, having a angular size of $\sim6\arcmin$ in the
radio, has good sampling of the {\sc uv}-plane
and thus this map is not affected by lack of short spacing measurements.

{\bf NGC 5055 (M63)} is a nearby flocculent spiral which has short
multiple arms. Although this galaxy lacks organized spiral arms in the
optical, polarization observation at 10.55 GHz shows regular, spiral
magnetic fields with radial component due to dynamo action
\citep{knapi00}.  Radio continuum spectral index maps between  610
  MHz and 1417 MHz by \cite{humme82} showed spectral index of about
$-0.6$ in the central regions of the galaxy, which steepens to $-1$
towards the outer parts. The galaxy appears featureless in the above
study as well as in the high frequency 10.7 GHz map by \cite{klein81}.

The archival $18.5\arcsec\times12.5\arcsec$ 1696 MHz map from the
Westerbork {\sc sings} \citep{braun07}, after smoothing to $20\arcsec \times 20\arcsec$, was used 
for obtaining the spectral index with 333 MHz.
The ``maxi-short'' configuration of the WSRT at 1695 MHz allows us to image angular scales 
up to 10$\arcmin$. However, the {\sc uv} coverage is too sparse to detect flux from
structures above $\sim7\arcmin$. NGC 5055 has an angular extent $\gsim8\arcmin$,
and thus we believe this map has missing flux density.

{\bf NGC 5236 (M83)} is a large barred spiral galaxy and is seen
almost face-on.  The central bar extends up to 3$\arcmin$ ($\sim4$ kpc), showing
clear dust lanes. The galaxy is bright in radio continuum and the
central bar and the disk is in the 1465 and 4885 MHz
observation of \cite{ondre85}. \cite{sukum87} found
the total and nonthermal integrated spectral index to be
$-0.75$ and $-0.8$ respectively, obtained between 327, 1465 and 4750
MHz and a thermal fraction of 20\% at 4750 MHz. 

The archival data at 1452 MHz VLA CD array (project code: AS325) were
downloaded and analyzed to produce a $26\arcsec \times 14\arcsec$ map,
which was used for obtaining spectral index with our 333 MHz
observations.  The 1452 MHz dataset had shortest baseline of
$\sim150\lambda$ which is sensitive to a maximum angular size of
$\sim14\arcmin$.  The {\sc uv} coverage is sparse for structures above
$\sim10\arcmin$.  NGC 5236 is about $11\arcmin$ in size and hence
there may be a small amount of missing flux density in this
observation.

{\bf NGC 6946} is a large spiral galaxy with multiple arms.  The
galaxy has well separated arm and interarm regions and has been the
subject of several studies in the radio continuum.  There are several
star forming regions in the galaxy and one of them shows the first
likely detection of ``anomalous'' dust emission, due to spinning dust,
outside of the Milky Way \citep{murph10}.  Observations at 610 MHz,
1415 MHz and 5 GHz by \cite{kruit77} revealed a bright radio disk of
nonthermal origin. \cite{klein82}, based on observations at 10.7 GHz,
estimated a thermal fraction of $19\pm10$\%. The spectral index study
by \cite{beck07} between  1465 MHz and 8.5 GHz, clearly shows a
flatter spectral index of $-0.5$ in the arms which steepens to about
$-1$ in the interarm regions.

For obtaining the spectral index with 333 MHz, we used
an archival VLA $15\arcsec \times 15\arcsec$ map at 1465 MHz.
The VLA map was made by combining interferometric data from
C configuration \citep{beck07} and D configuration \citep{beck91}. 
\cite{beck07} noted the integrated flux density from the VLA map and 
single dish Effelsberg measurement had similar values and there was no 
indication of missing large angular scale structures.

We emphasize that the GMRT 333 MHz images of the galaxies do not
suffer from missing flux density, while some
of the higher frequency maps (especially NGC 1097 and NGC 5055) may
have missing flux density from large angular size structures.  Thus, the true
spectral index, particularly in the outer parts of the galaxies, is flatter 
than determined in this study.

\section{Observations and analysis}
\label{section3}

\begin{table*}
 \centering
\scriptsize
  \caption{Observational summary of 333 MHz observations of our sample galaxies. Time spent on source excludes calibration overheads. Column (5) gives the
synthesized beam achieved and column (6) gives the map rms ($\rm \sigma_{map}$). Column (7) gives the integrated flux of the source
inside the 4$\sigma_{map}$ contour.}
   \begin{tabular}{@{}lccccccccr@{}}
  \hline
     Name & Obs. & Time on      & ~~~~Calibrators  & used~~~~~~~~ & Synthesized beam          &   $\sigma_{map}$         & Integrated Flux \\
          & date &  source (hrs)&                  &              & ($\arcsec \times \arcsec$)& ($\mu$Jy/beam) &   (Jy)          \\
         (1) & (2) & (3)& (4)                  &              & (5)& (6) &  (7)          \\
 \hline
 NGC 1097 & 29-Oct-09 & 6  &{\it Flux Cal:}   & 3C48 & 16$\times$11 & 300 &  2.0$\pm$0.14  \\
          &	      &    &		      & 3C286&                  &     &     \\
 	  &           &    &{\it Phase Cal:}  & 0116-208&               &     &     \\
          &	      &	   &	              & 0409-179&               &     &     \\
 NGC 3034 & 31-Oct-09 & 4.5&{\it Flux Cal:}   & 3C48    & 22$\times$15 & 3000 & 14$\pm$1\\
 (M82)	  &     &&{\it Phase Cal:}   & 0834+555 & & & 	  \\
          &	&	&	    & 1459+716    & & & \\
 NGC 4736 & 04-Jul-09&7  &{\it Flux Cal:}    & 3C286  & 13$\times$12 & 250 &  0.9$\pm$0.06 \\	
 (M94)	  &   &&{\it Phase Cal:}   & 3C286  & & &\\	
 NGC 5055 & 02-Nov-09 &3 &{\it Flux Cal:}    & 3C286 & 17$\times$10 & 240 &   2.3$\pm$0.2 \\	
 (M63)	  &   &&{\it Phase Cal:}   & 3C286  & & &	\\
 NGC 5236 & 31-Oct-09 &5 &{\it Flux Cal:}	  & 3C286& 16$\times$12 & 500 & 7.4$\pm$0.5    \\	
 (M83)	  &   &&{\it Phase Cal:}	  & 1311-222   & & & 	  \\
 NGC 6946 & 29-Jul-09  &6.5&{\it Flux Cal:}    & 3C286 & 12$\times$11 & 300 &   4.3$\pm$0.24 	\\
 	  &   &	&		  & 3C147    & & &	\\
 	  &   &&{\it Phase Cal:}   & 1459+716    & & &	  \\
	  &	&&		  & 2350+646     & & &\\
\hline 
\end{tabular}
\label{table2}
\end{table*}

We conducted interferometric observations at 333 MHz for the six
galaxies in our sample with the Giant Meterwave Radio Telescope (GMRT)
near Pune, India \citep{swaru91}. The data with 16 MHz bandwidth
(corresponding to $\nu_{\rm RF}$ of 325 to 341 MHz, and a center
$\nu_{\rm RF}$ of 333 MHz), divided into 128 channels, were recorded 
following the usual protocol
of observing flux and phase calibrator interlaced with observation on
the source. At the beginning and end of the observing run, one of the flux
calibrators 3C48, 3C286 or 3C147 was observed for about $\sim$
10$-$20 minutes. At the beginning and end of the observing run, one of the flux
calibrators 3C48, 3C286 or 3C147 was observed for about $\sim$10$-$20 
minutes. Phase calibrators were observed every $\sim$30$-$35 
minutes for $\sim$4$-$5 minutes. Calibrators chosen for each of 
the observed sources are listed in Table~\ref{table2}.

Data reduction was done using the Astronomical Image
Processing System ({\sc aips}) following standard procedure. 
After editing the data for strong {\it radio frequency interference} (RFI),
standard flux and phase calibration were applied to the source.  
The \cite{baars77} absolute flux density scale was used to determine the
flux densities of the flux calibrators and then applied them to the phase 
calibrator and the source.

The task `{\sc flgit}' was
used to remove low level RFI at $4-6\sigma$ level across the frequency
channels. Even lower level RFI was subsequently removed using the tasks
`{\sc tvflg}', `{\sc wiper}' and/or `{\sc spflg}'.  The procedure was
iteratively done by obtaining the gain solutions after every RFI
removal stage. The final gain solution that was applied to the target
source when the closure errors were less than 1\% on the phase calibrators.  

The calibrated data were then used to obtain the deconvolved images
using the task `{\sc imagr}'. Prior to this, the data were condensed
in frequency by vector averaging 6 adjacent channels of 125 KHz each
(resulting in channel width of 750 KHz) using the task `{\sc
  splat}'. This ensured that the bandwidth smearing was less than the
size of one synthesized beam at 333 MHz.  To account for wide field
imaging with non-coplanar baselines, the technique of polyhedron
imaging was used in {\sc imagr} \citep{cornw92}, where the field of
view is subdivided into a number of smaller fields (facets). For our
purpose we used $7\times7 = 49$ facets covering the primary beam up to
the half power beam width (HPBW $\sim1.5^\circ$). Strong sources
outside the HPBW, if present, were included in additional facets.

\begin{figure*}
 \includegraphics[width=7.2cm,height=7.2cm,angle=0]{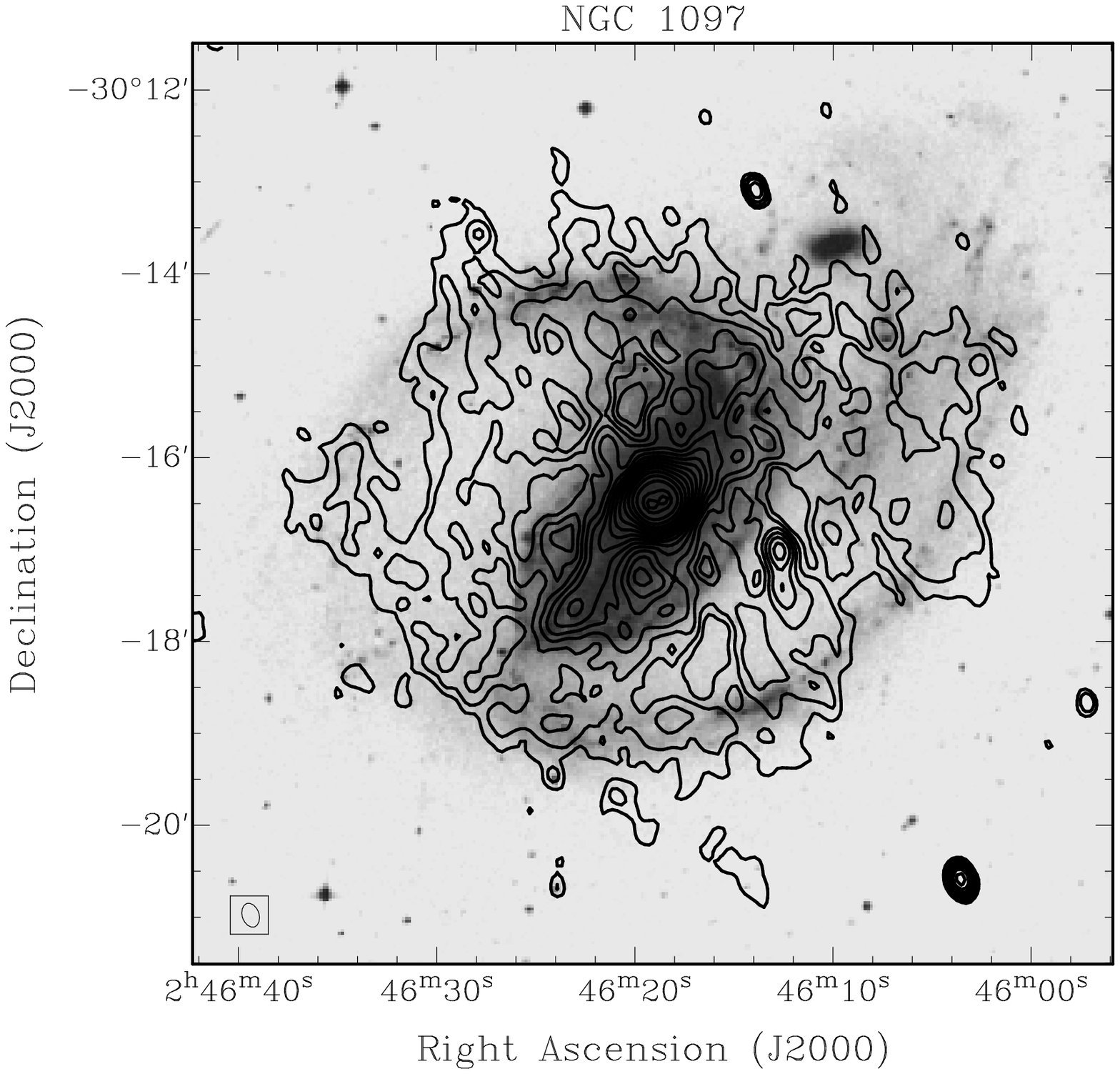}
 \includegraphics[width=7.2cm,height=7.2cm,angle=0]{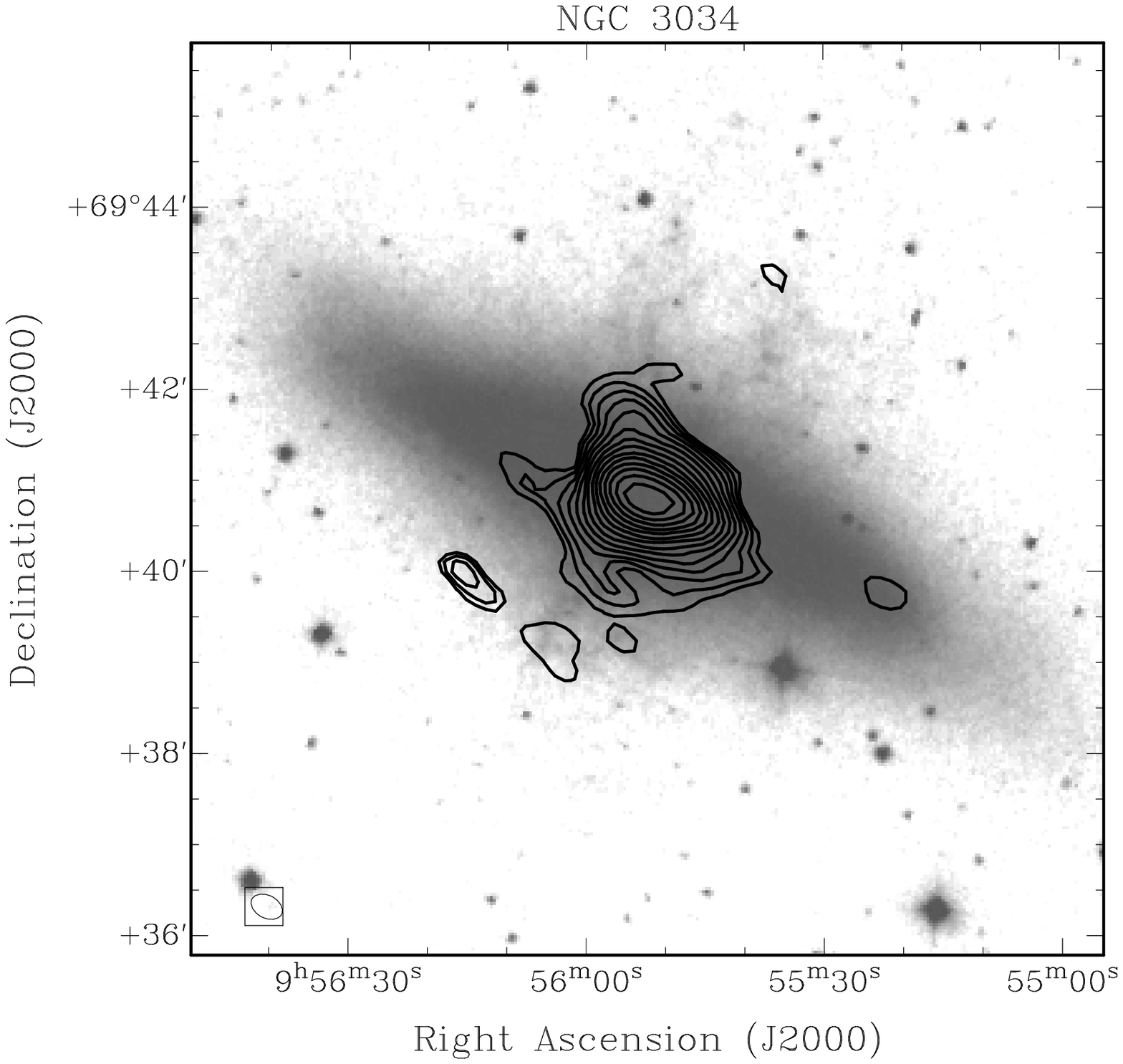}
 \includegraphics[width=7.2cm,height=7.2cm,angle=0]{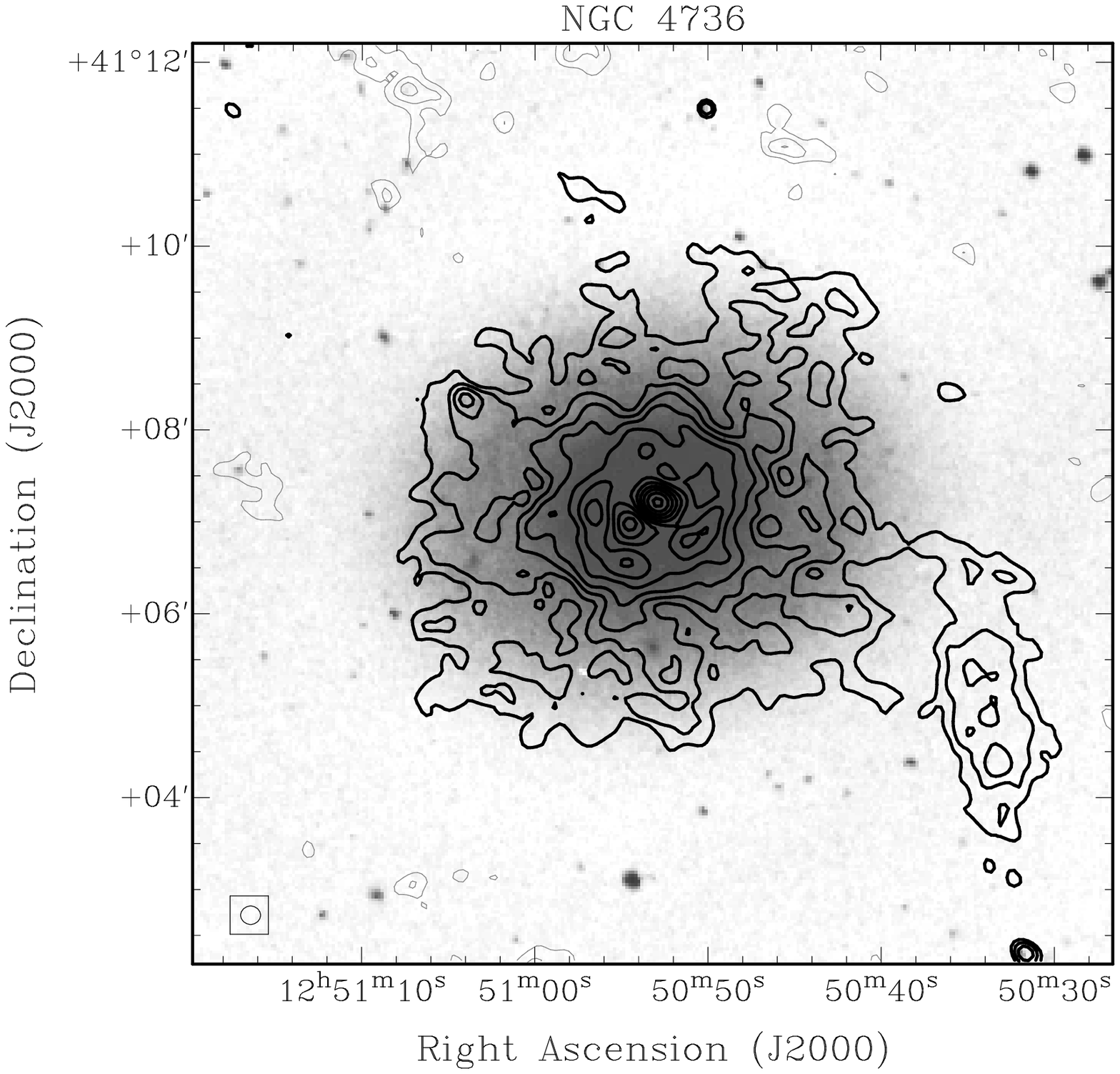}
 \includegraphics[width=7.2cm,height=7.2cm,angle=0]{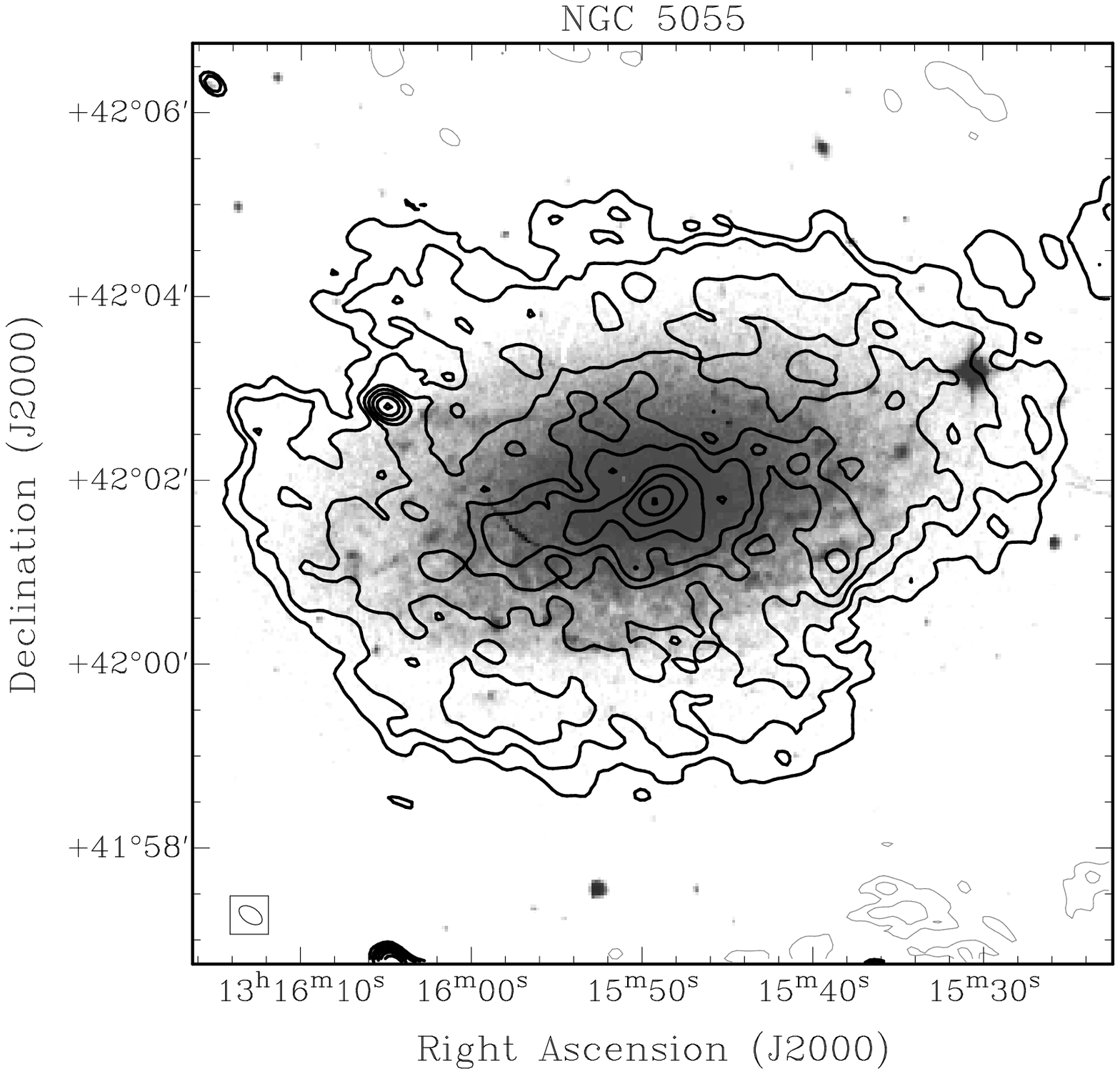}
 \includegraphics[width=7.2cm,height=7.2cm,angle=0]{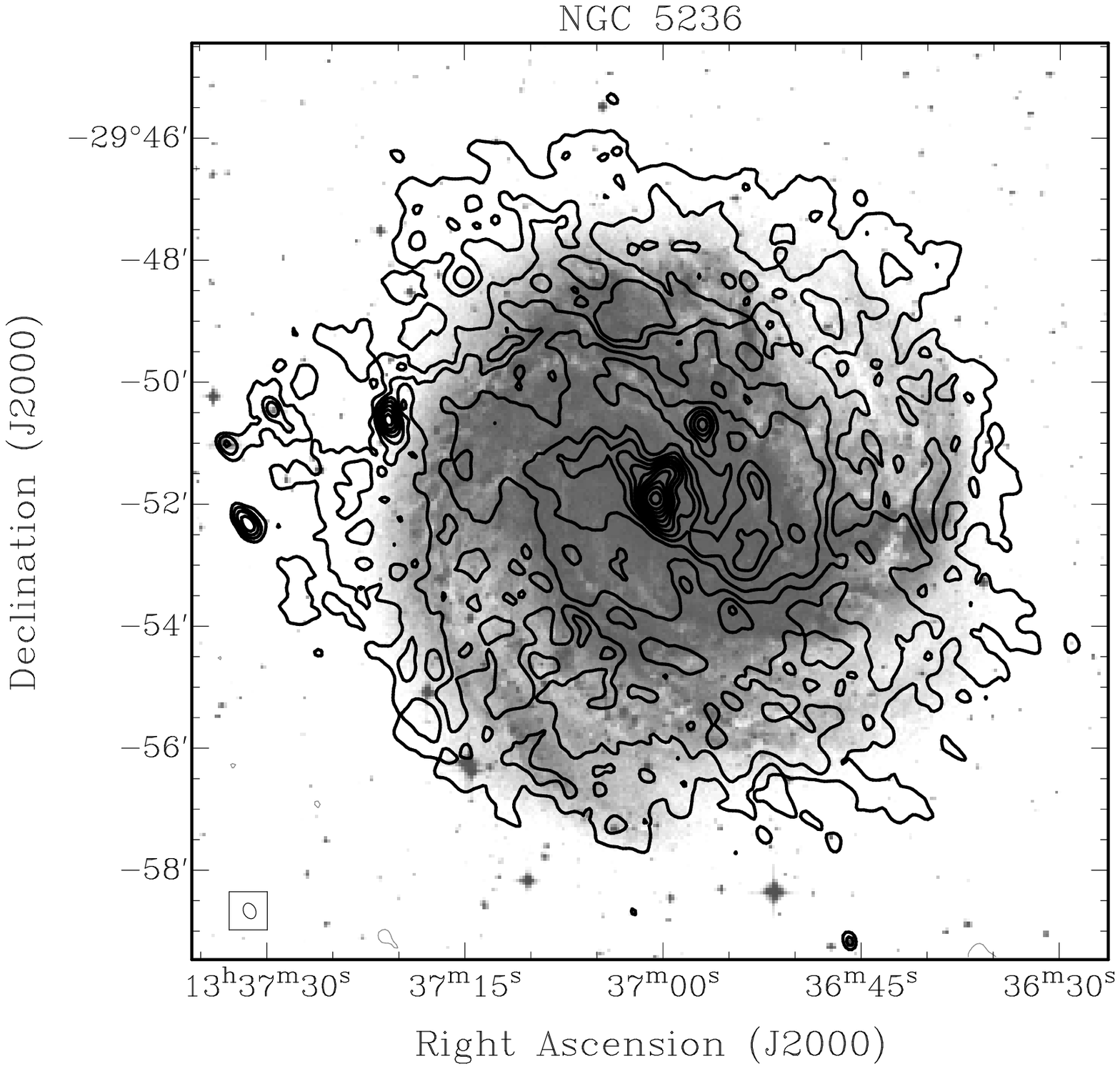}
 \includegraphics[width=7.2cm,height=7.2cm,angle=0]{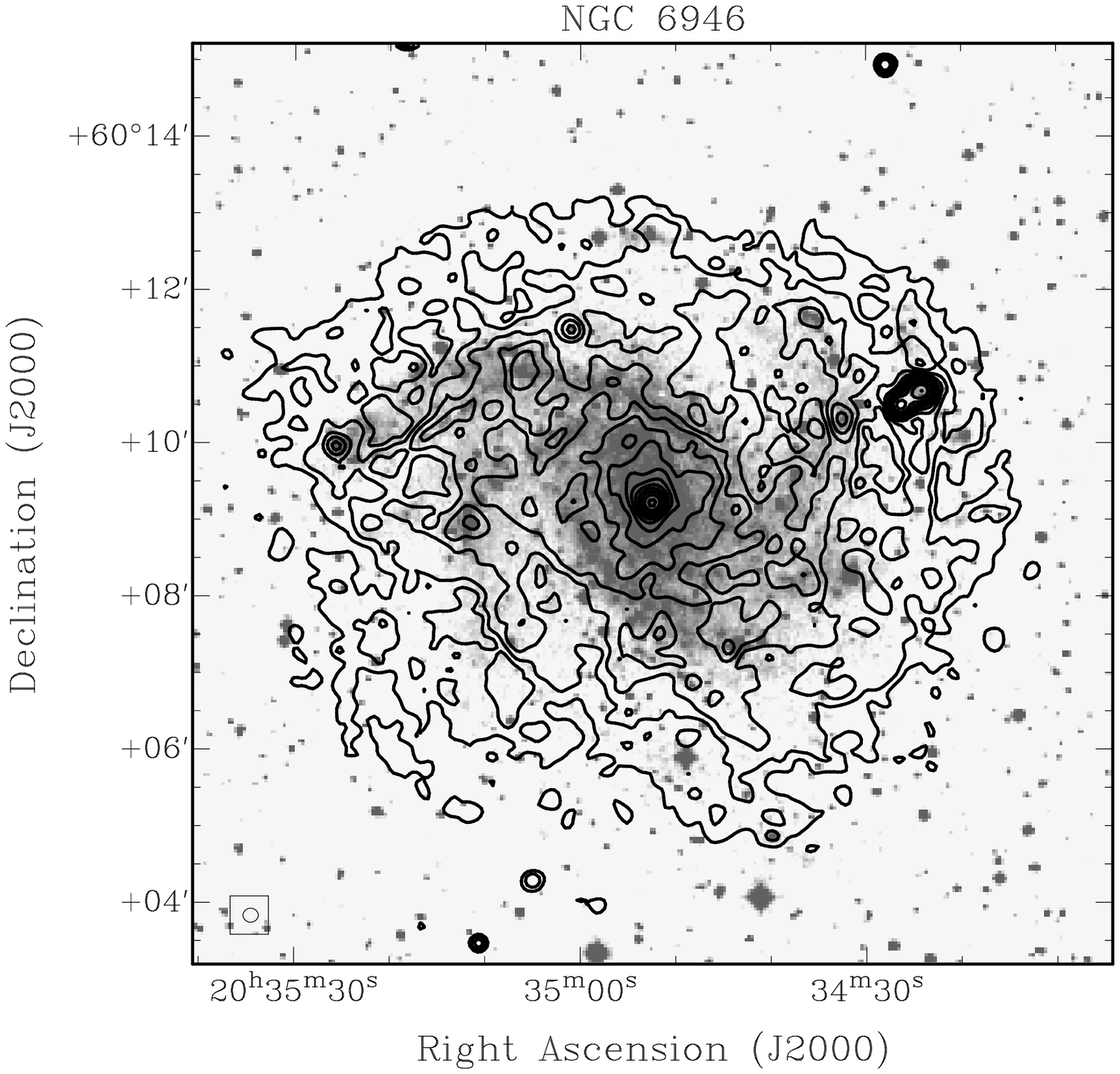}
 \caption{The contour maps are GMRT 333 MHz observations of the sample galaxies which are
overlaid on the optical DSS image in greyscale.
The contour levels starts from 4$\sigma$, increasing in multiples of
$\sqrt{2}$. The grey contours shows $(-2,-3,-4)\times\sigma$.}
\label{figure1}
\end{figure*}

Several rounds of {\it phase only} self-calibration were done
iteratively, by choosing point sources such that, the flux density within
one synthesized beam is more than 8$\sigma$.
Any stripe present in
the map due to bad data, were removed using a Fourier transform
method\footnote{ see
  http://www.ncra.tifr.res.in/$\sim$aritra/stripe\_removal.pdf}.  In
the last iteration, one round of {\it amplitude and phase} self
calibration was done. Final maps were made from full {\sc uv} coverage
and the {\sc uv} data were weighted using Briggs robust weighting of 0
\citep{brigg95}. To {\sc clean} the extended diffuse emission from the
galaxies we used {\sc sdi clean} algorithm \citep{steer84}. The final
full resolution images obtained are shown in Figure~\ref{figure1} and
their corresponding synthesized beams are listed in column 6 of
Table~\ref{table2}.

\begin{figure*}
\includegraphics[width=17cm,height=16cm]{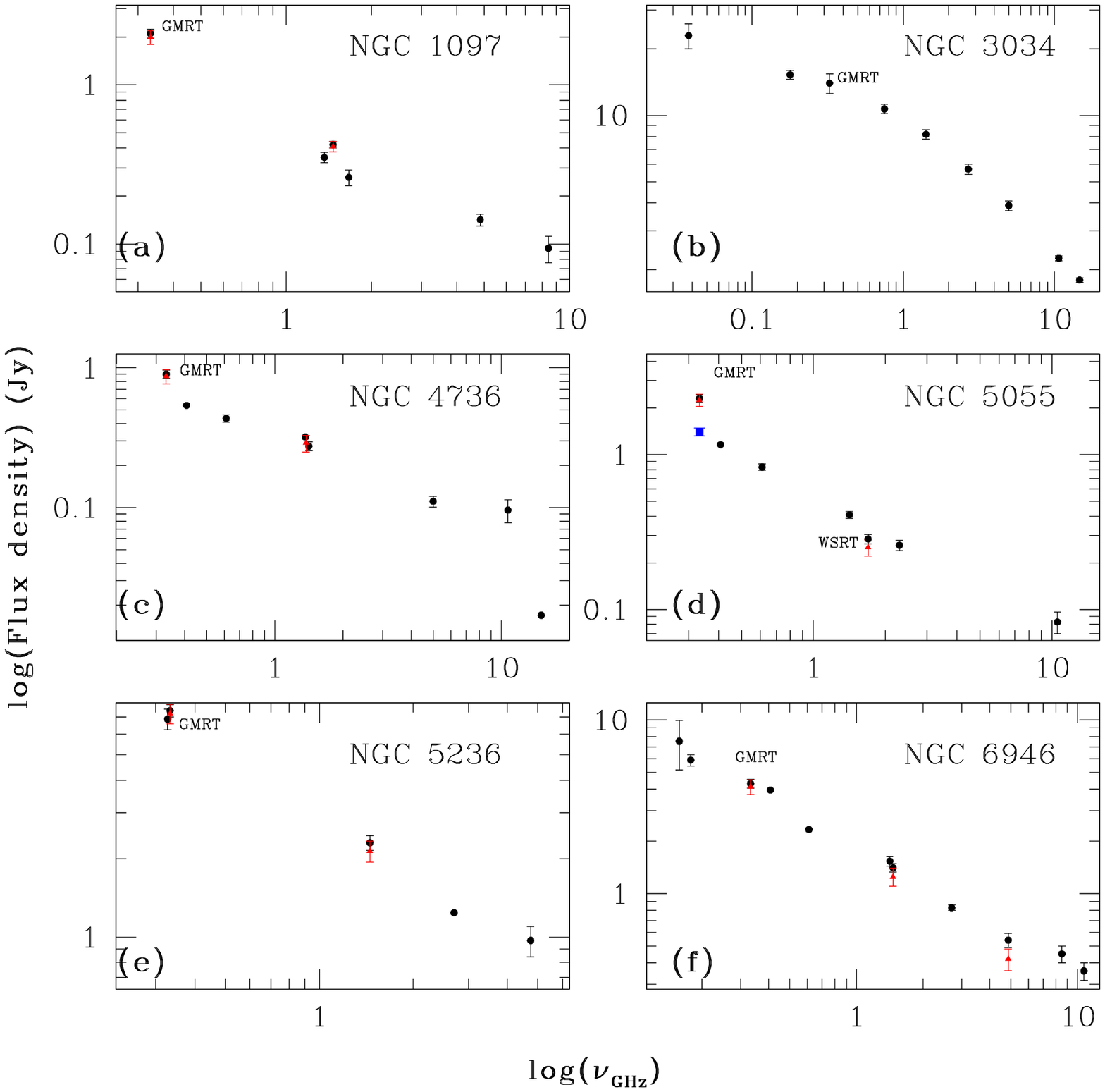}
\caption{Integrated flux density as a function of frequency. The black circles represent
the integrated flux densities without the thermal-nonthermal emission separation (see Table
\ref{table3}), while the red triangles represent the nonthermal flux densities of the
galaxies. The blue square for NGC 5055, is the flux density measured within $6.7\arcmin\times3.2\arcmin$
(see Section~\ref{section4}).}
\label{figure2}
\end{figure*}

The integrated flux density of the all the six galaxies were obtained by integrating within the
4$\sigma_{\rm map}$ (where $\sigma_{\rm map}$ is the map noise) contour
and is given in Table~\ref{table2}.
We compared the flux density obtained for each of our sample galaxies with measurements done at other radio
frequencies by other researchers, as given in
Table~\ref{table3}. Figure \ref{figure2} shows the integrated
broadband spectrum for our sample galaxies.  Our flux density measurements are
in good agreement with the interpolated flux densities from higher and lower
radio frequency observations reported in earlier studies.

The uncertainties in the estimated flux density ($S_{\rm source}$) depend on the
rms noise in the map as well as on errors associated with uncalibrated
system temperature ($T_{\rm sys}$) variations, which is about 5\% \citep{roy04}
at 333 MHz for the GMRT. The flux density error ($\delta S_{\rm source}$) can be
calculated as,
$$
\delta S_{\rm source} = \sqrt{\left(\frac{\delta T_{\rm sys}}{T_{\rm sys}}\times 
S_{\rm source}\right)^2 + \sigma_{\rm map}^2} 
$$
The flux density also has an systematic 
error of about 8\% associated with the absolute flux scale error
at 333 MHz.

\begin{table}
 \centering
  \caption{Multifrequency integrated flux density for our sample galaxies.}
\scriptsize
   \begin{tabular}{@{}lcccr@{}}
  \hline
     Source     & Frequency &  Flux density & Ref.\\
                &  (GHz)    & (Jy)   &\\
 \hline
 NGC 1097  &   0.333     & 2.1$\pm$0.14    & This paper\\
           &   1.365    & 0.350$\pm$0.025 & 1  \\
           &   1.465    & 0.42$\pm$0.02   & VLA CD array  \\
           &   1.665    & 0.262$\pm$0.030 & 1  \\
           &   4.850    & 0.142$\pm$0.012 & 1  \\
           &   8.450    & 0.094$\pm$0.018 & 1  \\
\hline
 NGC 3034  &   0.038      & 23$\pm$3   & 2\\
           &   0.178     & 14.6$\pm$0.7 & 2  \\
           &   0.333     & 14$\pm$1    & This paper  \\
           &   0.750     & 10.7$\pm$0.5 & 2  \\
           &   1.415    & 8.0$\pm$0.4 & 3  \\
           &   2.695    & 5.7$\pm$0.3 & 2  \\
           &   5.0    & 3.9$\pm$0.2 & 2  \\
           &   10.7   & 2.25$\pm$0.06 & 4  \\
           &   14.7   & 1.79$\pm$0.04 & 4  \\
\hline
 NGC 4736  &   0.333     & 0.9$\pm$0.06   & This paper\\
           &   0.408    & 0.539$\pm-$ & 5  \\
           &   0.610    & 0.435$\pm$0.026 & 7  \\
           &   1.365    & 0.320$\pm$0.01 & 6  \\
           &   1.415    & 0.276$\pm$0.02 & 7  \\
           &   4.995    & 0.111$\pm$0.01 & 7  \\
           &   10.7    & 0.096$\pm$0.018 & 8  \\
           &   15     & 0.017$\pm-$ & 9  \\
\hline
 NGC 5055  &   0.333     & 2.3$\pm$0.13   & This paper\\
           &   0.408     & 1.16$\pm$0.022    & 5 \\
           &   0.610    & 0.83$\pm$0.04 & 10  \\
           &   0.750    & 0.88$\pm$- & 11  \\
           &   1.417    & 0.409$\pm$0.02 & 10  \\
           &   1.696    & 0.285$\pm$0.02 & WSRT \\
           &   2.295    & 0.26$\pm$0.02 & 12  \\
           &   3.7      & 0.254$\pm-$ & 13  \\
           &   10.7     & 0.117$\pm$0.026 & 14  \\
\hline
 NGC5236   &   0.327    & 6.86$\pm$0.62 & 15  \\
	   &   0.333     & 7.3$\pm$0.4   & This paper\\
           &   1.452    & 2.3$\pm$0.15 & VLA CD array  \\
           &   1.465    & 2.2$\pm$0.11 & 15  \\
           &   2.7    & 1.24$\pm-$ & 16  \\
           &   4.75    & 0.97$\pm$0.13 & 15  \\
\hline
 NGC 6946 & 0.158  &    7.54$^{+2.40}_{-3.54}$  & 17  \\
          & 0.178  &     5.88$\pm$0.44               & 18  \\
          & 0.333  &     4.e$\pm$0.24       & This paper \\
          & 0.408  &     3.95                        & 5  \\
          & 0.610  &     2.34                        & 19 \\
          & 1.415  &     1.54$\pm$0.10       & 19 \\
          & 1.465  &     1.41$\pm$0.08       & VLA C$+$D array \\
          & 2.695  &    0.83$\pm$0.03           & 20   \\
          & 4.86    &    0.54$\pm$0.05          & VLA$+$Effelsberg\\
%         & 5.0    &    0.625$\pm$0.05          & 8\\
          & 8.50   &   0.45$\pm$0.05         & 21\\
          & 10.70  &   0.359$\pm$0.042       & 8\\
\hline
\end{tabular}

$^1$\cite{beck02},
$^2$\cite{kelle69},
$^3$\cite{humme80},
$^4$\cite{klein88},
$^5$\cite{gioia80},
$^6$\cite{braun07},
$^7$\cite{bruyn77},
$^8$\cite{klein81}, 
$^9$\cite{nagar05},
$^{10}$\cite{humme82},
$^{11}$\cite{jong66},
$^{12}$\cite{arp73},
$^{13}$\cite{kuril67},
$^{15}$\cite{sukum87},
$^{16}$\cite{wrigh90},
$^{17}$\cite{brown61},
$^{18}$\cite{caswe67},
$^{19}$\cite{kruit77},
$^{20}$\cite{jong67},
$^{21}$\cite{kuril70}.
\label{table3}
\end{table}

\begin{figure*}
\begin{center}
\begin{tabular}{|c|c|}
{\mbox{\includegraphics[width=7.5cm,height=7.4cm]{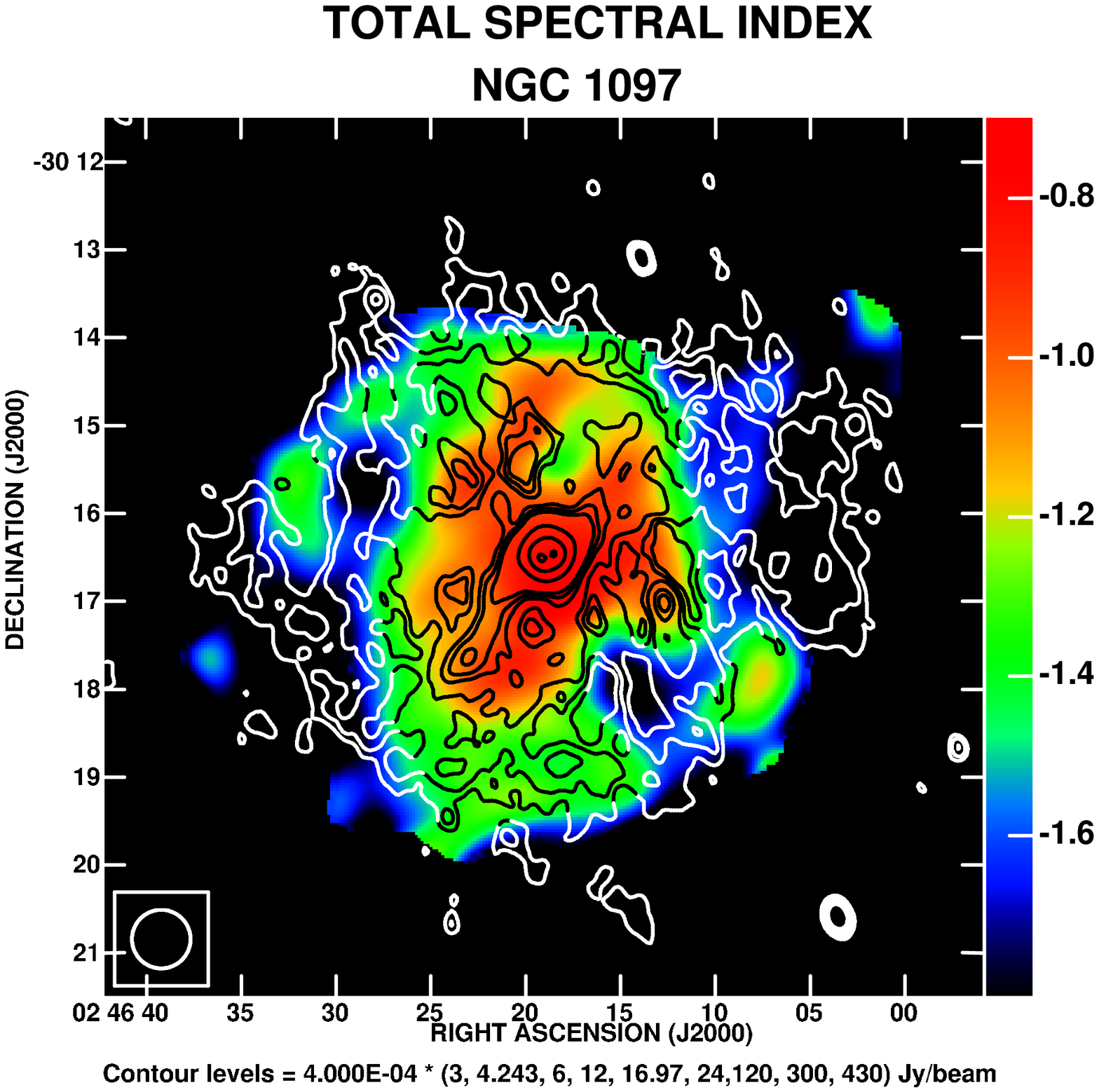}}} &
{\mbox{\includegraphics[width=7.5cm,height=7.4cm]{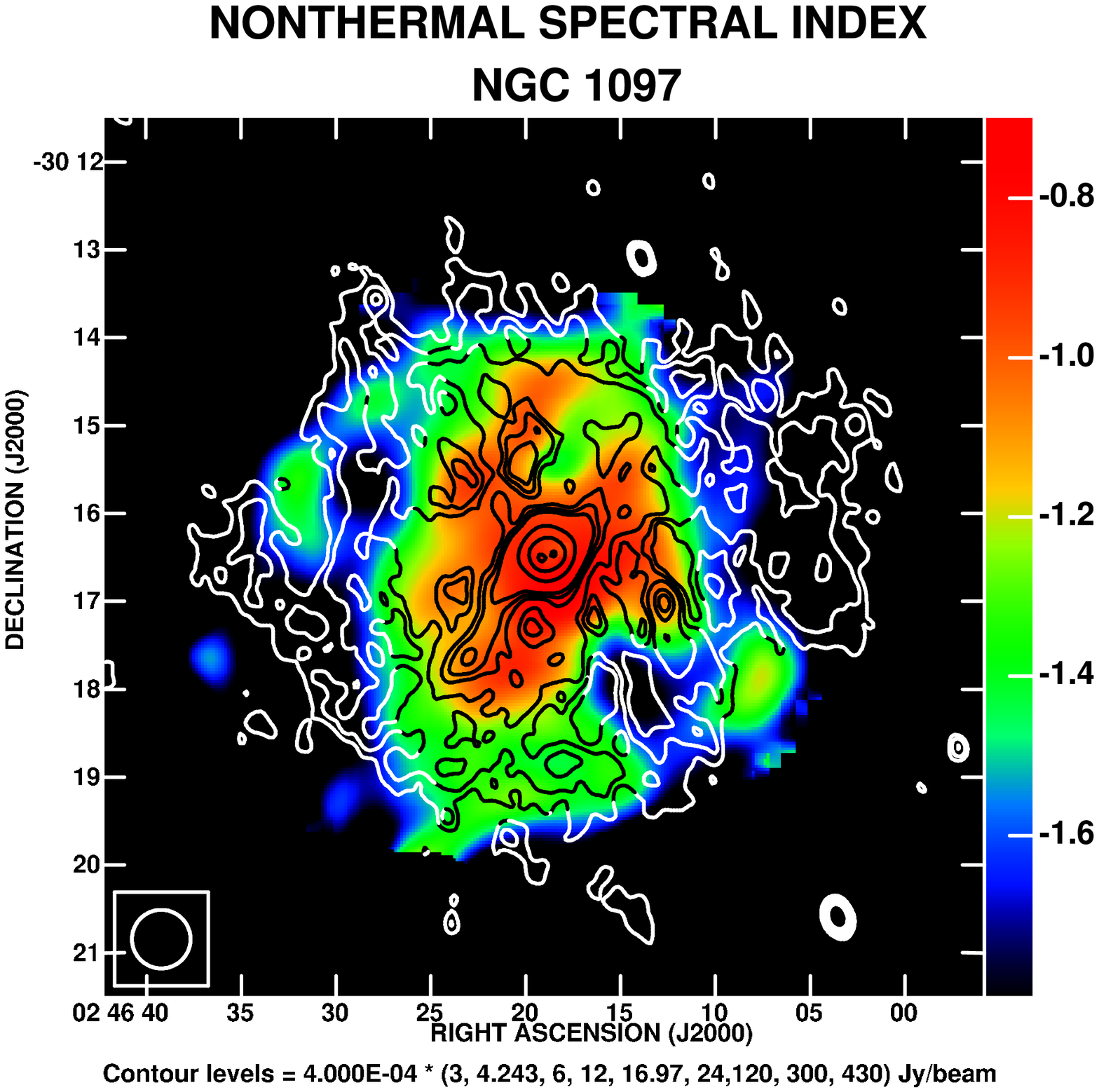}}} \\
{\mbox{\includegraphics[width=7.5cm,height=7.4cm]{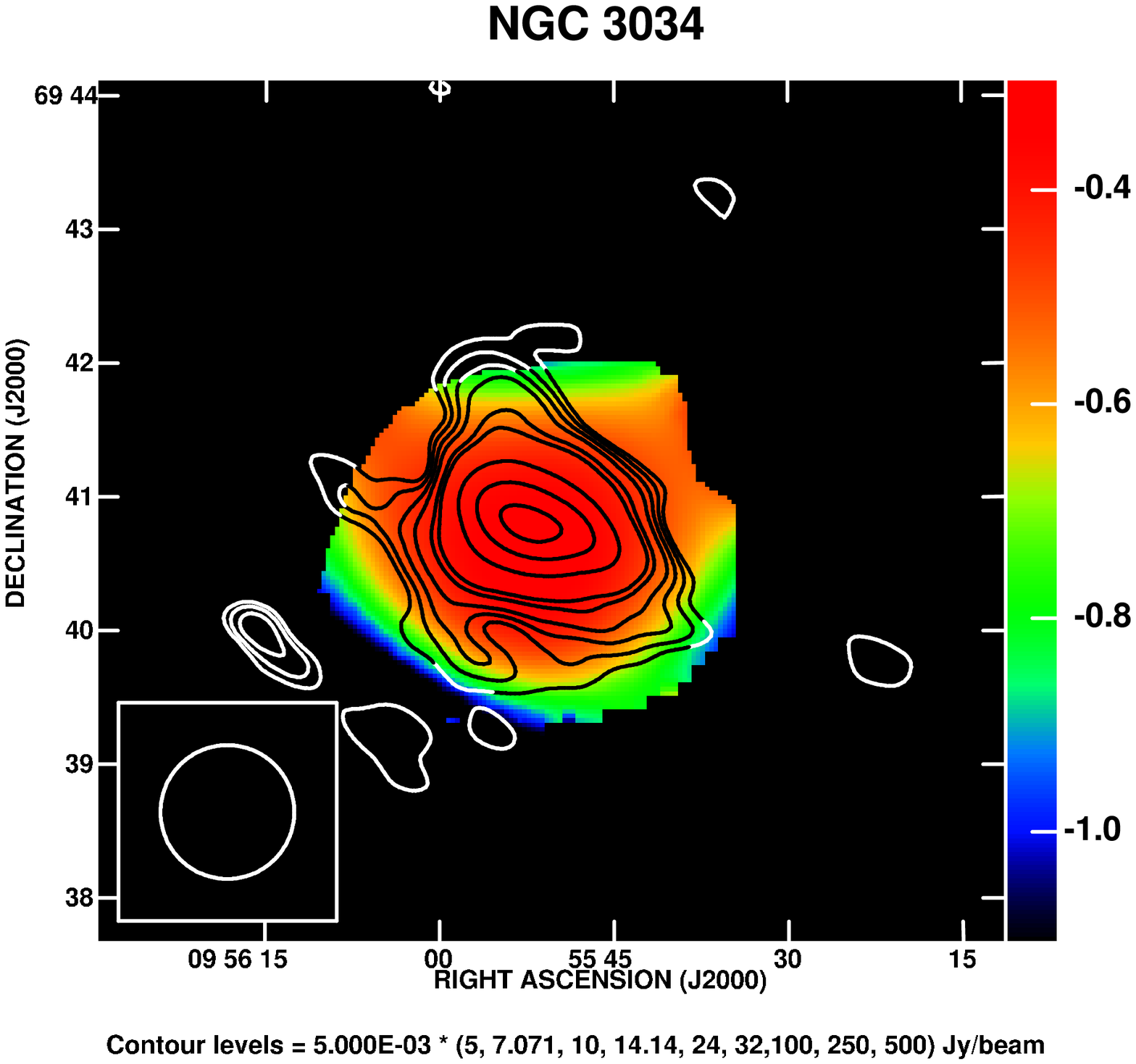}}} &
                                                                              \\
{\mbox{\includegraphics[width=7.5cm,height=7.4cm]{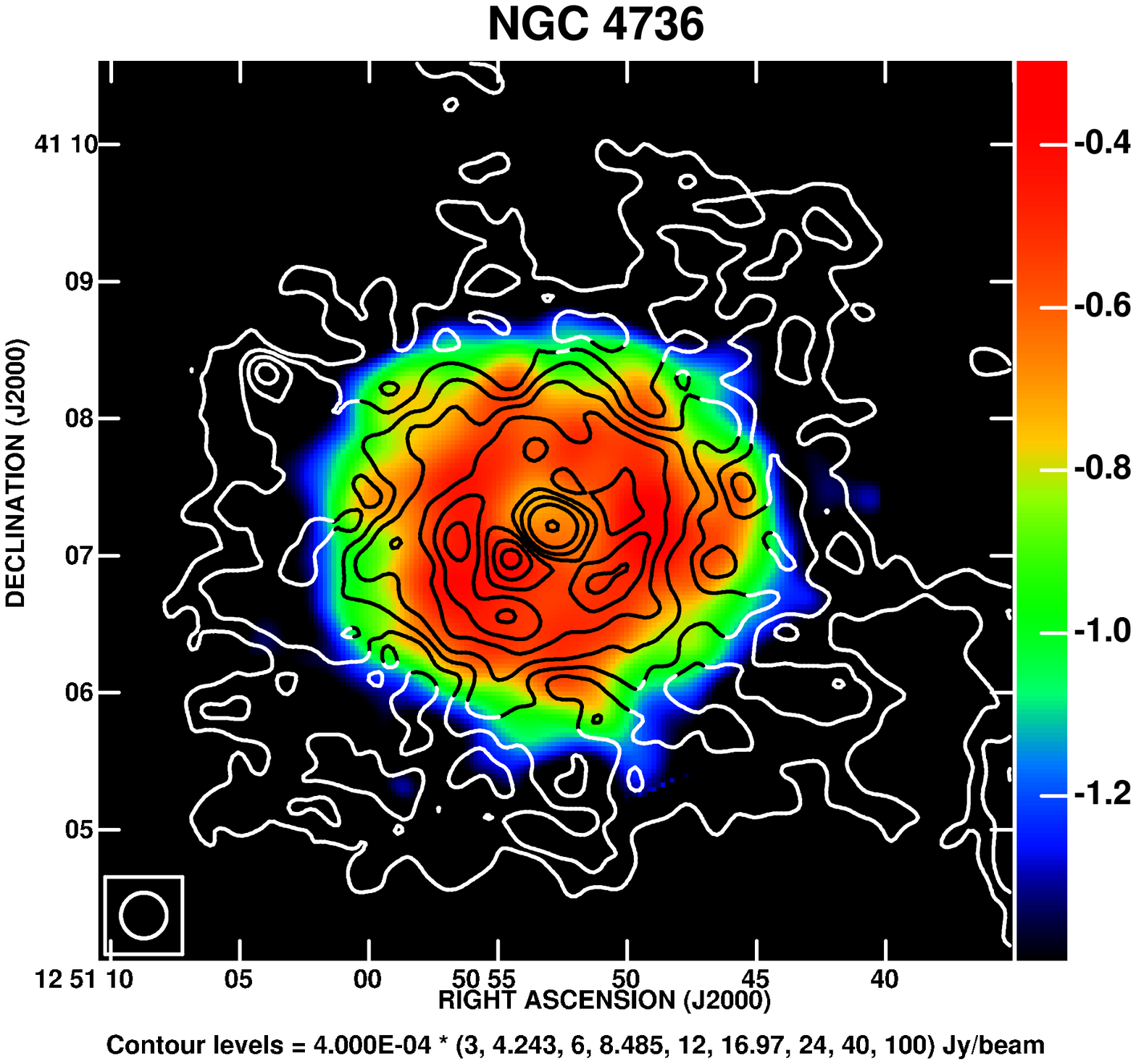}}} &
{\mbox{\includegraphics[width=7.5cm,height=7.4cm]{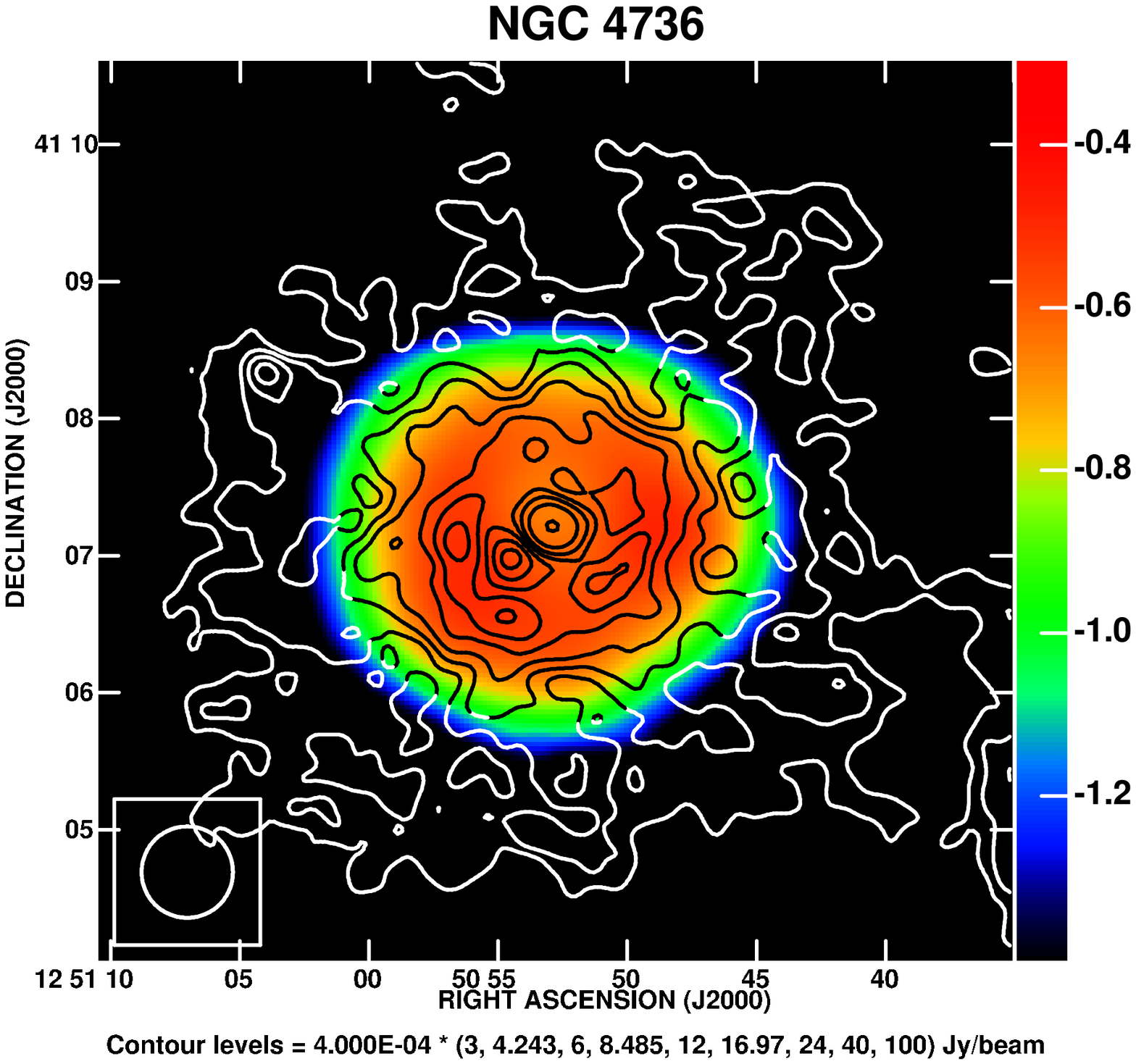}}} \\
\end{tabular}
\caption{{\it Left column:} The total spectral index ($\alpha$) maps between GMRT 333 MHz and
near 1 GHz. The top row is NGC 1097, middle row is NGC 3034 and the bottom row is NGC 4736.
The $\alpha$ maps for NGC 1097, NGC 3034 and NGC 4736 have a resolution of $40\arcsec\times40\arcsec$,
$60\arcsec\times60\arcsec$ and $20\arcsec\times20\arcsec$ respectively.
{\it Right column:} The nonthermal spectral index ($\ant$) between 333 MHz and near 1 GHz
at resolution of $40\arcsec$. Overlaid are the 333 MHz contours. Contour levels
are indicated below each figure.}
\label{figure3}
\end{center}
\end{figure*}

\addtocounter{figure}{-1}

\begin{figure*}
\includegraphics[width=7.5cm,height=7.5cm]{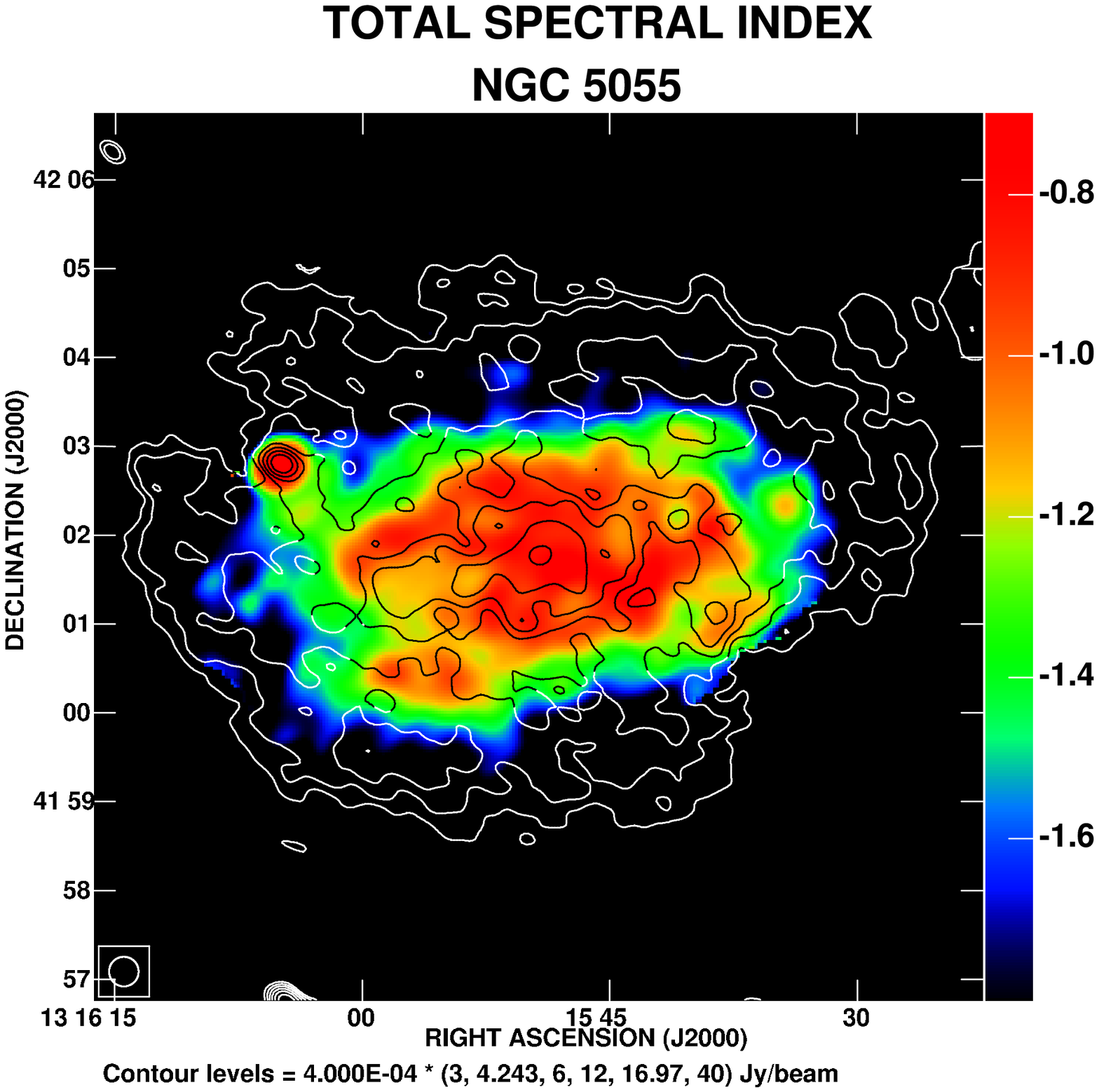}
\includegraphics[width=7.5cm,height=7.5cm]{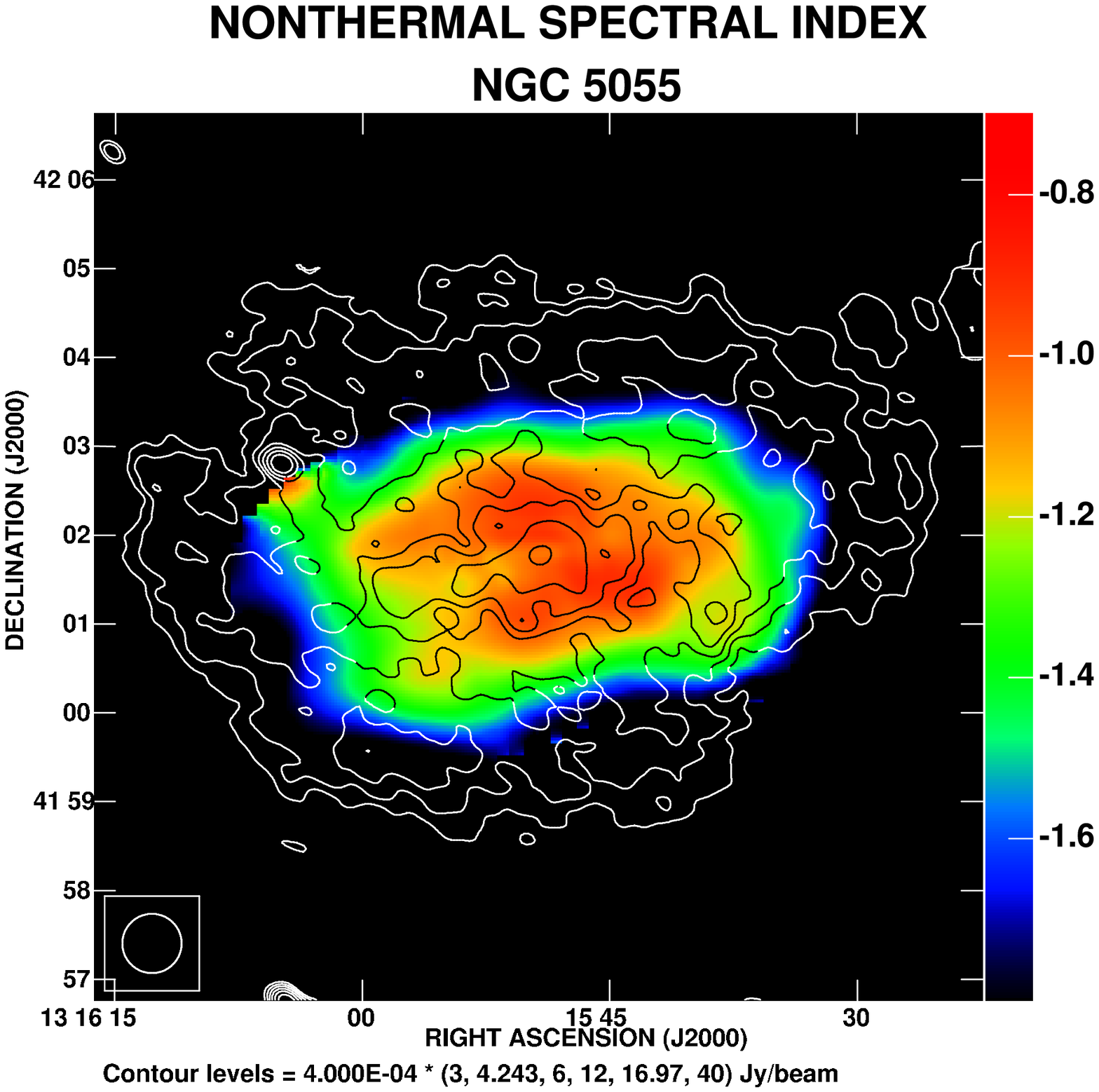}
\includegraphics[width=7.5cm,height=7.5cm]{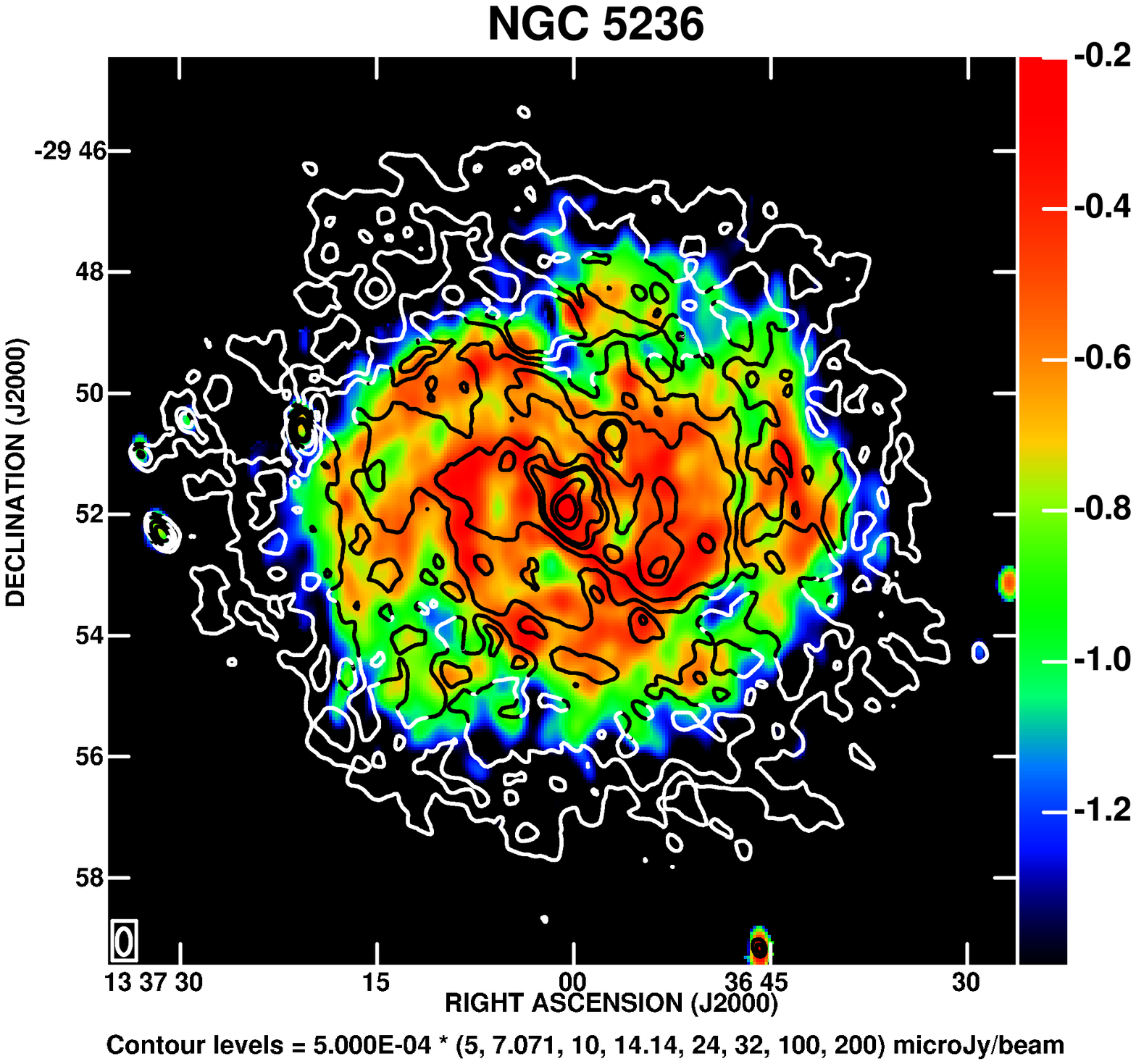}
\includegraphics[width=7.5cm,height=7.5cm]{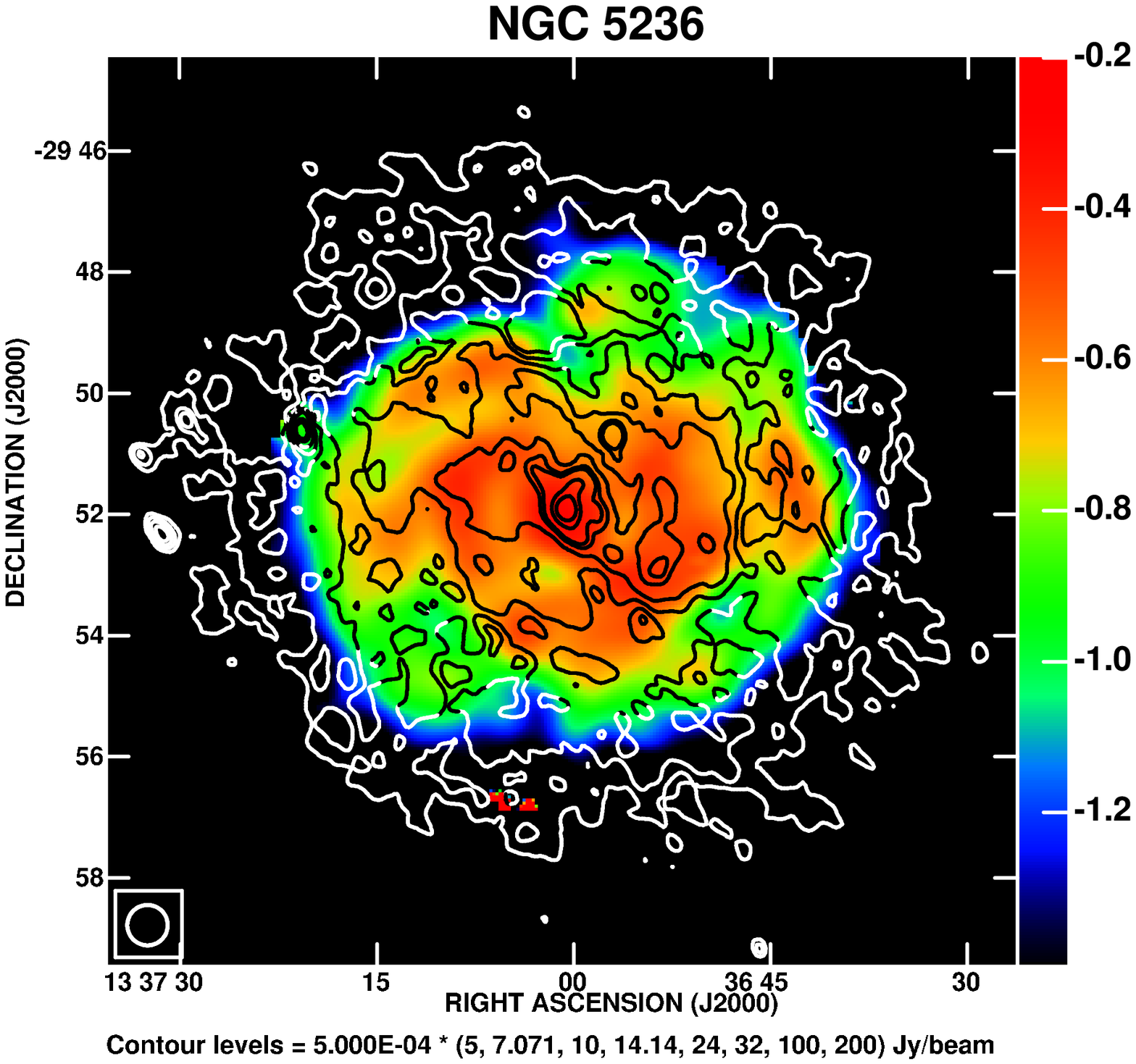}
\includegraphics[width=7.5cm,height=7.5cm]{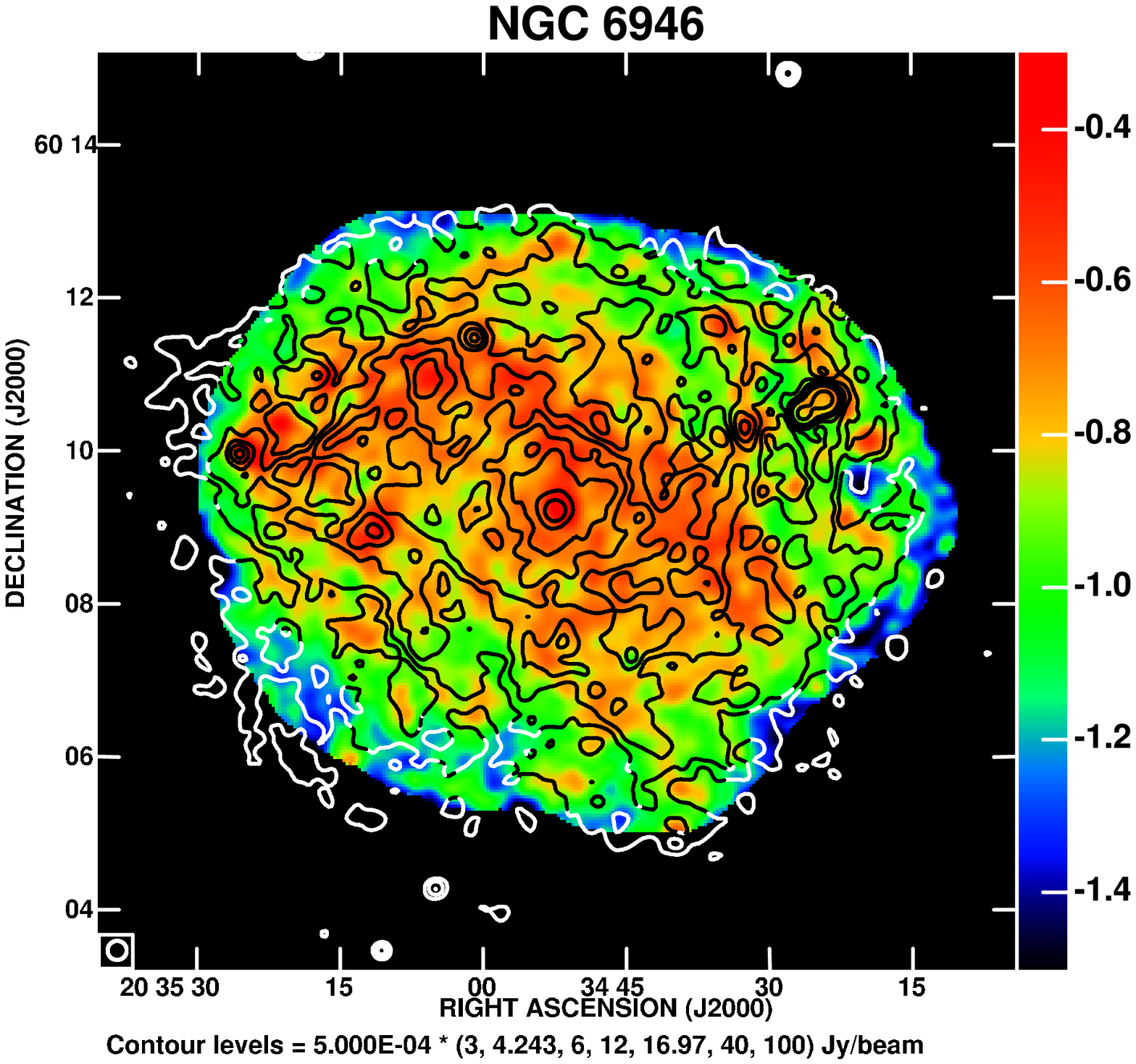}
\includegraphics[width=7.5cm,height=7.5cm]{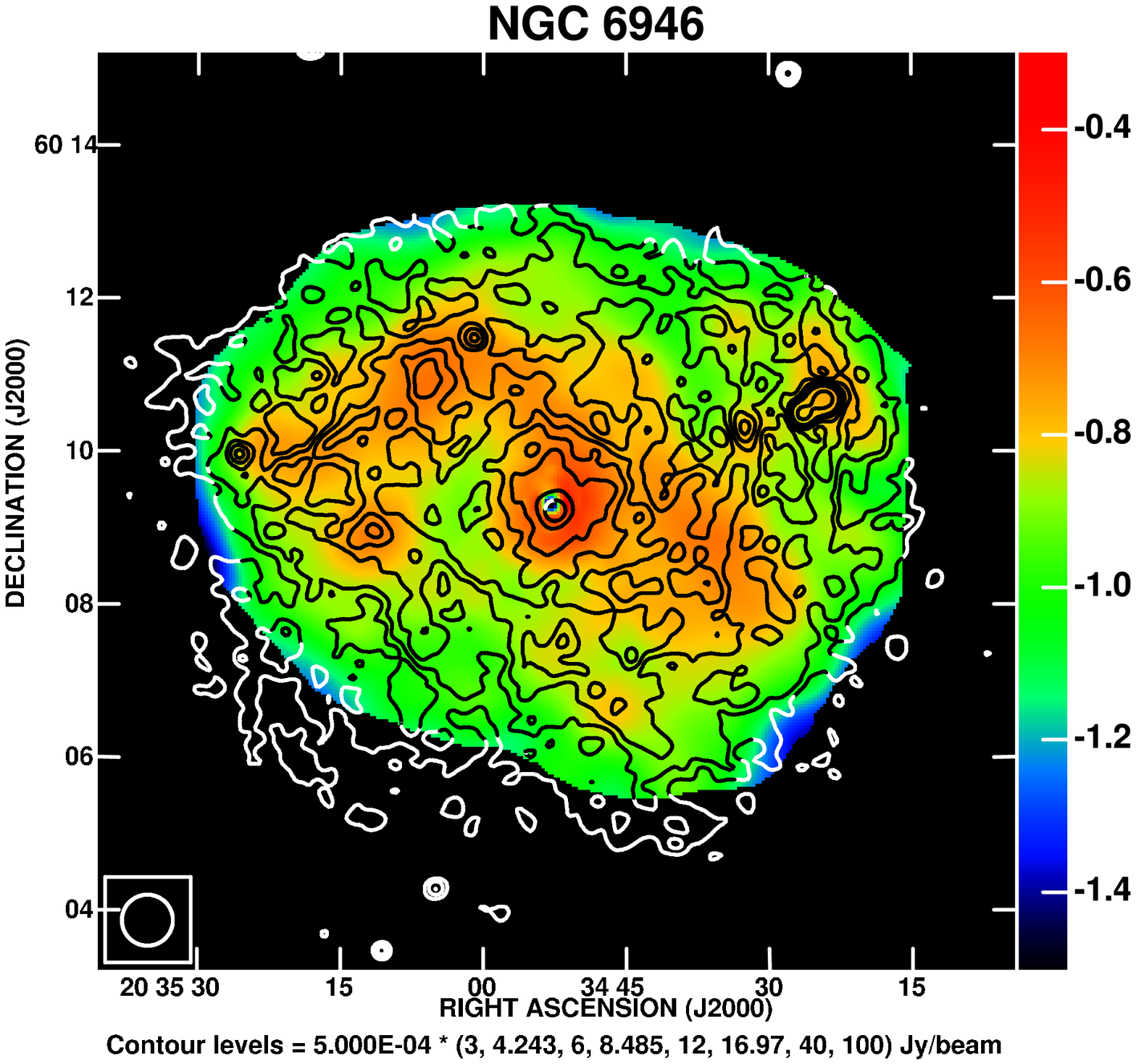}
\caption{{\it contd...} {\it Left column:} The total spectral index  ($\alpha$) maps between GMRT 333 MHz and
near 1 GHz. The top row is NGC 5055, middle row is NGC 5236 and the bottom row is NGC 6946.
The $\alpha$ maps for NGC 5055, NGC 5236 and NGC 6946 have a resolution of $20\arcsec\times20\arcsec$,
$26\arcsec\times14\arcsec$ and $15\arcsec\times15\arcsec$ respectively.
{\it Right column:} The nonthermal spectral index ($\ant$) between 333 MHz and near 1 GHz
at resolution of $40\arcsec$. Overlaid are the 333 MHz contours. Contour levels
are indicated below each figure.}
\label{figure3c}
\end{figure*}

\begin{figure*}
\includegraphics[width=5.5cm,height=4.2cm]{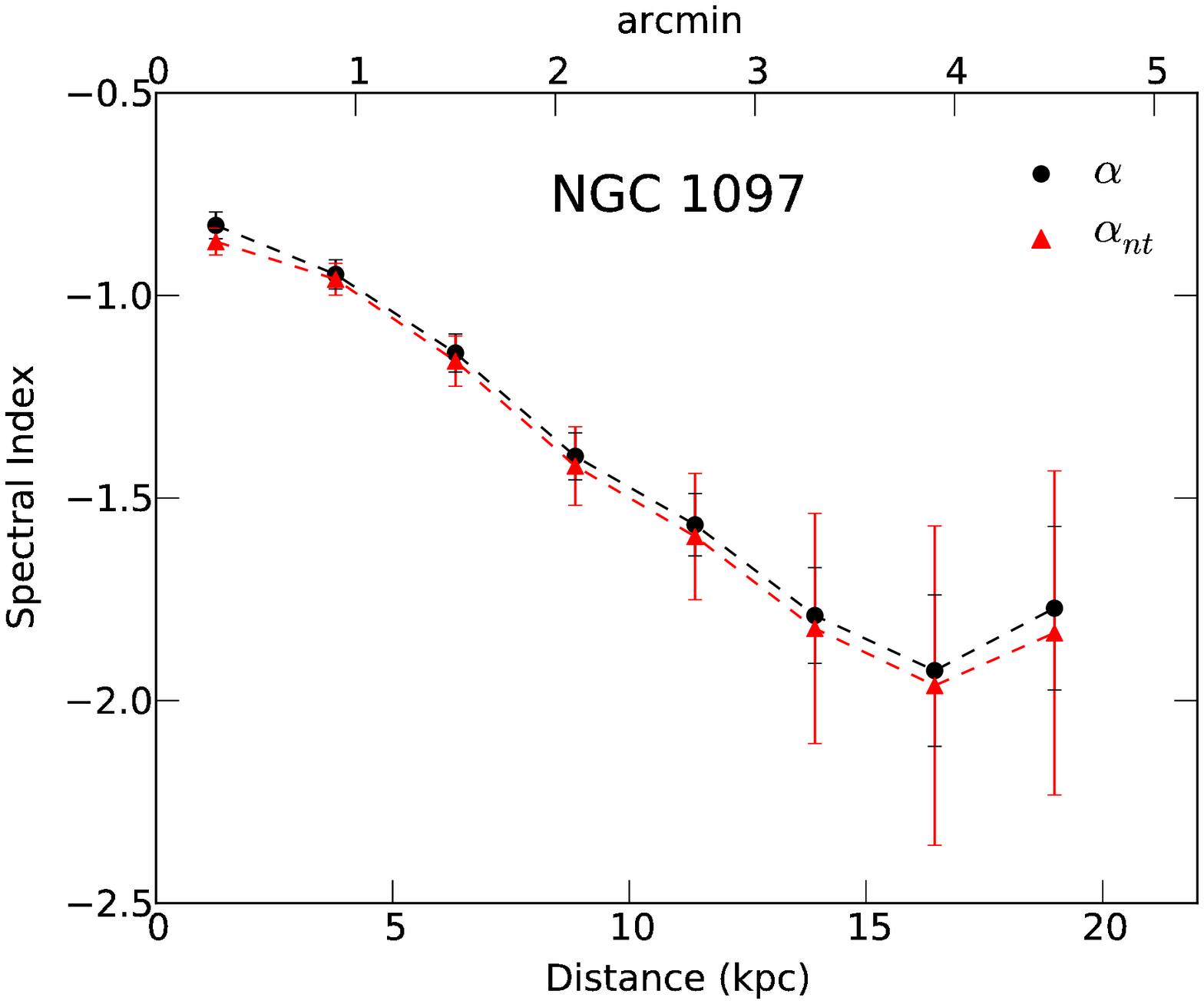}
\includegraphics[width=5.5cm,height=4.2cm]{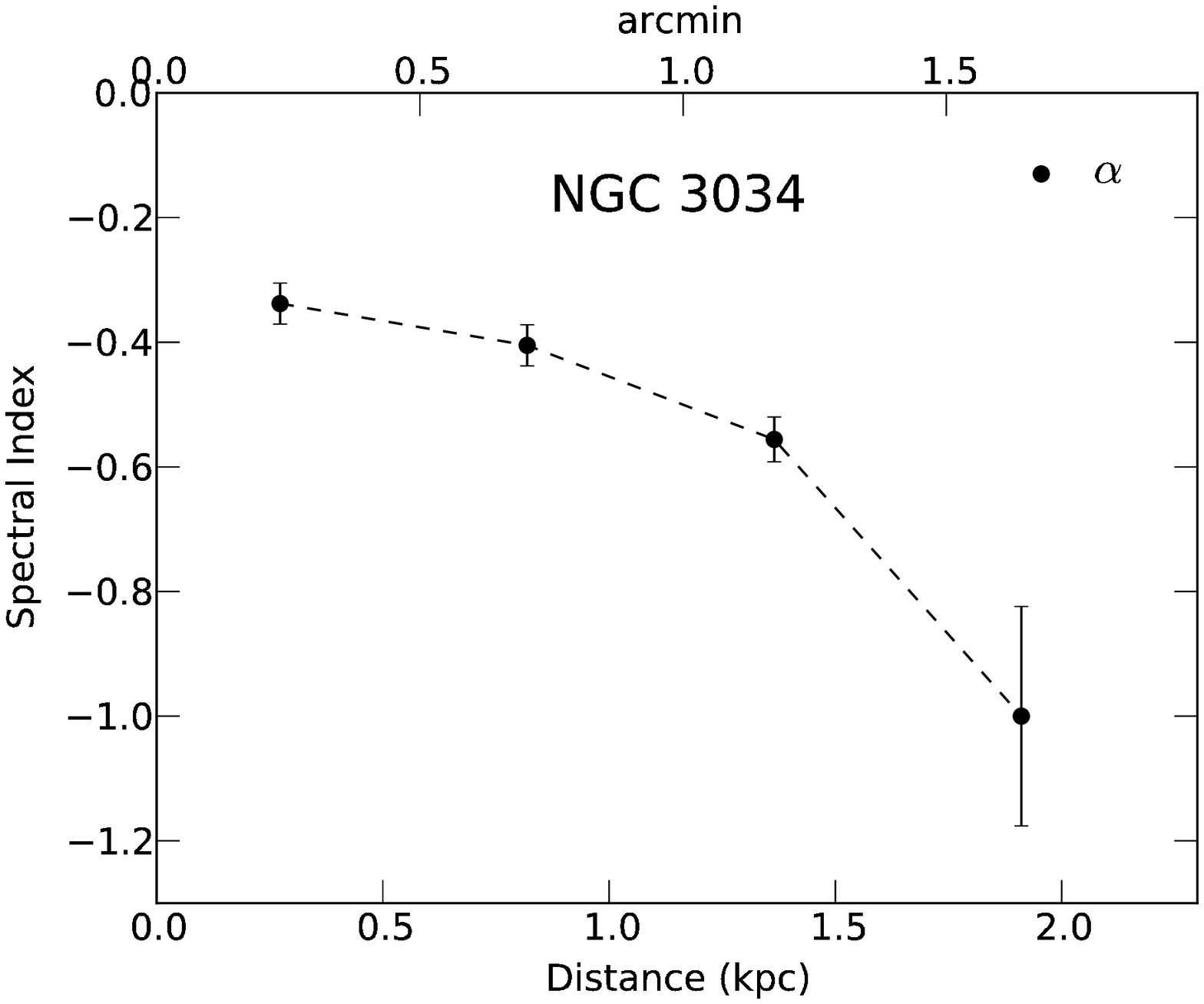}
\includegraphics[width=5.5cm,height=4.2cm]{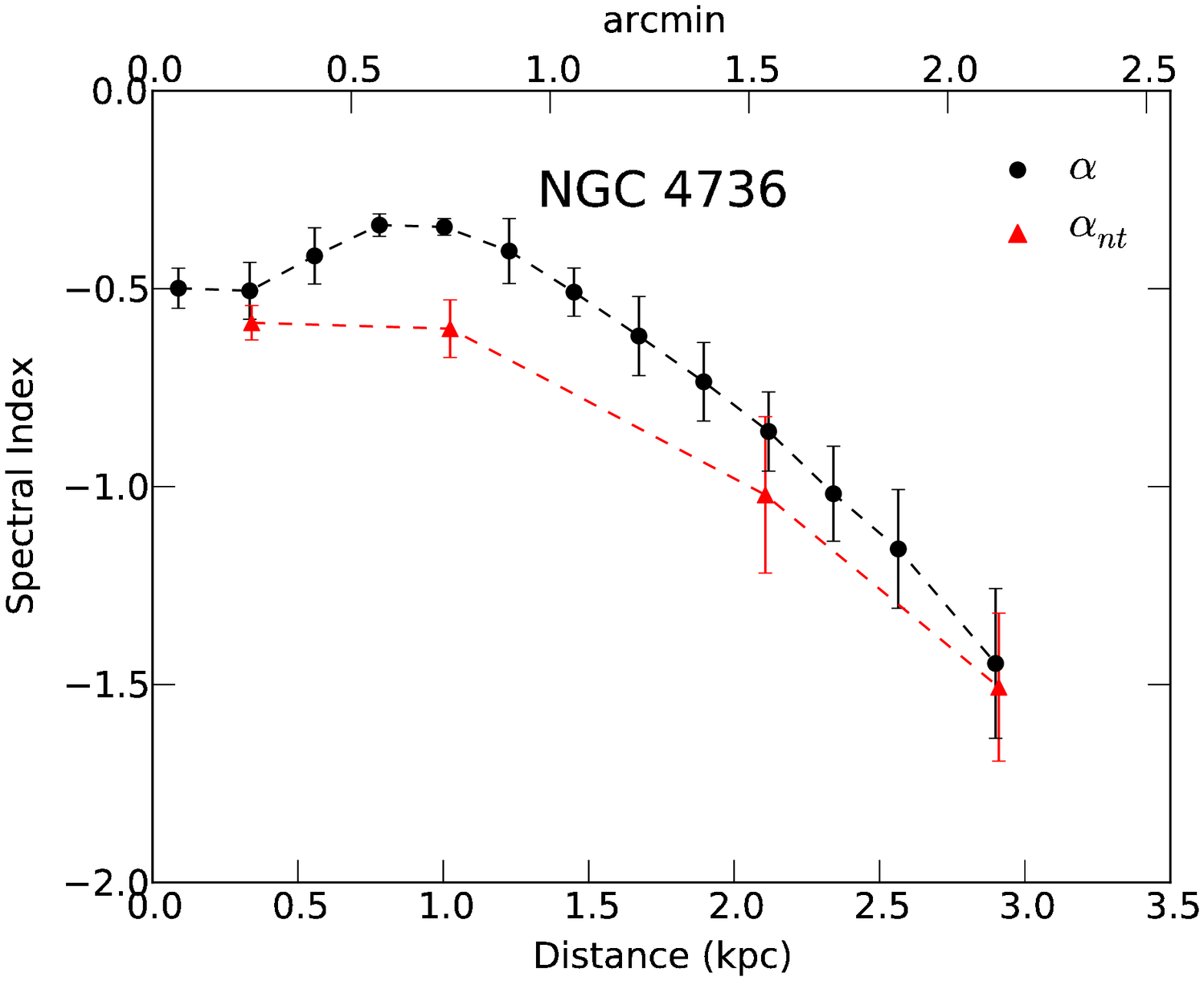}
\includegraphics[width=5.5cm,height=4.2cm]{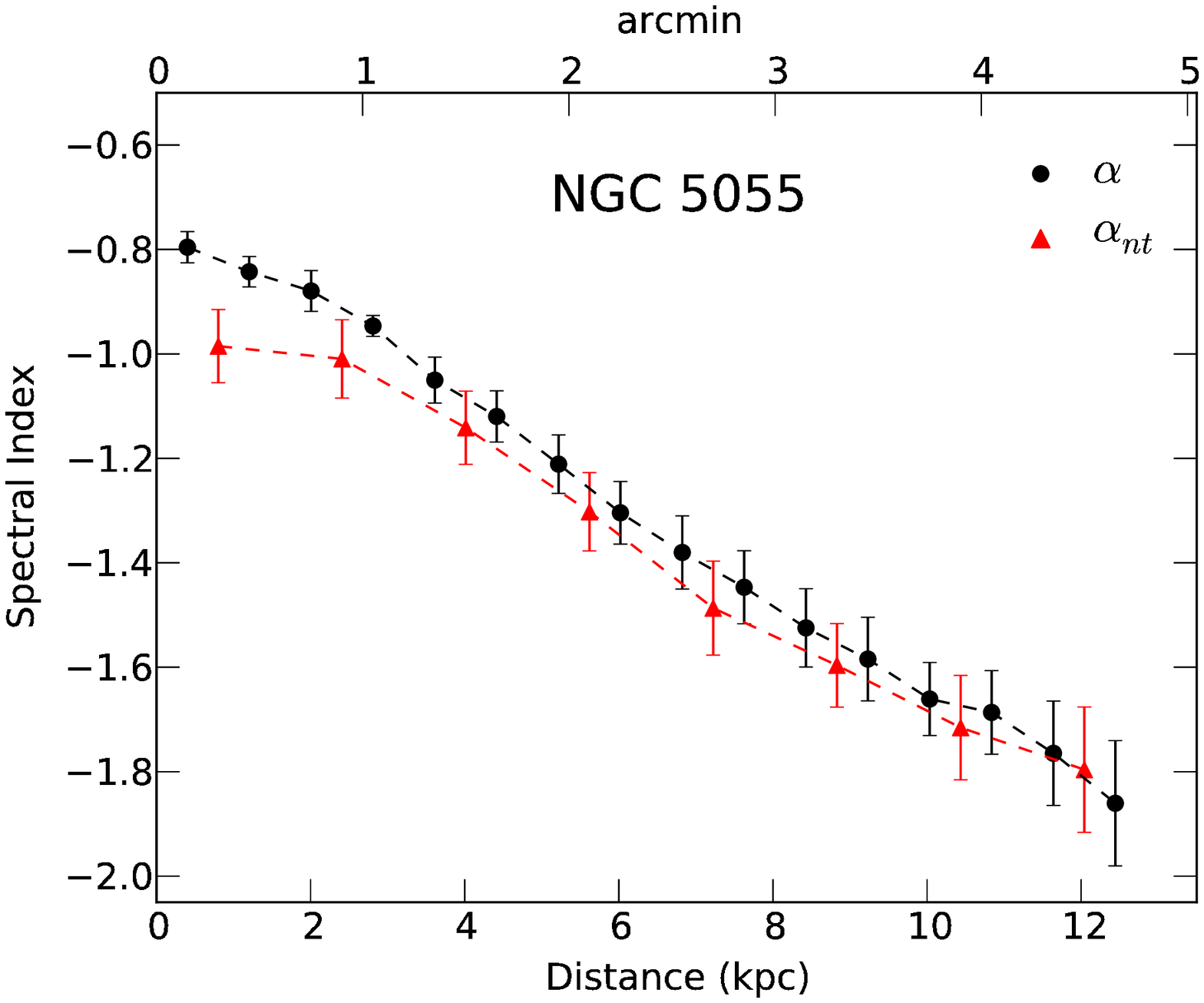}
\includegraphics[width=5.5cm,height=4.2cm]{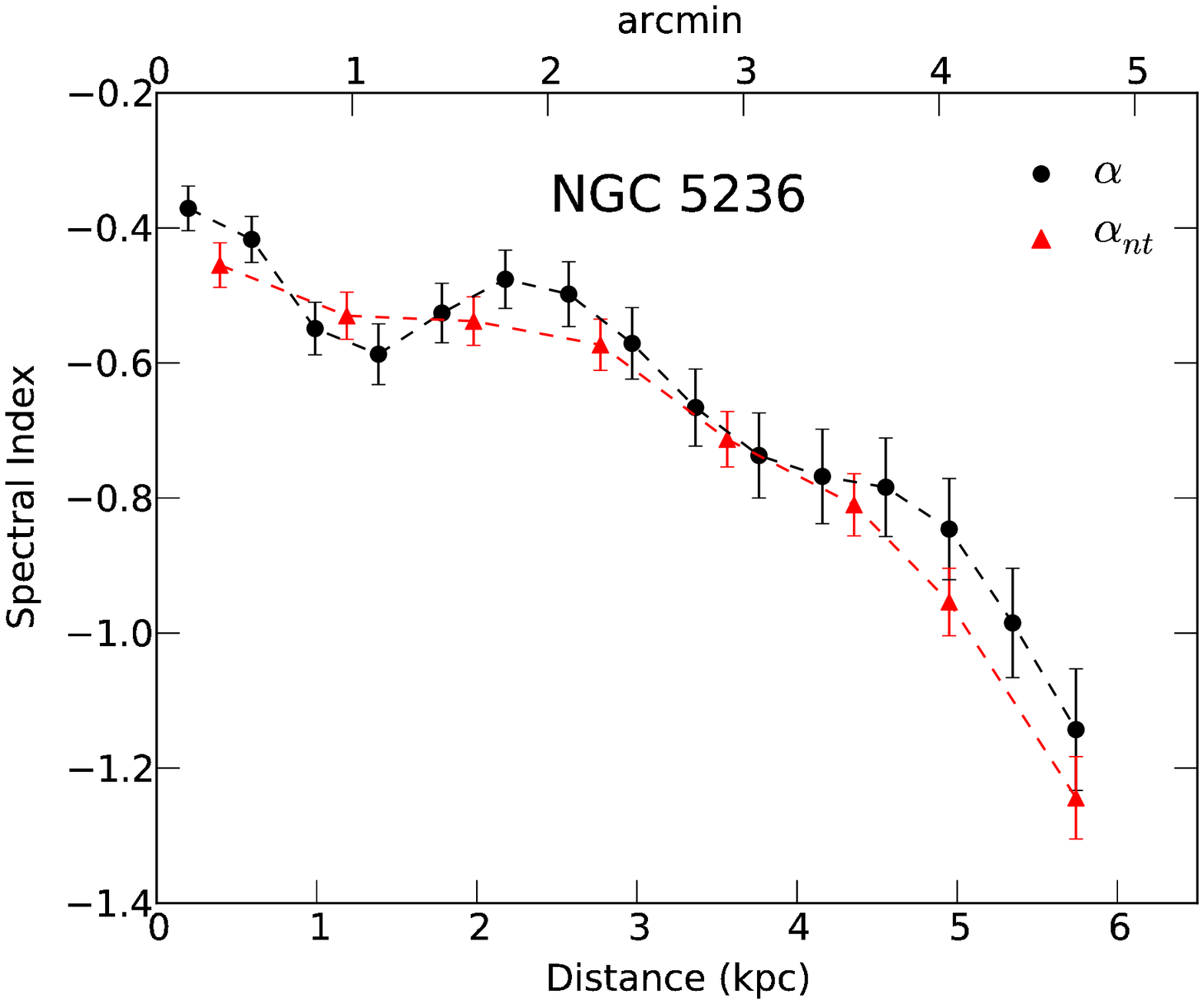}
\includegraphics[width=5.5cm,height=4.2cm]{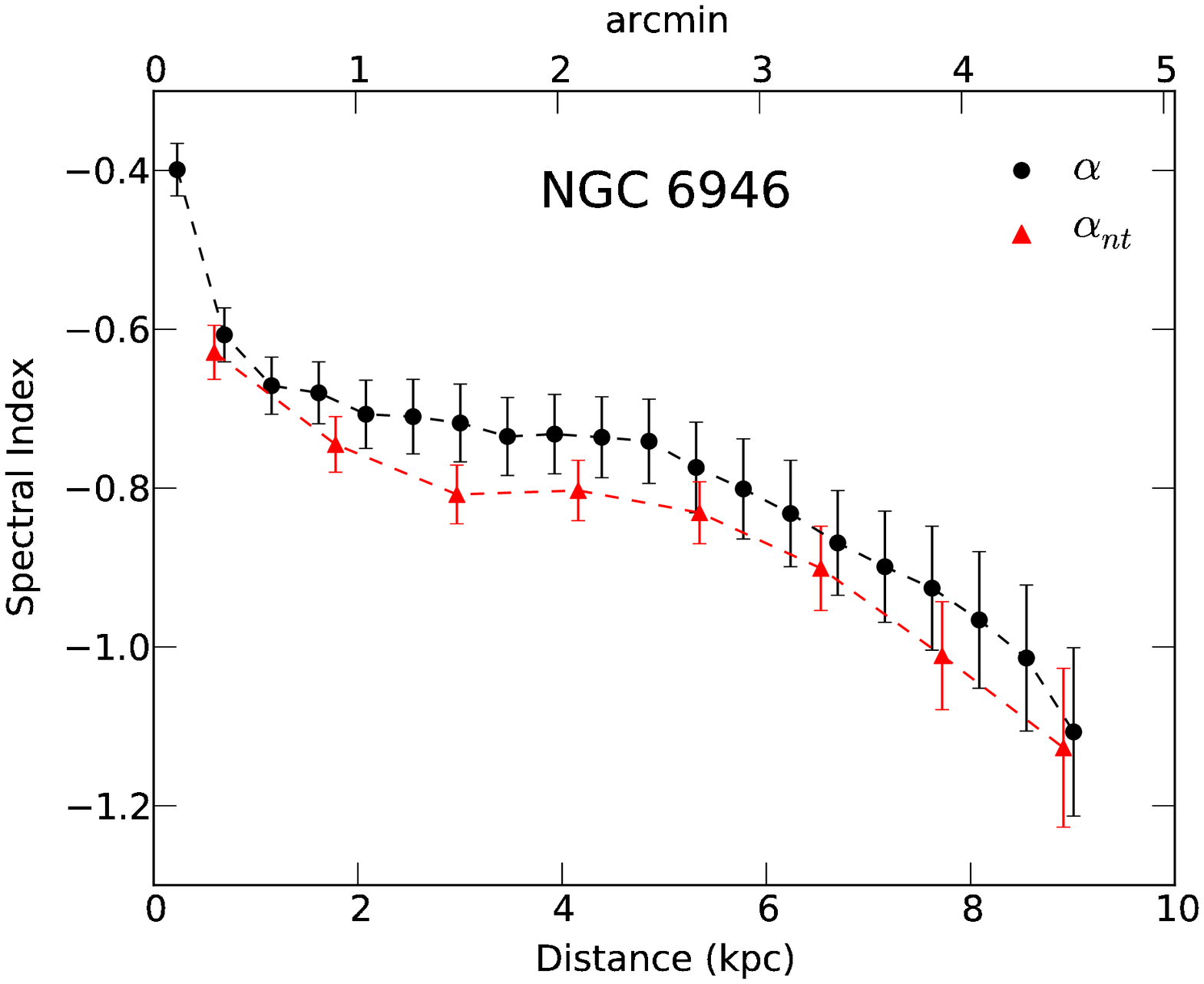}
\caption{Azimuthally averaged spectral index determined
within annuli of one beam width. The black circles are the total
spectral index ($\alpha$), while the red triangles are the nonthermal spectral index ($\ant$). The top x-axis shows the angular distance from the center in arcmin.
For NGC 3034, the averaging was done for half the synthesized beam
and therefore adjacent points are not independent.}
\label{figure4}
\end{figure*}

\section{Nonthermal spectral index ($\ant$)}
\label{section4}

The nonthermal spectral index $\ant$, is the main parameter of
  interest because it is used to model CRe generation and
  propagation. However, it is not possible to measure it directly. The
  quantity which is measured is the total spectral index $\alpha$
  which is contaminated by thermal free-free emission. This
  contamination is significant for spatially resolved regions of the
  galaxy where the thermal fraction\footnote{Thermal fraction is
  defined as: $\fth = {S_{\nu,\rm th}}/{S_{\nu,\rm tot}}$, where, $S_{\nu,\rm th}$ is the
  flux density of the thermal emission. In the text we express $\fth$
  as percentages.}  (henceforth, $\fth$) is high. In this
section, we assess the effect of $\fth$ on $\ant$ obtained between 333
MHz and near 1.4 GHz in spatially resolved regions of the galaxy. To
do this, we use the method of thermal-nonthermal separation developed
by \cite{tabat07} and apply it to five of our galaxies (except NGC 3034).  
The method uses dust extinction corrected
H$\alpha$ maps to get a template for the free-free emission across the
galaxy, and extrapolates it to the desired radio frequency to obtain
spatially resolved thermal radio emission maps. The extinction
correction is achieved by estimating the dust temperature and optical
depth using the far infra-red maps at $\lambda$70$\mu$m and
$\lambda$160$\mu$m (see Appendix~\ref{appendixa1} for details).  
The far infrared $\lambda$160$\mu$m map has an
angular resolution of $40\arcsec\times40\arcsec$, and hence the final
nonthermal spectral index maps have this angular resolution, which is
significantly coarser than the total spectral index (hereafter 
$\alpha$) maps. The usefulness of this method lies in the fact that it can be used in
spatially resolved parts of the galaxy. Further, \cite{tabat07}
demonstrate the robustness of their method over several other existing
thermal-nonthermal separation techniques (see references therein and
\citealt{broad89,gioia82,humme82}).

The mean $\fth$ for all our sample galaxies was found to be less than 5\%
at 333 MHz, although the value can go up to 10\% in some specific
bright \hii regions and in the spiral arms.  Applying the same method to
the available higher frequency data at or close to 1.4 GHz, we found
the mean $\fth$ to be less than 12\%. The distribution of the $\fth$
at 333 MHz and near 1.4 GHz for five galaxies is shown in
Figure~\ref{figurea1} of the appendix.

The nonthermal 333 MHz and higher frequency maps were further used to
obtain $\ant$ variation for the five galaxies. The left panels of
Figure~\ref{figure3}, show the $\alpha$ maps and the right panels show
the $\ant$ maps.  Figure~\ref{figure4} shows the variation of $\alpha$
and $\ant$ with radius for the galaxies. Here, the spectral index was
estimated by azimuthally averaging over annuli of one beam width.  The
radial profiles clearly reveal that the nonthermal spectral index is
nominally steeper than the uncorrected values, and further steepens
towards the outer parts of the galaxy.  However, we note that towards
the outer parts of the galaxies, the steepening of both the total and
nonthermal spectral indices may be caused due to ``missing flux density''
problems.

\begin{figure}
\begin{centering}
\includegraphics[width=8cm,height=8cm]{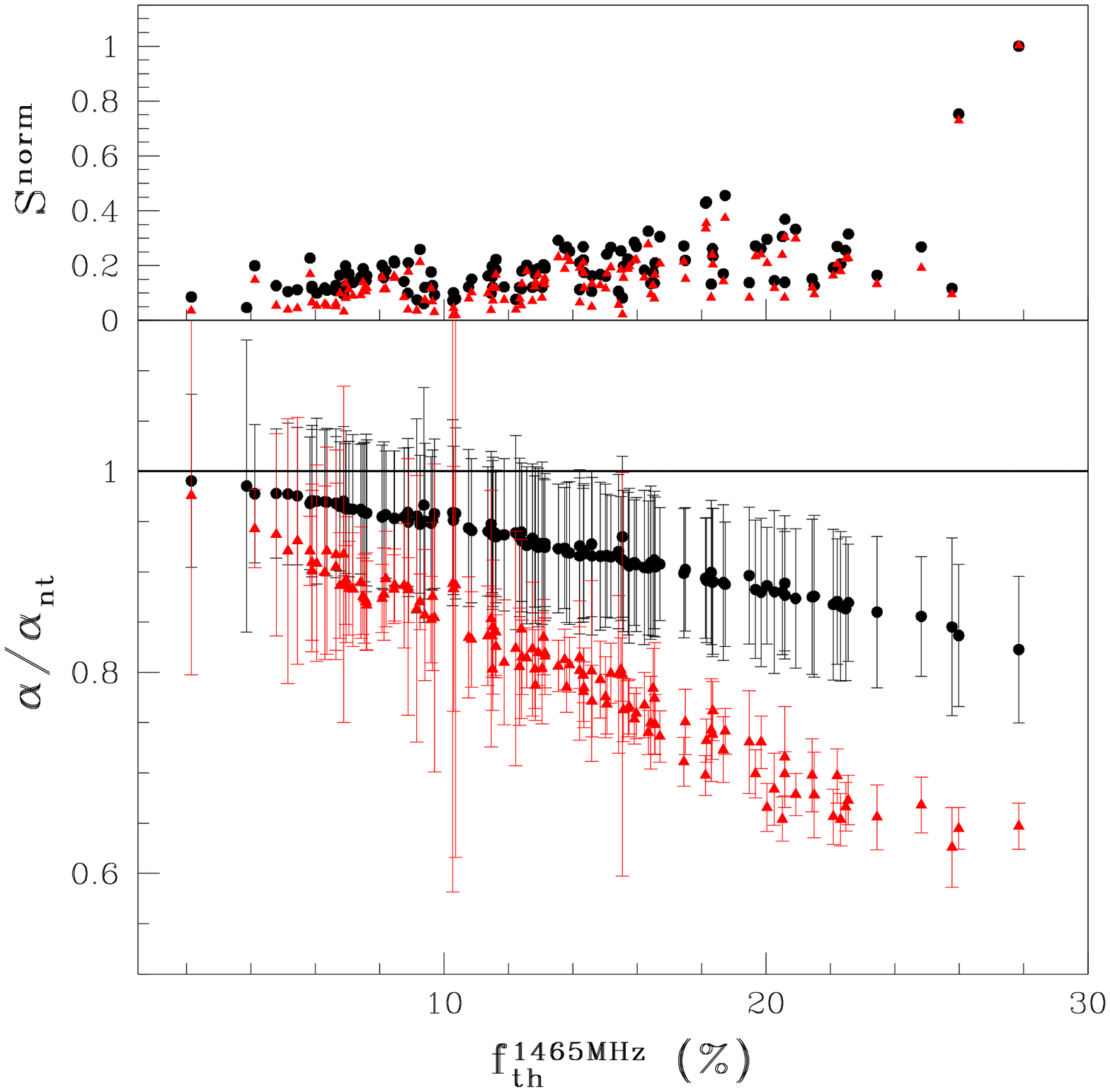}
\caption{ The bottom panel of the plot shows the 
ratio $\alpha/\ant$ as a function of $\fth$ at 
1465 MHz for NGC 6946 obtained at spatially resolved regions
of the galaxy.
The black circles corresponds to $\alpha/\ant$ between 333 MHz and 
1465 MHz, while the red triangles shows
between 1465 MHz and 4860 MHz. The top panel shows the 
flux density normalized to the peak value at 333 MHz (in black) 
and 1465 MHz (in red) respectively. Note that generally higher
flux density regions have high thermal fractions.}
\end{centering}
\label{figure5}
\end{figure}

We are now in a position to quantify the importance
of thermal-nonthermal separation, for robust estimates of $\ant$
between 333 MHz and near 1 GHz. The
uncertainty in $\ant$ can be estimated
by propagating the various sources of errors in obtaining the
nonthermal maps. Three primary sources of error affect the $\ant$
measurement, namely: the rms in the two
radio maps ($\rm \sigma_{\nu,map}$), the error associated with the uncalibrated system
temperature of the instrument $\rm \sigma_{\nu,Tsys}$, and the error in
thermal fraction, $\rm \sigma_{T}$.
$\rm \sigma_{\nu,Tsys}$ is about 5\% at 333 MHz for the GMRT and assumed
about 2\% for VLA and WSRT near or at 1.4 GHz maps. 
The error $\rm \sigma_{T}$ is estimated to be about 10\% at these frequencies by
\cite{tabat07} (also see Appendix A). Incorporating these errors, we
have computed the ratio $\alpha /\ant$
in regions by dividing the galaxy into rectangular
grids of size of approximately 40$\arcsec$.  We find that for thermal
fraction less that 5\% at 333 MHz and less than about 10\% near
1.4 GHz, $\ant \sim \alpha$. This effect is
illustrated for NGC 6946 in the bottom panel of Figure~6, where the
$\alpha /\ant$ is plotted (in black) as
a function of thermal fraction near 1.4 GHz. Within the error bars the
ratio is consistent with unity for thermal fraction less than 10\%. For 
higher thermal fraction (predominantly in the arms or \hii regions)
the ratio drops below unity, and correction to $\alpha$, to obtain
$\ant$, could be as high as 20\%.
The thermal-nonthermal
separation is even more important for $\ant$ calculated between higher
frequency pairs like 1.4 GHz and 4.8 GHz. The ratio $\alpha /\ant$ 
for NGC 6946 obtained between 1465 MHz and 4860 MHz (using archival map
downloaded from NED, \citealt{beck07}) is plotted in red in Figure~6, where 
we clearly see that the ratio is significantly below unity
for the whole range of thermal fractions.
Note that since there is also a systematic uncertainty associated with the
absolute calibration scale for calibrating radio flux densities, the absolute
value of the spectral index can vary by about 10\%.

\section{Discussion on Individual Galaxies}
\label{section5}

{\bf NGC 1097:} Our 333 MHz map (see Figure~\ref{figure1}) has an
angular resolution of 16$\arcsec\times$11$\arcsec$, which corresponds
to a spatial resolution of $\sim$ 1.1 kpc, if the galaxy is at a
distance of 14 Mpc. The only other low frequency radio continuum study
of this galaxy was done by \cite{ondre83} and \cite{ondre89} at 1465
MHz. They reported strong radio emission coincident with the narrow
dust lanes in the bar.  The central bar region is also clearly visible
in our 333 MHz map, and is coincident with dust emission inferred from
the far infra-red $\lambda$24$\mu$m images from {\it Spitzer}, as seen
in Figure 7.  The bar pattern rotates in the clockwise direction (see
\citealt{ondre89}), and as expected prominent shock fronts are visible
on the leading side of the bar with respect to the sense of rotation.
The $\fth$ in the bar region at 333 MHz is estimated to be $<$1\%,
hence the radio emission at this frequency is entirely nonthermal in
origin. These observations fit well with the explanation that the
radio emission from the bar results due to shock compression of gas,
dust and magnetic field, as was concluded by \cite{ondre83}. The total
emission from the bar and the central component (the region shown in
Figure 7) is 835$\pm$57 mJy, out of which 750$\pm$52 mJy arises from
the central compact region (25$\arcsec\times$21$\arcsec$). For a
similar sized central region \cite{ondre83} quote a flux density of
260 mJy at 1465 MHz, based on which we get a $\alpha$ of
$\sim-0.75$. The $\ant$ within a 40$\arcsec$ region based on our
analysis is $-0.85\pm0.1$. 
This $\ant$ close to the nuclear region is
steeper when compared with other galaxies in our sample.  This
indicates possibility of radiative cooling by inverse
Compton/synchrotron processes.

\begin{figure}
\begin{centering}
 \includegraphics[width=8cm,angle=0]{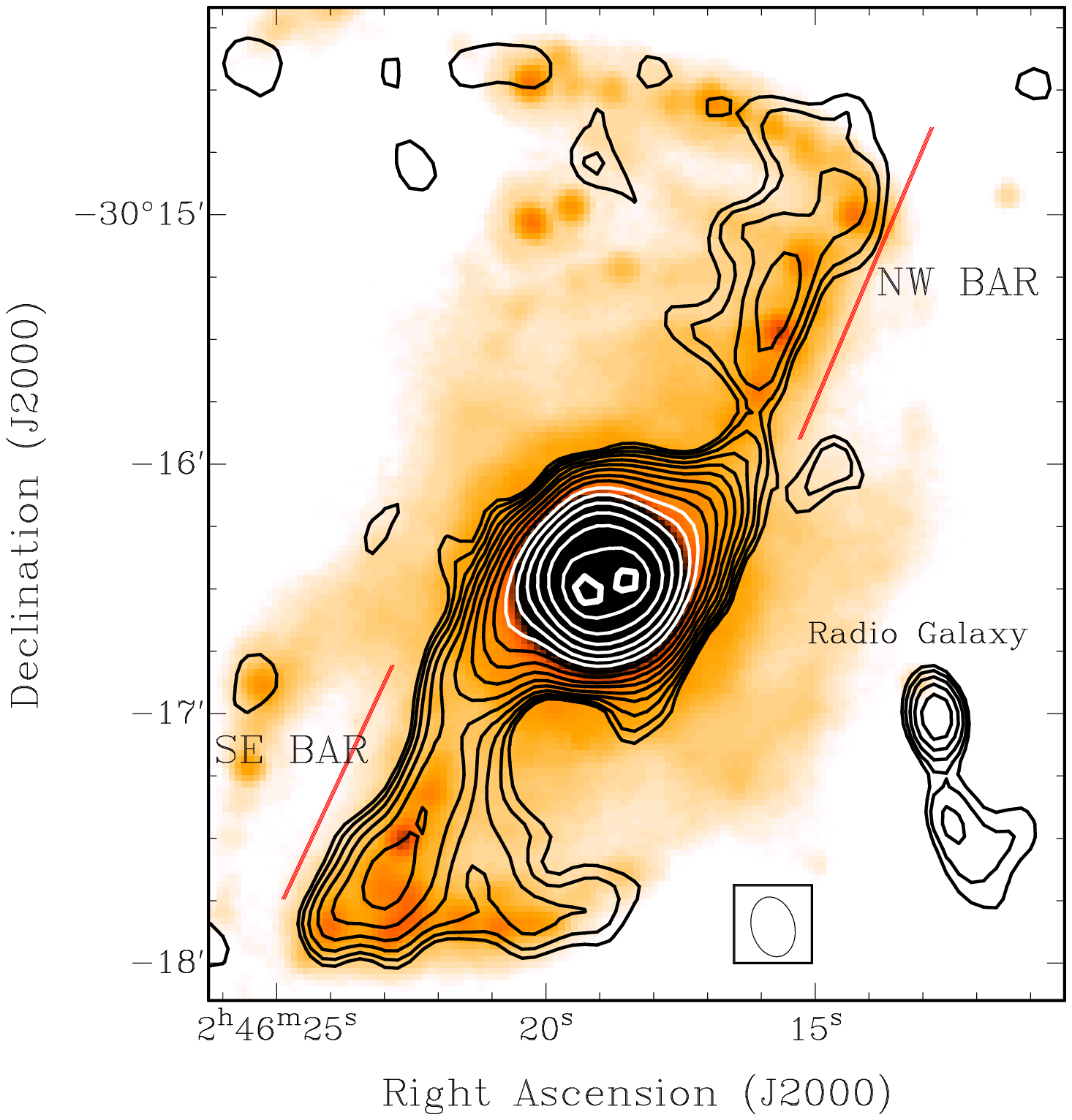}
 
\caption{The bar in the galaxy NGC 1097. In shades of orange is the
  24$\mu$m {\sc mips} image (resolution $\sim$6$\arcsec$), overlaid is
  the high resolution 333 MHz map made using baselines more than 1
  kilo$\lambda$ with a resolution $14.5\arcsec\times10\arcsec$.  The
  contours start from 1 mJy, and subsequent contours are in multiples
  of $\sqrt{2}$. The highest intensity contour is at 150 mJy/beam.}

\label{figure6}
\end{centering}
\end{figure}

The southeastern (SE) side of the bar has a flux density of $50\pm4$
mJy and the northwestern (NW) side has a flux density of $30\pm2$ mJy,
measured within the 2 mJy closed contours.  The red lines in
Figure~\ref{figure6} show the SE bar and NW bar.  In the optical, the
bar region continues into two prominent arms however the radio
emission is highly diffuse.  We find the thermal fraction of the
interarm region at 333 MHz and 1465 MHz to be 0.2\% and 2.5\%
respectively whereas in the spiral arm regions it is 1\% and 4.5\%
respectively.  In the arm regions, the $\langle\ant\rangle$ is
$-1.4\pm0.09$ while it gets significantly steeper ($-1.8\pm0.12$) in
the interarm regions which may be caused due to missing flux density
  problems.  The radial profile of the $\alpha$ and $\ant$ in
  Figure~\ref{figure4}, gets steep towards the outer parts of the
  galaxy possibly due to missing flux density.  We emphasize that the overall thermal
fraction in this galaxy is very low, with $\langle\fth\rangle$ of
$\sim$0.6\% at 333 MHz and $<$4.8\% at 1465 MHz (see
Figure~\ref{figurea1}).

At 333 MHz, we detect the radio galaxy discovered by \cite{beck05}, located $\sim$90$\arcsec$
from the nucleus of NGC 1097 towards the southwest at a PA of $-110^\circ$ 
(see Figure~7). The radio galaxy has a total flux density of $9.5\pm0.5$ mJy, with 
$4.8\pm0.25$ mJy in the northern component ($\rm{RA} =
02\rm{h}46\rm{m}12.7\rm{s}$, $\rm{dec} = -30\rm{d}17\arcmin01\arcsec$
(J2000)) and $3.8\pm0.2$ mJy in the southern component ($\rm{RA} =
02\rm{h}46\rm{m}12.5\rm{s}$, $\rm{dec} = -30\rm{d}17\arcmin26.6\arcsec$
(J2000)).

{\bf NGC 3034 (M82)}: The 333 MHz map (see Figure~\ref{figure1}) of
this galaxy has an angular resolution of 22$\arcsec \times 15
\arcsec$, corresponding to a spatial resolution of about 0.5 kpc for a
distance of 3.9 Mpc.  We find the integrated flux density at 333 MHz
to be 14 Jy which is consistent with flux densities measured at other
frequencies. The rms of the image, which is 3 mJy/beam, is dynamic
range limited since the peak flux density in the galaxy is 3
Jy/beam.  Morphologically the galaxy is featureless at our
  resolution, although the northern extension ($\rm RA = 09h55m56s,
  ~dec = +69d41\arcmin48\arcsec$) is coincident with the base of the
  H$\alpha$/optical horn (see Figure~\ref{figure1}), which is likely
  due to synchrotron emitting particles convected outwards by the
  nuclear wind, tracing the H$\alpha$ horns. The galaxy was observed
between 330 MHz to 4835 MHz by \cite{seaqu91}, and a similar horn was
seen at $\sim$1515 MHz. Their study also revealed a nonthermal radio
halo surrounding the galaxy, however our observations do not show this
due to dynamic range limitations.

For this galaxy, the thermal-nonthermal separation was not possible, but we have
computed the $\alpha$ map and radial profile as shown in
Figure~\ref{figure3} and \ref{figure4}. The 333 MHz GMRT map was convolved to
the resolution of VLA map at $60\arcsec$.
This is a good approximation for $\ant$ index variation across the
galaxy since the thermal fraction of this galaxy has been estimated to
be 15\% at 32 GHz \citep{klein88}, corresponding to $\sim3$\% and $2$\%
at 1.4 GHz and 333 MHz respectively.
The spectral index is seen to be very flat with
$\alpha = -0.35\pm0.03$ towards the center and steepens to $-1\pm0.2$
at a distance of $\sim$2 kpc from the center. Similar radial
variation of the spectral index was also reported by \cite{seaqu91}.

\begin{figure}
\centering
 \includegraphics[width=8cm,angle=0]{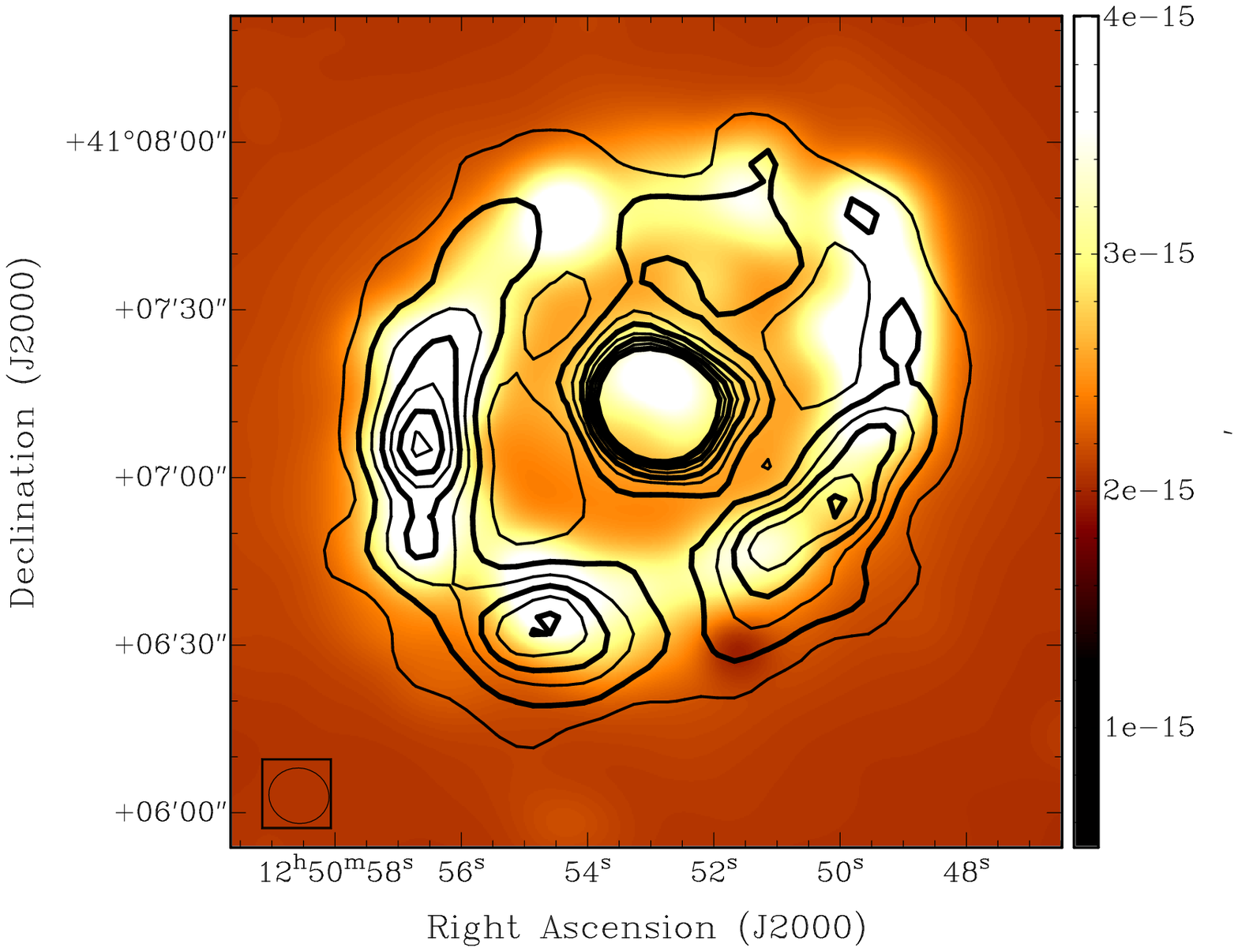}
 \caption{The star forming inner circumnuclear ring in NGC 4736. The contours are the high resolution
333 MHz GMRT map  with synthesized beam of 10.8$\arcsec\times$9.9$\arcsec$
and the grey scale shows the H$\alpha$ image (in $\rm erg~s^{-1}cm^{-2}$)
from the Jacobi Kapteyn Telescope with the 6570\AA\   
filter \citep{knape04}, smoothed to the same resolution as 
the 333 MHz map. The lowest contour is at 3 mJy increasing in steps of 1 mJy.}
\label{figure7}
\end{figure}

{\bf NGC 4736 (M94)}:
The 333 MHz map (see Figure~\ref{figure1}) of this galaxy has an
angular resolution of 13$\arcsec \times 12 \arcsec$, corresponding to
an spatial scale of about 0.3 kpc for a distance of 4.7 Mpc. The other
low frequency study of this galaxy exist at 610 MHz \citep{bruyn77},
observed with much lower angular resolution of 56$\arcsec \times 85
\arcsec$.  At our angular resolution we clearly detect emission from
the inner star forming circumnuclear ring of the galaxy as shown in
Figure 8 overlaid on the H$\alpha$ map. The ring is
located at about 50$\arcsec$ (major axis), corresponding to 1.1 kpc,
from the center, and is $\sim$19$\arcsec$ (420 pc) wide (up to the
half power) having a flux density
of 240$\pm$13 mJy. The estimated thermal fraction of the ring 
at 333 MHz and 1374.5 MHz is $5\pm1$\% and $10\pm4$\% respectively,
and hence the ring at 333 MHz is largely nonthermal in origin.  
The ring is seen to be coincident with the H$\alpha$ ring, which is shown in
Figure~\ref{figure7} after convolving to the same radio resolution.

Radio emission from the ring is also seen in the higher frequency 8.46 GHz
observations made by \cite{chyzy08}, however at
these frequencies the emission is a mixture of thermal and nonthermal
components. The ring has relatively high thermal
fraction which leads to a noticeable change in spectral index (see 
Figure~\ref{figure4})
: $\langle\alpha\rangle$ is $-0.45\pm0.02$, while the $\langle\ant\rangle$ is $-0.6\pm0.1$. 
The ring has massive on going star formation \citep{smith91} and $\ant$
value indicates that both CRe generation and escape happens on timescales
shorter than the radiative timescale.
The compact nucleus lies within the central 0.5 kpc 
(25$\arcsec\times$22$\arcsec$) with integrated flux density of 110$\pm$6 mJy and
a peak flux of 45 mJy/beam and $\ant = -0.6\pm 0.05$. 
Outside the ring extending to the edge of the galaxy there 
is no thermal emission observed leading to $\ant \sim \alpha$,

{\bf NGC 5055 (M63)}: The 333 MHz observation (see
Figure~\ref{figure1}) has an angular resolution of
17$\arcsec\times$10$\arcsec$ corresponding to a spatial scale of 0.7
kpc at a distance of 9 Mpc. As is evident from the figure, the extent
of the galaxy is larger at 333 MHz than the optical.  In an earlier
study by \cite{humme82} at 610 MHz and 1410 MHz at the much lower
angular resolution of 58$\arcsec\times$87$\arcsec$, the galaxy appears
featureless which is also the case in our higher resolution 333 MHz
map. However, 10.55 GHz polarization observations by \cite{knapi00}
show regular spiral structure.  The overall radio emission at 333 MHz
at a level of 4$\sigma$ is $\sim$8.6$\arcmin\times$5.3$\arcmin$ in
extent, which has an integrated flux density of 2.3$\pm$0.13 Jy. This
value is higher than what one would obtain using a spectral index of
$-0.78$ estimated between 610 and 1417 MHz \citep{humme82}. The reason
for this is \cite{humme82} estimate the integrated flux density within
a region 6.7$\arcmin\times$3.2$\arcmin$ at 610 MHz, which is smaller
than our detected size at 333 MHz. However, if a similar size region
is considered, we obtain a flux density of 1.4$\pm$0.1 Jy which is
consistent with a spectral index of $-0.78$ as seen in
Figure~\ref{figure2}(d).  NGC 5055 is highly inclined, with
inclination angle, $i = 59^\circ$, and the extended low frequency
radio emission can result due to escape of low energy ($<$ 2 GeV) CRe
from the disc which travel larger distances ($\sim$ 4.5
kpc) into the galactic halo.

The mean thermal fraction $\langle\fth\rangle$ at 1696 MHz
($\lambda$18cm) and 333 MHz is 12\% and 1.3\% respectively. The $\ant$
in the central region is $\sim$ $-1$ which steepens to $-1.7$ towards
the outer parts of the galaxy\footnote{Note that \cite{humme82} using
  observations at 610 and 1417 MHz quote a spectral index
  variation of $-0.6$ to $-1$ between the central and outer parts of
  the galaxy, while we find much steeper values. The WSRT 1696 MHz map
  has a flux density of 284 mJy within the same region of
  6.7$\arcmin\times$3.2$\arcmin$, which when extrapolated using a
  spectral index of $-0.78$, has a much lower value (see
  Figure~\ref{figure2}(d)).  This may be due to missing flux density issues as
  discussed in Section~\ref{section2}.  Thus the true spectral index
  should be flatter than what is observed by us.} (see
Figure~\ref{figure4}). This result supports the conclusion drawn by
\cite{humme82} that the spectral index steepening is mainly due to
energy losses of CRe and decrease in number of relativistic electrons
with increasing galactic radius.

\begin{figure}
\begin{centering}
\includegraphics[width=9cm,height=9cm,angle=0]{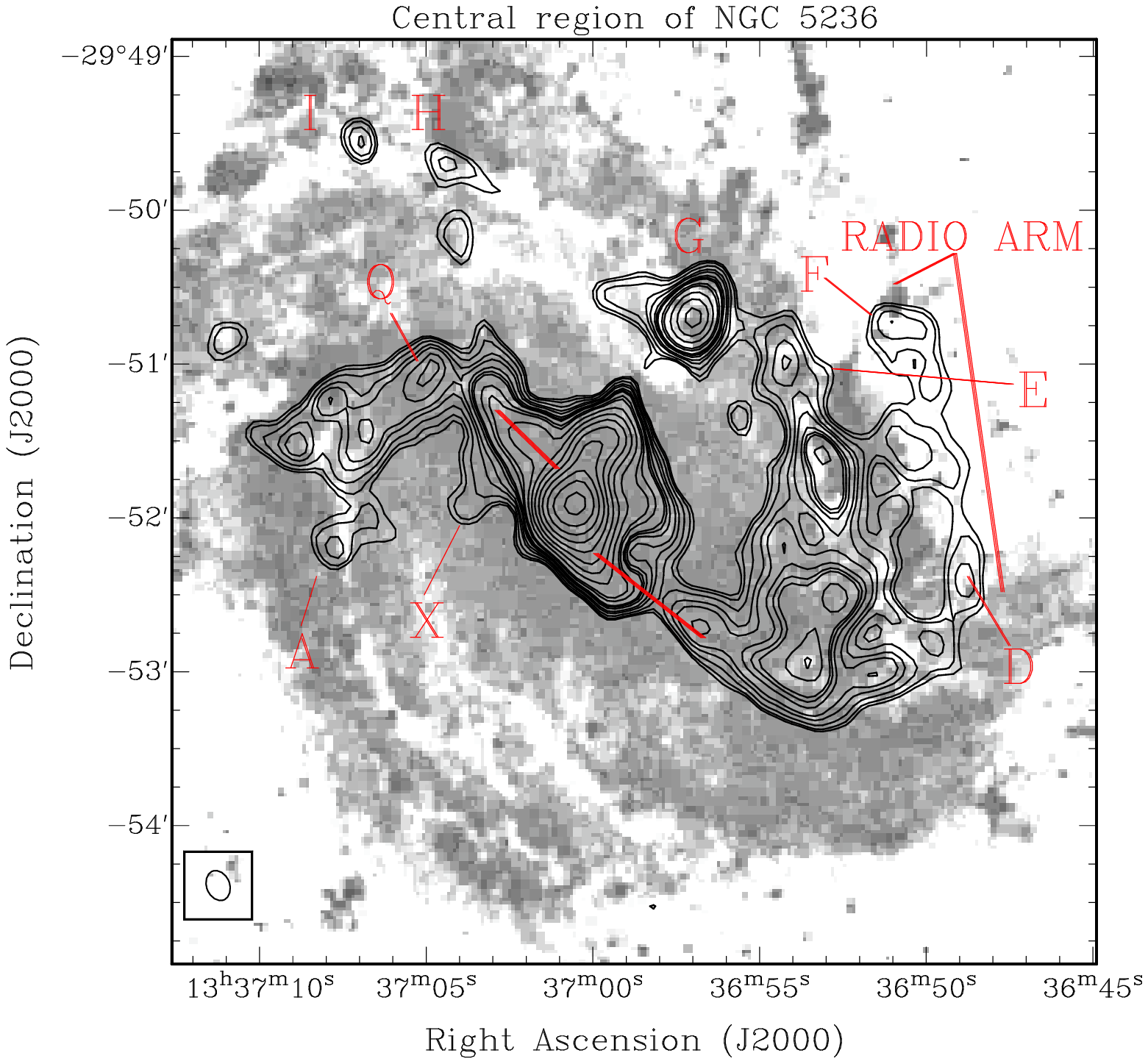}
\caption{The central region in NGC 5236. Greyscale is the DSS image
at a resolution of $1\arcsec$, overlaid are the 333 MHz contours at 
resolution $13\arcsec\times10\arcsec$. 
The marked regions are as per Ondrechen (1985).
The flux densities of the sources are as follows; A : $12.3\pm0.7$ mJy, 
D : $11.5\pm0.7$ mJy, F : $13.5\pm0.8$ mJy, G : $55\pm3$ mJy, H : $10.2\pm0.6$ mJy,
I : $9.2\pm0.5$, X : $20.0\pm1.2$ mJy}
\label{figure8}
\end{centering}
\end{figure}

{\bf NGC 5236 (M83):} The 333 MHz observation (see Figure~\ref{figure1}) has an angular
resolution of 16$\arcsec\times$12$\arcsec$, which correspond to a
linear scale of about 0.4 kpc at a distance of 4.2 Mpc.  An early
study of this galaxy by \cite{sukum87} at 327 MHz using the Ooty
Synthesis Radio Telescope (OSRT) quotes integrated flux density of
6.86$\pm$0.62 Jy at a resolution of $53\arcsec\times33\arcsec$, while
our GMRT observations has a slightly higher integrated flux density
of 7.4$\pm$0.4 Jy.

\cite{ondre85} has reported several point sources in this galaxy. To
verify the presence of these sources at 333 MHz, we made a high
resolution image of $13\arcsec\times10\arcsec$ as shown in
Figure~\ref{figure8}. Comparing this with \cite{ondre85} we detect the
central bar (shown in red lines) and the \hii regions named as A, F, H
and I, and nonthermal polarized point sources D, G and X. Note that
source G is identified as a background galaxy by \cite{maddo06}.  The
region E was identified as a shocked region in the arm, and has a flux
density of $33\pm2$ mJy, while the point source Q is a \hii region and
has flux density of 5 mJy/beam.  However, we were unable to detect
point sources B and C.  The flux densities of the point sources are
given in Figure~\ref{figure8}.  In addition to this we see an extended radio arm,
which runs along the faint narrow dust lane visible in the infrared
$\lambda$70$\mu$m map.

The galaxy has a low mean thermal fraction of about 3\% at 333 MHz and
about 7\% at 1452 MHz, and hence $\ant \sim \alpha$ as seen in
the radial spectral index profile of the galaxy in
Figure~\ref{figure4}.  In the arms, at 333 MHz, the thermal fraction
is estimated to be $8-10$\%, where $\alpha$ ranges from $-0.3\pm0.04$
to $-0.5\pm0.05$ which gets modified to $\ant$ as $-0.4\pm0.05$
to $-0.65\pm0.05$.  $\ant$ along the arms is seen to change
significantly from $-0.4$ within the central 3 kpc to $-0.65$ beyond 3
kpc.  Thermal fraction at 333 MHz in the interarm regions is $1-4$\%
and beyond the central 4.5 kpc is less than 1\%, indicating purely
nonthermal emission. In the interarm regions, $\ant$ values lie
in the range $-0.7\pm0.06$ to $-1.2\pm0.09$.

{\bf NGC 6946:} 
The 333 MHz observation (see Figure~\ref{figure1}) has an angular
resolution of 12$\arcsec\times$11$\arcsec$ which corresponds to a
spatial resolution of 0.4 kpc at a galaxy distance of 6.8 Mpc.
The integrated flux density at 333 MHz is 4.5$\pm$0.24 Jy. Using our
333 MHz and archival 1465 MHz VLA maps, we estimate the mean thermal
fraction to be about 4.3\% and 11\% respectively.

Spectral index study by \cite{klein82} between 610 MHz and 10.7 GHz
showed variation from about $-0.55$ near the center to about $-0.9$
towards the edge. The spectral index between GMRT 333 MHz map and VLA
1465 MHz map shows a similar trend, with spectral index of $-0.45$ at the
center and $-1$ towards the edge. Recently, \cite{beck07} studied the
spectral index between 4860 MHz and 1465 MHz which showed flatter
spectral index in the arm than the interarm due to higher thermal
fraction.  In our studies, thermal fraction is higher in the arm
(5--10\%) than in the interarm (1--5\%).  In the arm, $\alpha$
lies in the range $-0.55$ to $-0.65$, while the
$\ant$ is steeper $-0.7$ to $-0.85$.  In the
interarm regions the average spectral index is steeper with
$\langle\alpha\rangle = -0.9\pm0.06$, while the $\ant$ does not change 
significantly, $\langle\ant\rangle = -0.95\pm0.11$.

The 333 MHz map prominently shows the giant \hii regions identified by
\cite{kruit77}.  The \hii region at $\rm{RA} =
20\rm{h}35\rm{m}06\rm{s}$, $\rm{dec} = +60\rm{d}10\arcmin56\arcsec$
(J2000) (region C in \cite{kruit77}), is about 35$\arcsec$ (1.2 kpc)
in size, and can be seen in the $40\arcsec$ low resolution nonthermal
maps. This region is a suitable candidate to investigate the CRe
injection near high star forming sites.  We find the thermal fraction
to be $10\pm3$\% at 333 MHz and $15\pm4$\% at 1465 MHz.  The $\ant$ in
this region is $-0.6\pm0.08$, whereas before the separation the
spectral index is $-0.4\pm0.05$.

The central 35$\arcsec$ region (at $\rm{RA} = 20\rm{h}34\rm{m}52\rm{s}$,
$\rm{dec} = +60\rm{d}09\arcmin14\arcsec$ (J2000))
has a flux density of 226$\pm$13 mJy at 333
MHz while at 1465 MHz the same region has a flux density of 117$\pm$6 mJy.  This
corresponds to an $\alpha$ of $-0.44\pm0.05$ at a
resolution of $15\arcsec\times15\arcsec$.  However, after the
separation, similar region has a $\ant$ of
$-0.54\pm0.1$ (see Figure~\ref{figure3}).  In the nuclear region about
15\% of the total emission is due to thermal emission at 333 MHz,
higher than the rest of the galaxy, where the thermal emission lies within 10\%.

\section{Summary}
\label{section6}

The 333 MHz GMRT observations of six nearby galaxies (greater than
10$\arcmin$ in size) presented in this paper are by far the highest
angular resolution maps for these galaxies at this frequency which are
sensitive to both small ($\sim 0.5$ kpc) and large ($\gsim10$ kpc)
scale structures.  The
hybrid array of GMRT enables us to recover emission from both the
diffuse and small scale emission in these sources, which we have
adequately verified by comparing our estimated integrated flux
densities with those available in the literature. Our observations
have comparable resolution as that attained at higher frequencies, and
probe linear scales of about 0.4--1 kpc.

At 333 MHz the galaxies appear smoother than at higher frequencies,
where spiral arms are often discernable.  The thermal emission (which
is primarily associated with ionized structures observed in H$\alpha$
maps) for all the galaxies was found to be less than 5\%
of the total emission, and hence at these frequencies, the emission is essentially 
nonthermal. The smooth appearance is consistent with the
conjecture that nonthermal emission at 333 MHz primarily results from
a population of old ($> 10^7$ yrs) CRe, which diffuses away from their
formation sites filling up a volume of radius $\gsim$ 1 kpc, without
loosing much energy.  This scale is larger than the width of the
gaseous spiral arms and comparable to the width of the inter-arm
region (see  e.g. \citealt{condo92}), hence reducing any contrast
in intensity across the galaxy.  The prominent radio structures
observed at higher frequencies ($>$ 1 GHz) are visible solely due to the
increased thermal fraction.

\begin{figure}
\begin{centering}
\includegraphics[width=8cm,height=8cm,angle=-90]{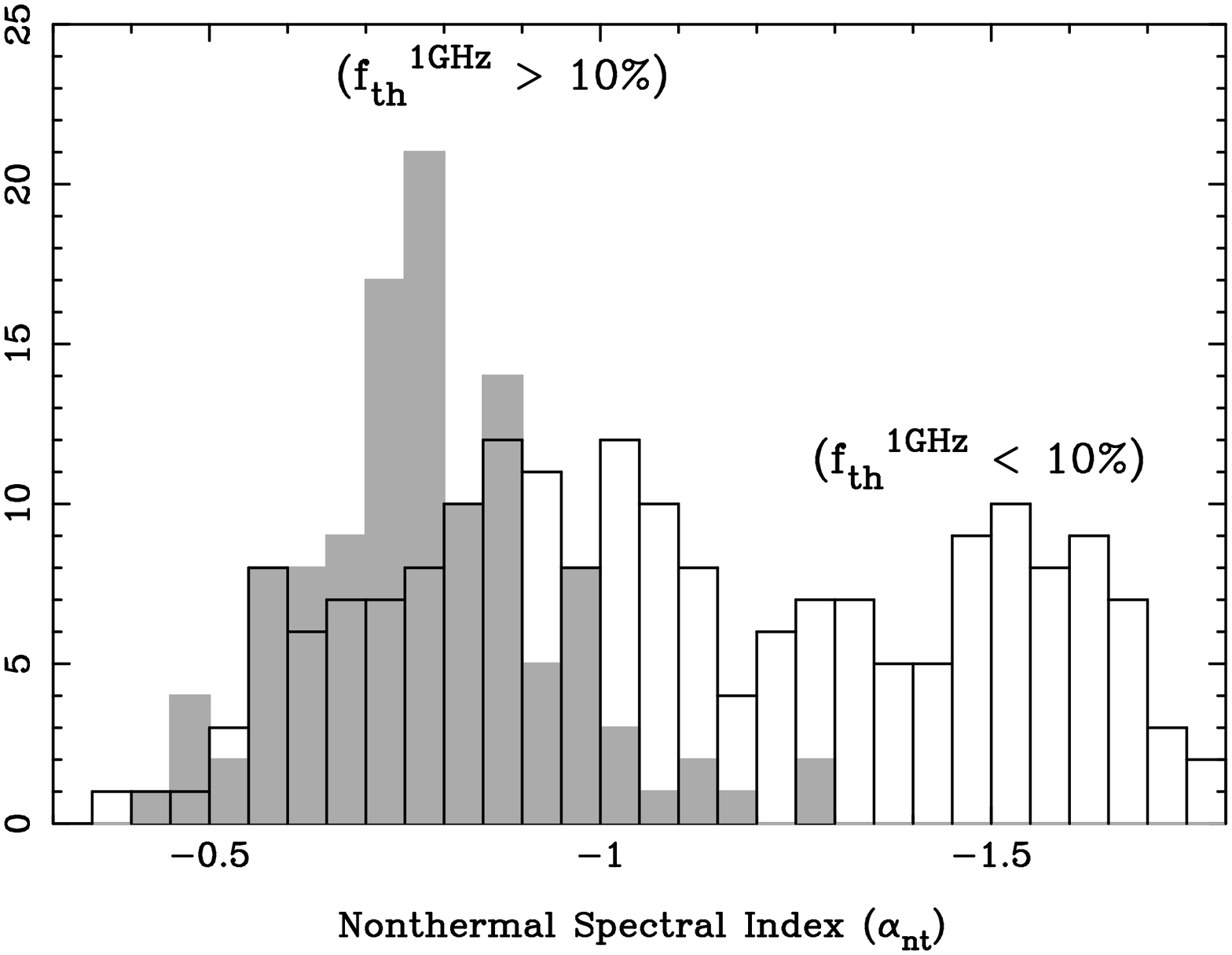}
\caption{ The above plot shows two histograms of nonthermal spectral 
index for 4 galaxies for two range of thermal fractions. The filled grey histogram corresponds
to thermal fraction $\fth^{1.4 \rm{GHz}} >10\%$ and the unfilled histogram is for $\fth^{1.4 \rm{GHz}}< 10\%$.}
\label{figure9}
\end{centering}
\end{figure}

We have determined robust estimates of spatially resolved $\ant$ (see
Section~\ref{section3}) for five galaxies in our sample, with a
resolution of 40$\arcsec$, corresponding to a linear scale of 1--3
kpc, which also corresponds to the diffusion scale of the CRe. The
observed $\ant$ ranges from $-0.3$ to $-1.8$ in various regions of the
galaxies. Generally CRe are thought to be accelerated at SNR where
both the theoretical prediction \citep{bell78} and observation based
on Galactic SNR suggest mean $\ant\sim-0.5$ \citep{green98,kothe06}.
The SNR are short lived ($\sim 10^5$ yrs), while the CRe diffuse away
from their acceleration sites losing their energy through several
physical mechanisms for typically 10$^8$ yrs, and hence their initial
energy spectrum gets distorted leading to a change in $\ant$. If the
energy losses are due to ionization, $\ant$ flattens
\citep{longa11}. Synchrotron radiation and inverse Compton scattering
loss leads to a steepening of the spectra, while adiabatic cooling and
bremsstrahlung keep the spectra unchanged.  To assess the behaviour of
$\ant$ in spatially resolved parts of the galaxy we plot $\ant$
distribution (as shown in Figure~\ref{figure9}) for two regimes of
thermal fraction: $\fth^{1\rm GHz} > 10\%$ (filled grey histogram) and
$\fth^{1\rm GHz} < 10\%$ (unfilled histogram), using data from 4
galaxies\footnote{We did not use NGC 5055, since we think that there
  is a missing flux density problem at 1690 MHz and hence the
  derived $\alpha$ and $\ant$ might be steeper than the
actual value.}.  To do this, we have computed $\ant$ by dividing the galaxy
in rectangular grids of 40$\arcsec$. The histograms are binned with
step size 0.05 in $\ant$.  The high thermal fraction corresponds to
bright regions in H$\alpha$ which traces gas ionized by massive OB
stars in star forming regions. These regions host the CRe generation
and acceleration sites. The distribution of $\ant$ for such high
thermal fraction regions has a Gaussian shape, with mean $\ant \sim
-0.78$ and narrow spread of 0.16. It is interesting that this narrow
distribution arises from multiple spatially resolved regions from an
assorted set of galaxies with very different star formation histories.
This is clearly indicative of a generic process of energy loss of CRe
by synchrotron radiation and inverse Compton scattering as they
propagate from their acceleration sites.  The flatter part of $\ant$
($> -0.55$) primarily arises from NGC 5236, indicating that ionization
losses are dominant \citep{longa11}.  \cite{nikla97}, based on
integrated flux density of 74 galaxies found the average
$\ant\sim-0.83$ which is comparable to the $\ant$ value that we have
obtained. The galaxy integrated spectral index is an average over tens
of kpc where the flux density is highly dominated by bright regions
which are mostly spiral arms or \hii regions. Our analysis shows that
the contribution to the average spectral index primarily arises at
least on kpc scales.

The unfilled histogram, corresponding to low thermal 
fraction ($\fth^{1\rm GHz}<10\%$) regions, has a much larger spread ranging from
$\ant \sim -0.4 ~\rm{to} -1.8$. The steepening in the range $-1.3 < \ant < -1$ arises from the
interarm regions, and indicates energy losses through synchrotron emission and inverse Compton
scattering. $\ant<-1.3$, is mostly from low flux density regions, primarily in the 
outer parts of the galaxies, and may be affected due to missing flux density in the
higher radio frequency maps.
Note that we have averaged in a region 40$\arcsec$, which
for all these galaxies corresponds to $\sim$ 1 kpc to 3 kpc, and hence
the flatter part of $\ant$ distribution may be affected due to
bright regions. As mentioned earlier, the absolute value of
$\ant$ can change by $\sim$10\% due to uncertainty in flux scale at
333 MHz. A recent study by \cite{palad09} for NGC 628, NGC 3627 and
NGC 7331, using infrared emission ($\lambda$70$\mu$m) as a tracer of
star forming regions, found that $\alpha$ obtained using VLA observations at 325 MHz
and near 1.4 GHz, is flatter for infrared bright regions. 
This is similar to what we observe in our set of galaxies. 
The high angular resolution 333 MHz
observations presented in this paper can be used 
to study in detail the spatially resolved FIR
-- radio and CO -- FIR correlations, which will be reported in a future
paper.

\appendix
\section{Thermal Nonthermal Flux Separation}
\label{appendixa1}

The \cite{tabat07} method uses an H$\alpha$ map of the galaxy as a
template of thermal emission. Since H$\alpha$ emission is affected by
extinction, far-infrared data at $\lambda$70$\mu$m and
$\lambda$160$\mu$m are used to obtain a model for extinction, which is
then used to correct the H$\alpha$ data. The corrected H$\alpha$ image
is then used to predict the thermal emission at radio frequencies of
interest (the whole process is described in detail in sections 3 to 6
of \citealt{tabat07}). In this section, we summarize the various steps and the
intermediate results that we obtained in the separation process.

\begin{itemize}
\item {\bf Data sets:} The far infrared images at $\lambda70\mu$m and
$\lambda160\mu$m were obtained from the publicly available data from
the {\it Spitzer Infrared Nearby Galaxy Survey} ({\sc sings}; \citealt{kenni03}). All
$\lambda70\mu$m images have a pixel size of 4.5$\arcsec$ and a {\it
point spread function} (PSF) of about 16$\arcsec$. At
$\lambda160\mu$m, each pixel is 9$\arcsec$ in size and has a PSF of
40$\arcsec$. Both the $\lambda70\mu$m and $\lambda160\mu$m maps are
calibrated in surface brightness units of $\rm MJy~sr^{-1}$.

The continuum subtracted H$\alpha$ images for NGC 1097 (1.5m CTIO,
filter: CT6586) and NGC 6946 (KPNO 2m, filter: KP1563) were
obtained from the ancillary data at the {\sc sings} website. The maps
were in units of $\rm {DN~s^{-1}pixel^{-1}}$, which was converted to
$\rm{erg~s^{-1}cm^{-2}}$ using the calibration provided in the {\sc sings} Fifth
Data Delivery documentation\footnote{http://data.spitzer.caltech.edu/popular/sings/20070410
\_enhanced\_v1/Documents/sings\_fifth\_delivery\_v2.pdf}.

The continuum subtracted H$\alpha$ images for NGC 4736, observed
with the 1m JKT at La Palma, filter: Ha6570, \citep{knape04}, NGC 5055
observed with the 2.3m telescope at KPNO, filter: 6580 and NGC 5236 observed 
with the 0.9m telescope at CTIO,
filter:6563  were downloaded from the NED.

The counts per second (cps), in these maps were converted to apparent magnitude, 
$m_{\rm AB}$, using the zero-point given in the FITS header. The apparent 
magnitude $m_{\rm AB}$ was further converted to
specific intensity, using $f_\nu (\rm{erg~s^{-1}cm^{-2}Hz^{-1}}) =
10^{-(m_{\rm AB}+48.6)/2.5}$. The flux in units of $\rm{erg~s^{-1}cm^{-2}}$ was
computed using, $f = f_\nu {\rm d\nu} = f_\nu c\frac{\rm d\lambda}{\lambda^2}$.

All the $\lambda70\mu$m and continuum subtracted H$\alpha$ maps were
convolved to the coarser 40$\arcsec$ PSF of $\lambda160\mu$m, and
pixels were re-gridded to 9$\arcsec$. All the subsequent calculations
were done on a pixel by pixel basis.
 
\item {\bf Dust Temperature and Optical Depth:} 

Following \cite{tabat07}, as a first step for extinction correction of
H$\alpha$ maps we need to estimate the optical depth due to obscuring
dust.  For this the colour temperature of the dust ($T_{\rm dust}$)  is
found by fitting a black body spectrum incorporating dust absorption
efficiency (see equation 1 of \citealt{tabat07}) to the far infrared maps
at $\lambda$70$\mu$m and $\lambda$160$\mu$m.  The distribution of the
dust temperature for all the five galaxies is shown in the left 
panel of Figure~\ref{figurea1}. The mean temperature estimated for all
the galaxies is between 19$-$22 K.  Note that a single dust
temperature model has been assumed.  The optical depth at 160$\mu$m
($\rm \tau_{160\mu m}$) was then obtained by using equation 2 of
\cite{tabat07} using the estimated $T_{\rm dust}$.  The H$\alpha$ optical
depth was obtained as $\tau_{\rm H\alpha} \sim f_{\rm d}\times2200\times\tau_{160\mu
 \rm m}$ \citep{kruge03} with H$\alpha$ filling factor $f_{\rm d} = 0.33$ \citep{dicki03}.
For all the five galaxies the $\tau_{\rm H\alpha}$ lies in the range 0.03 -- 0.6 at linear 
scales $>$ 1 kpc, with the highest values being towards the center of the galaxies.

\item {\bf Emission Measure:} 

Extinction corrected H$\alpha$ maps was used to estimate the emission
measure ($EM = \int n_{\rm e}^2 ~{\rm d}l$ pc cm$^{-6}$, where $n_{\rm e}$ is the
thermal electron density) using equation 3 and 4 of \cite{tabat07} (also see
\citealt{valls98}) and assuming an electron temperature of $T_{\rm e} =
10^4$~K.  The emission measure radial profile for all the five
galaxies is shown in the middle panel of Figure~\ref{figurea1}. The
emission measure towards the center of the galaxy lies in the range of
$\sim 10^3-10^4$~pc cm$^{-6}$ and falls off to $\sim 10-10^2$~pc
cm$^{-6}$ towards the edge of the galaxy.  These range of emission
measure values are consistent with those observed in the Milky way (see
e.g. \citealt{berkh06}).

\item {\bf Thermal emission at Radio frequencies:} 

Equation 5 and 6 of \cite{tabat07}, were used to determine the radio
continuum optical depth and the brightness temperature ($T_{\rm B}$) using
the Rayleigh-Jeans approximation. The thermal flux density
($S_\nu,{\rm th}$) at a radio frequency $\nu$ was obtained from the $T_{\rm B}$
using,
$$
\frac{S_{\nu,\rm th}}{\rm Jy~beam^{-1}} = 8.18\times10^{-7} \left(\frac{\theta_{\rm maj}}{\rm arcsec}\right)\left(\frac{\theta_{\rm min}}{\rm arcsec}\right)\left(\frac{\nu}{\rm GHz}\right)^{2}\frac{T_{\rm B}}{\rm K}
$$
Here, $\theta_{\rm maj}$ and $\theta_{\rm min}$ size of the restoring beam in
the radio map (in this case 40$\arcsec$).

For all the five galaxies, the pixel-wise distribution of the thermal
fraction ($\fth = S_{\nu,{\rm th}}/S_{\nu,{\rm tot}}$) at 333 MHz and near 1 GHz
are shown in the rightmost panel of Figure~\ref{figurea1}. The mean
$\fth$ at 333 MHz and near 1 GHz for all the galaxies were found to
be less than 5\% and less than 12\% respectively. However, in certain
giant \hii regions, the thermal fraction can go up to 10\% at 333 MHz
and 30\% near 1 GHz. \cite{tabat07} pointed out that the primary
source of error in the thermal fraction arises from the unknown value
of the $T_{\rm e}$. Using similar arguments, we estimate the uncertainty in
the thermal fraction at 333 MHz and near 1 GHz to be $\sim$ 10\% and
$15$\% respectively.

\end{itemize}

\begin{figure*}
 \includegraphics[width=5.5cm,height=5.5cm,angle=0]{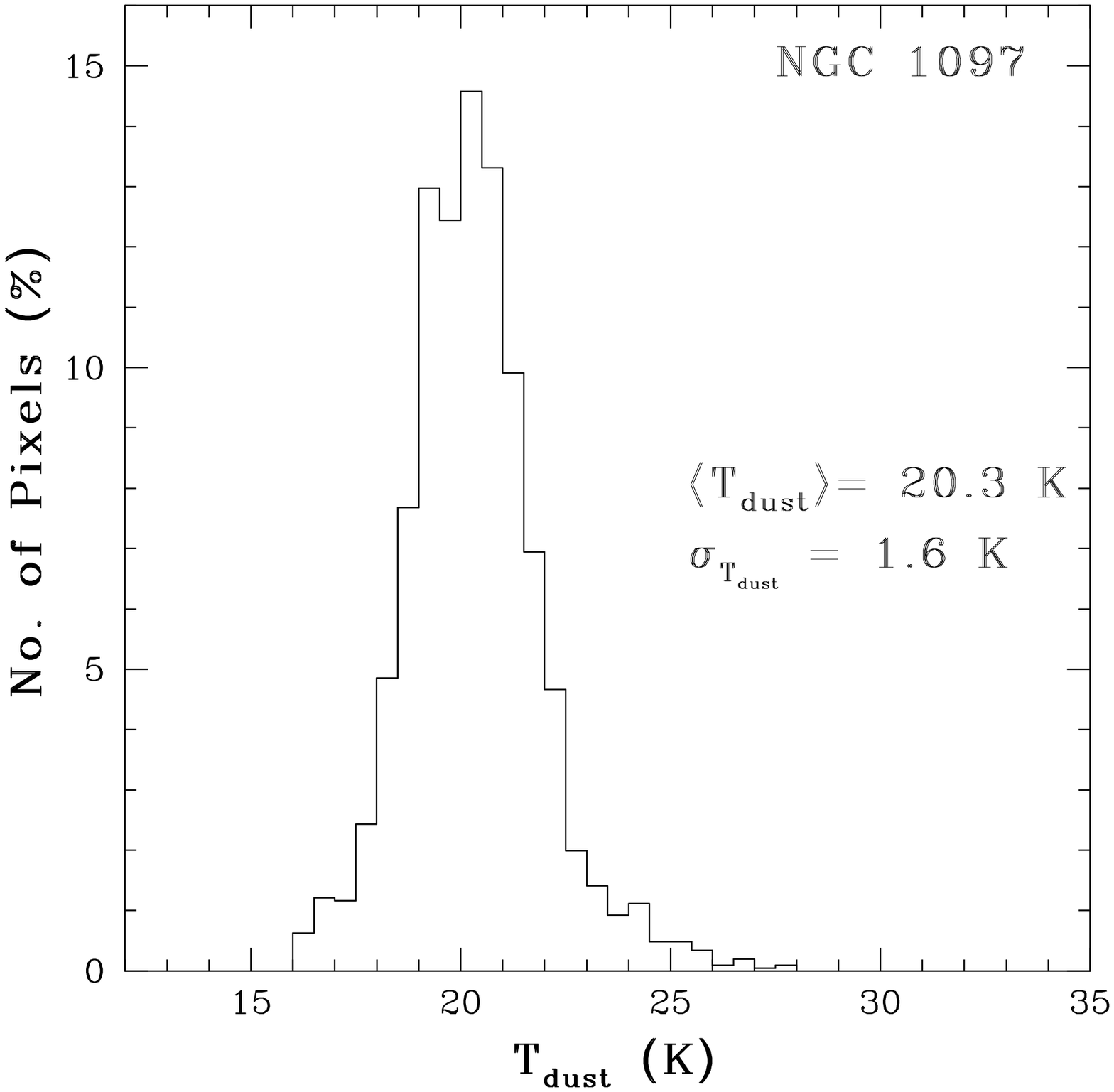}
 \includegraphics[width=5.5cm,height=5.5cm,angle=0]{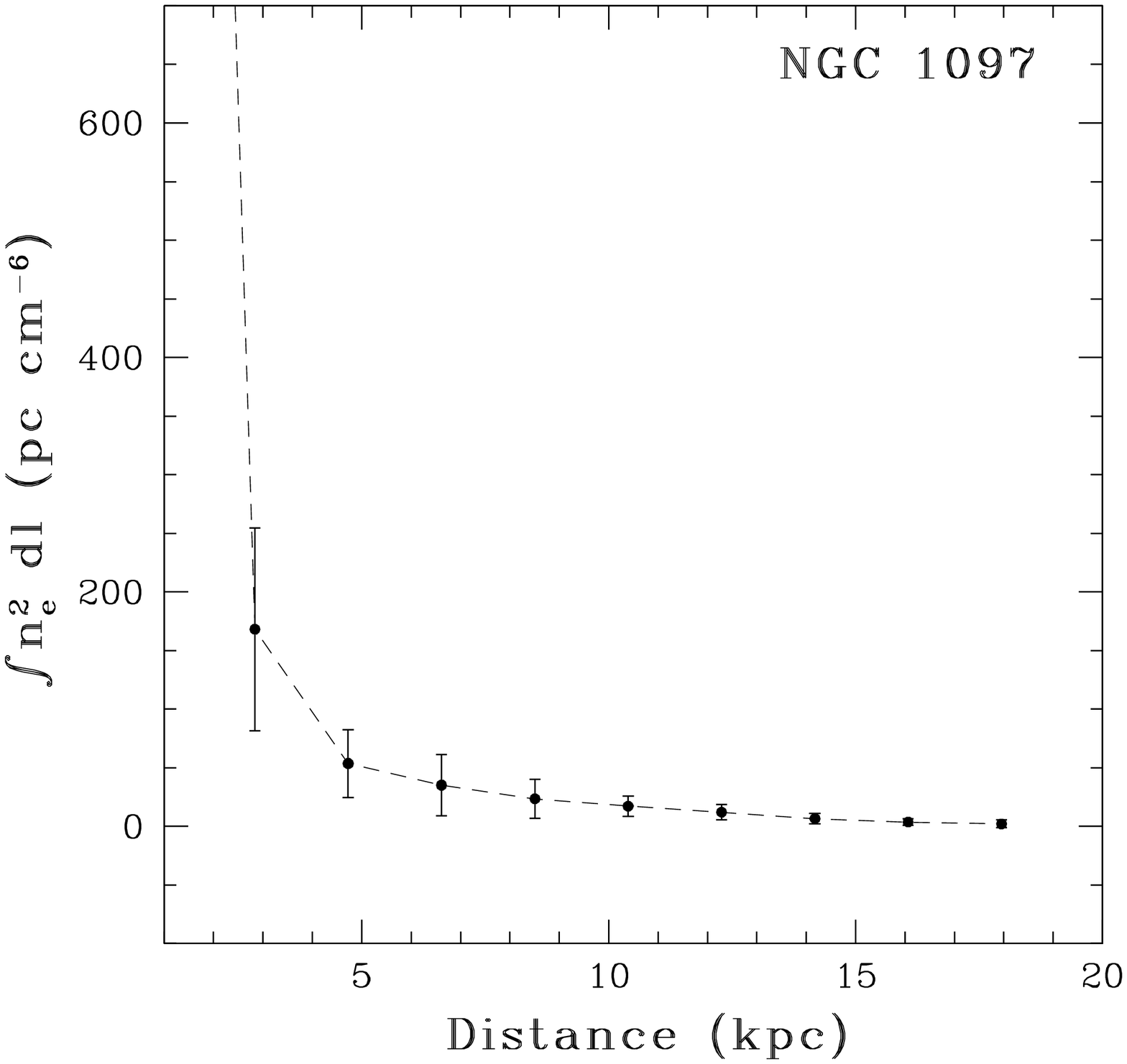}
 \includegraphics[width=5.5cm,height=5.5cm,angle=0]{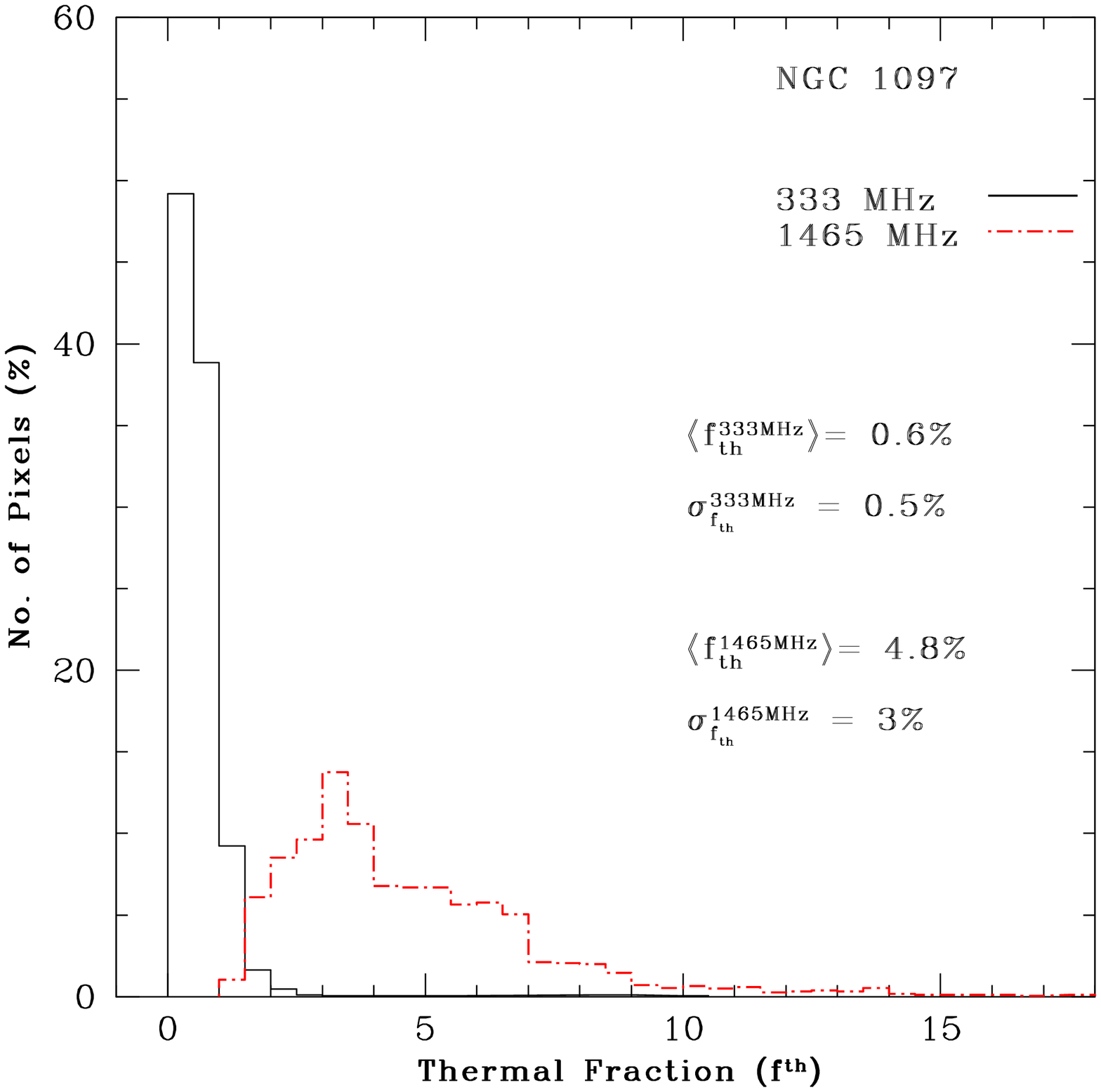}
 \includegraphics[width=5.5cm,height=5.5cm,angle=0]{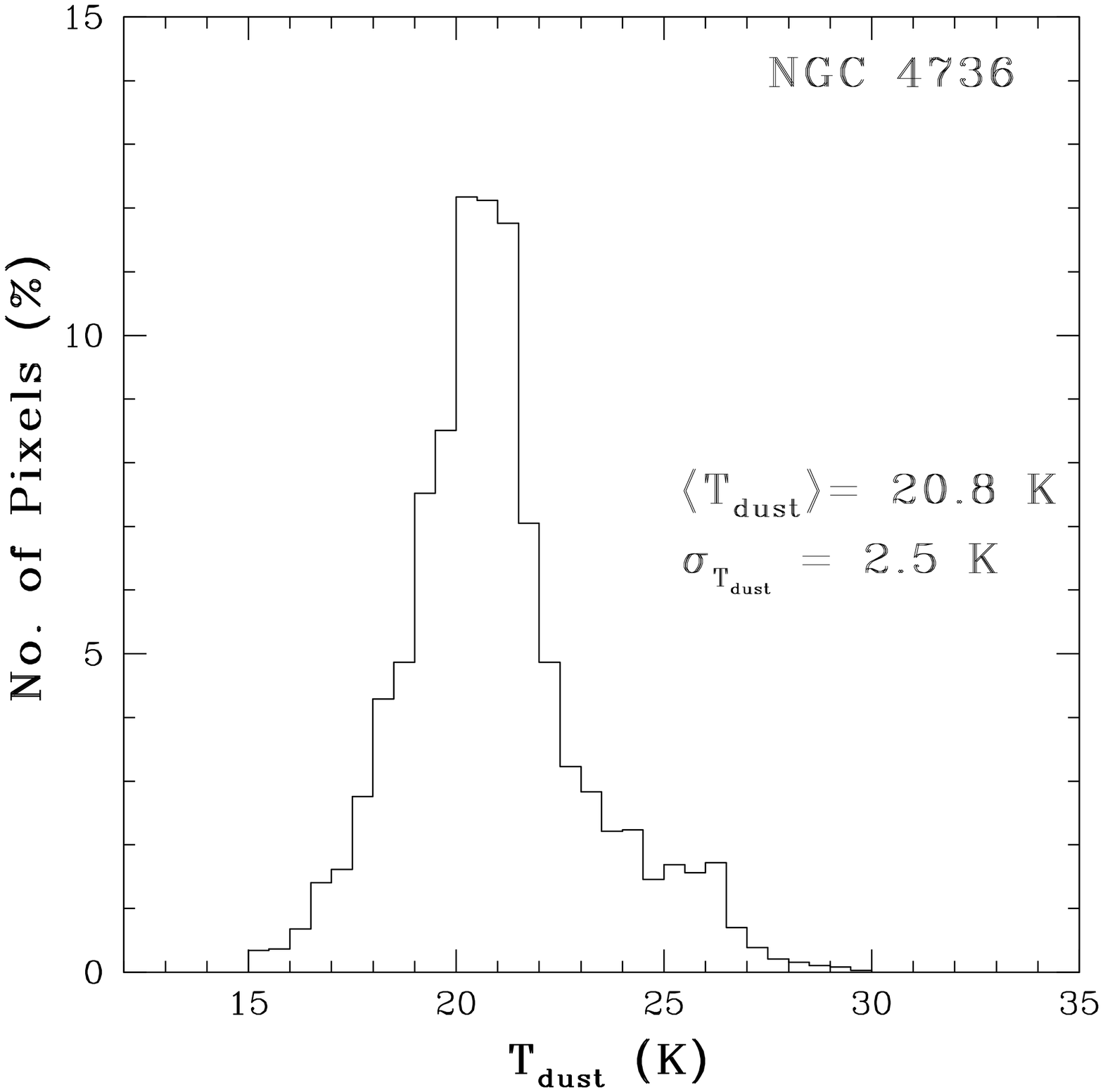}
 \includegraphics[width=5.5cm,height=5.5cm,angle=0]{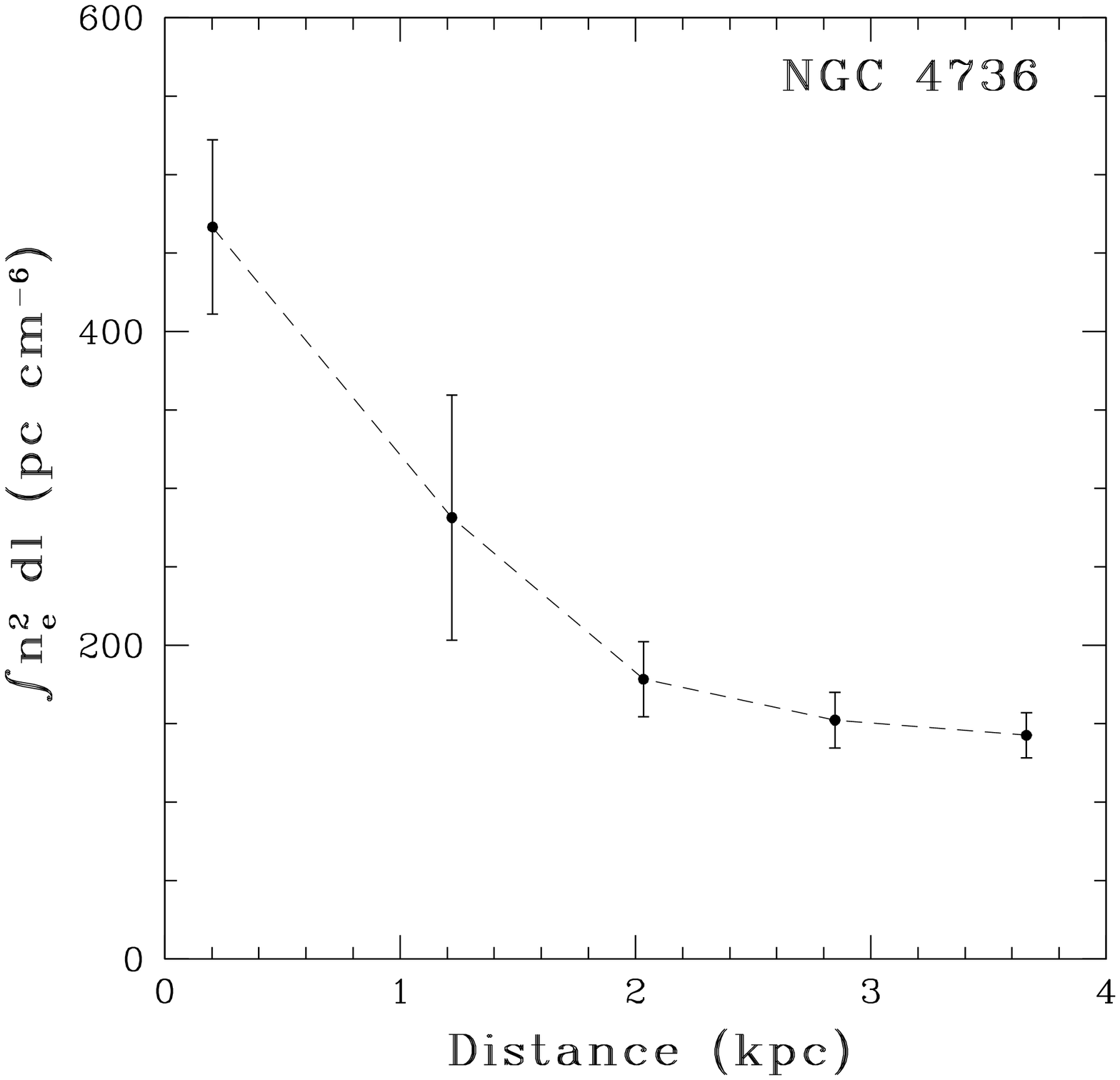}
 \includegraphics[width=5.5cm,height=5.5cm,angle=0]{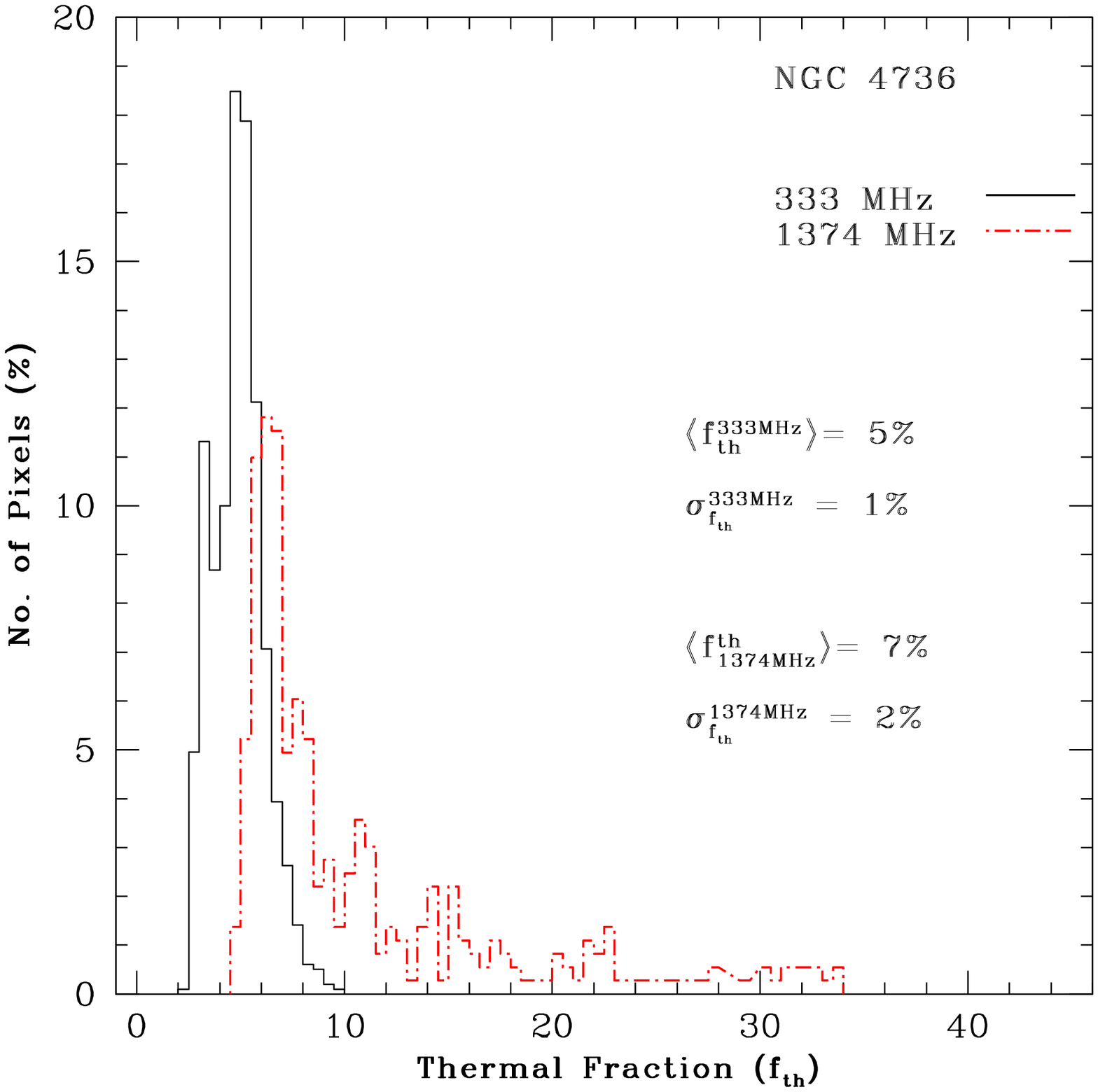}
 \includegraphics[width=5.5cm,height=5.5cm,angle=0]{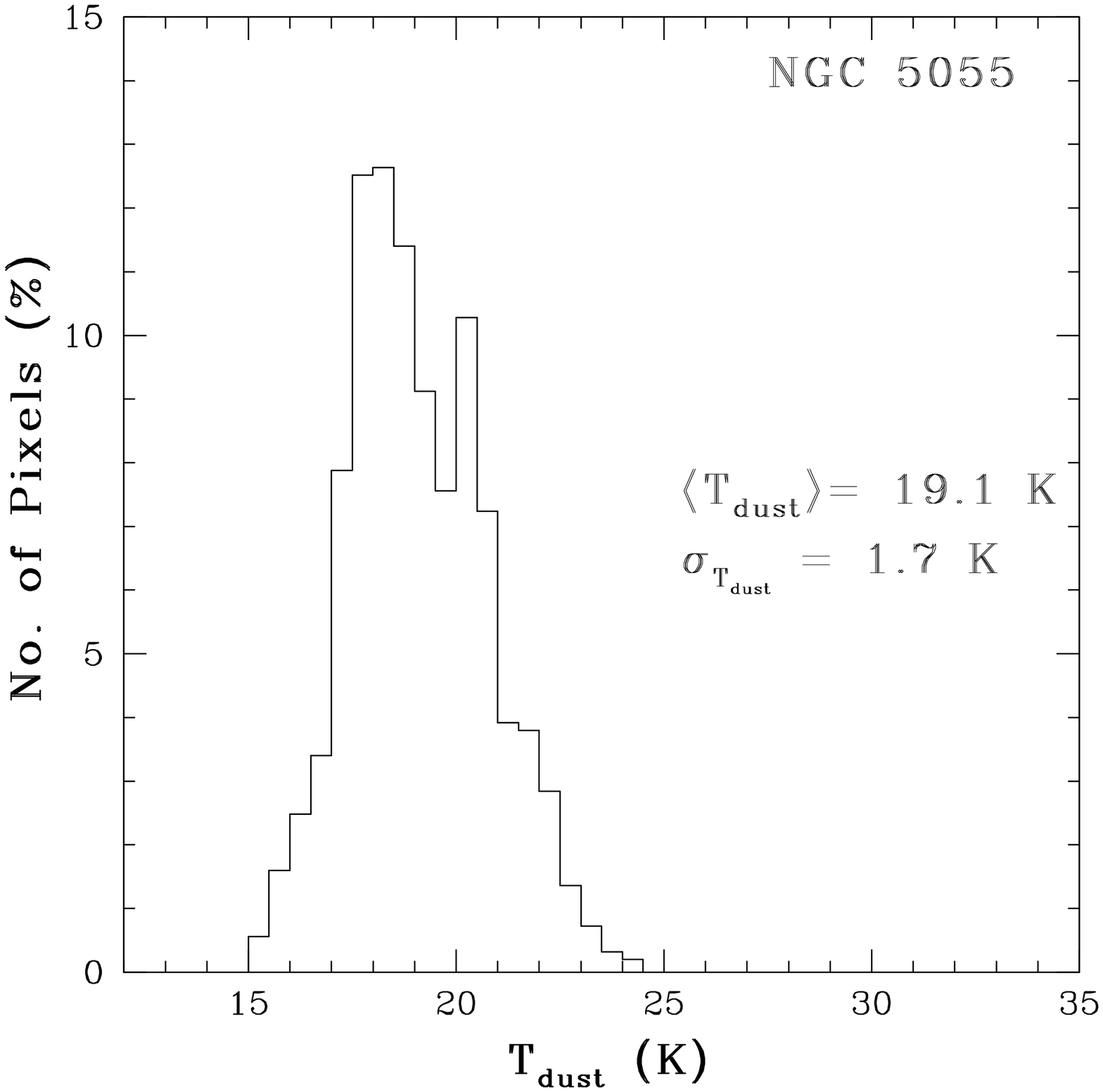}
 \includegraphics[width=5.5cm,height=5.5cm,angle=0]{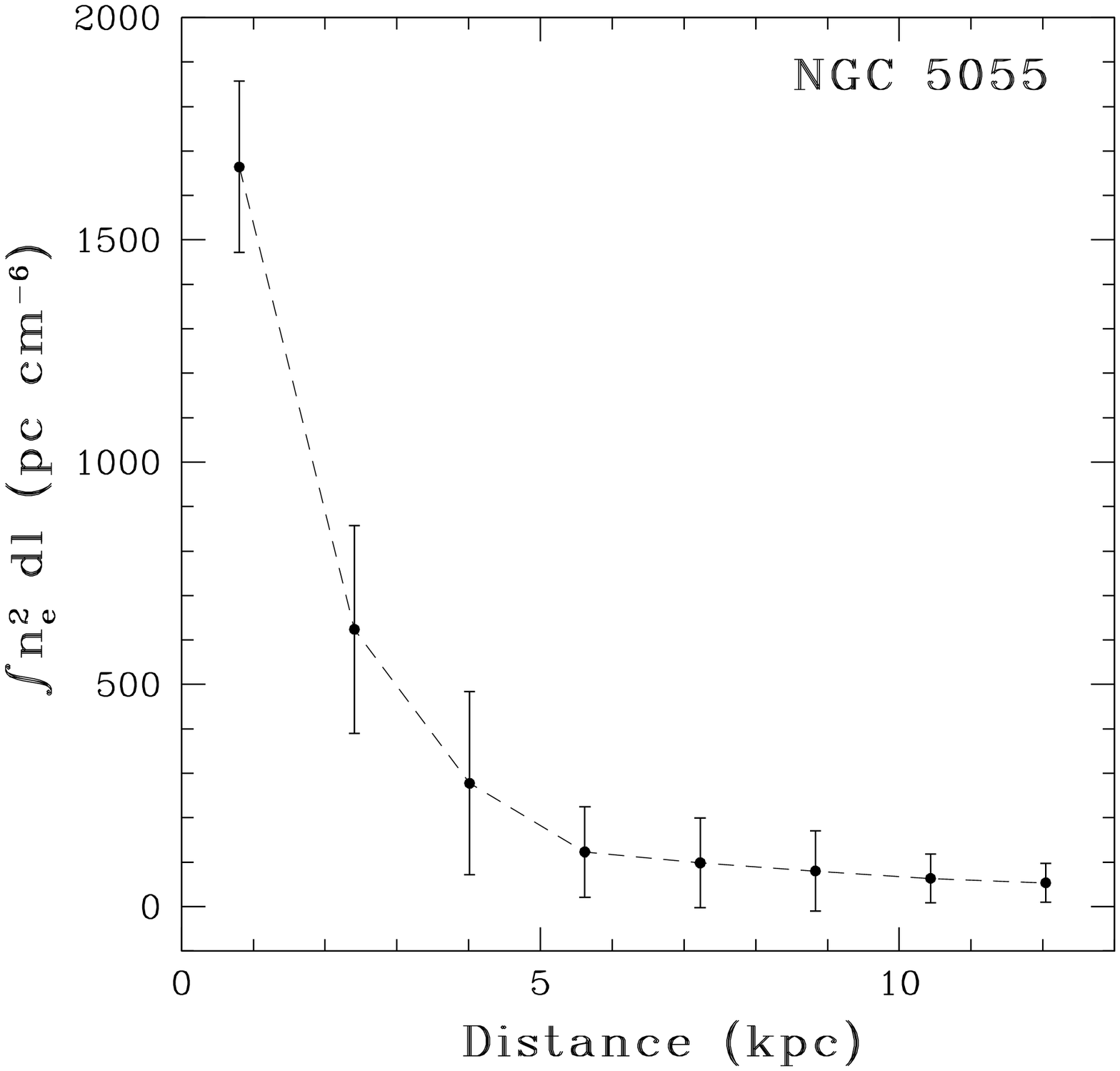}
 \includegraphics[width=5.5cm,height=5.5cm,angle=0]{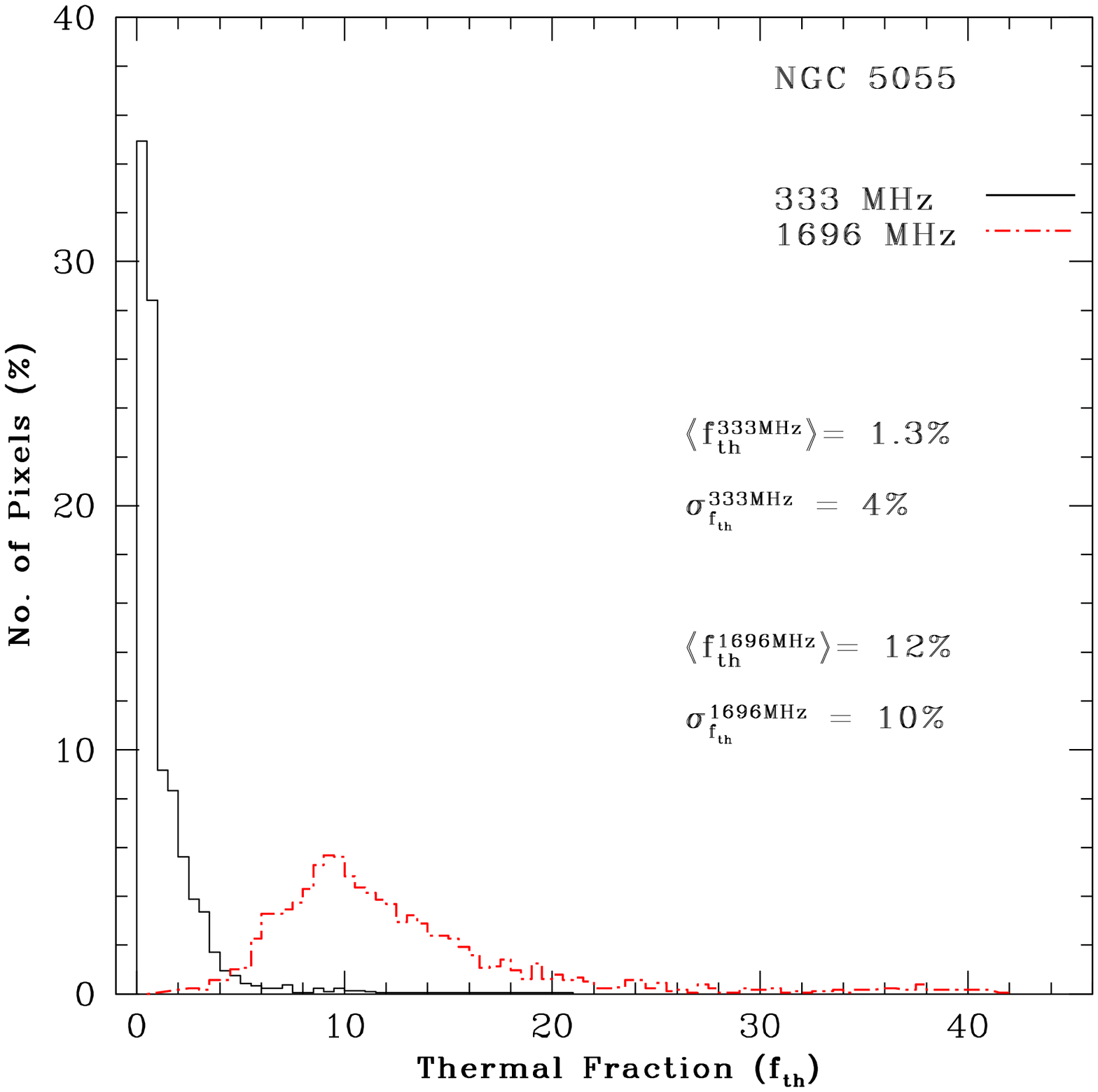}
 \caption{{\it Left panel:} The pixel-wise distribution of the estimated dust temperature ($\rm T_{dust}$).
The bins are at intervals of 0.5 K.
{\it Middle panel:} The radial profile of the estimated emission measure ($EM = \int n_{\rm e}^2~{\rm d}l$ $~~\rm pc~cm^{-6}$).
The {\it EM} is azimuthally averaged over annuli of one beamwidth, i.e, $40\arcsec$.
{\it Right panel:} The pixel-wise distribution of the thermal fraction ($\fth$) with a 
bin size of 0.5\%. The solid black histograms show the distribution at 333 MHz, 
while the dashed red histograms show the distribution near 1.4 GHz. The {\it top}, 
{\it middle} and 
{\it bottom} rows are for NGC 1097, NGC 4736 and NGC 5055 respectively.
Each pixel is of size 10$\arcsec$, corresponding to a physical 
scale of about 700 pc, 400 pc and 340 pc respectively.}
\label{figurea1}
\end{figure*}

\addtocounter{figure}{-1}
\begin{figure*}
 \includegraphics[width=5.5cm,height=5.5cm,angle=0]{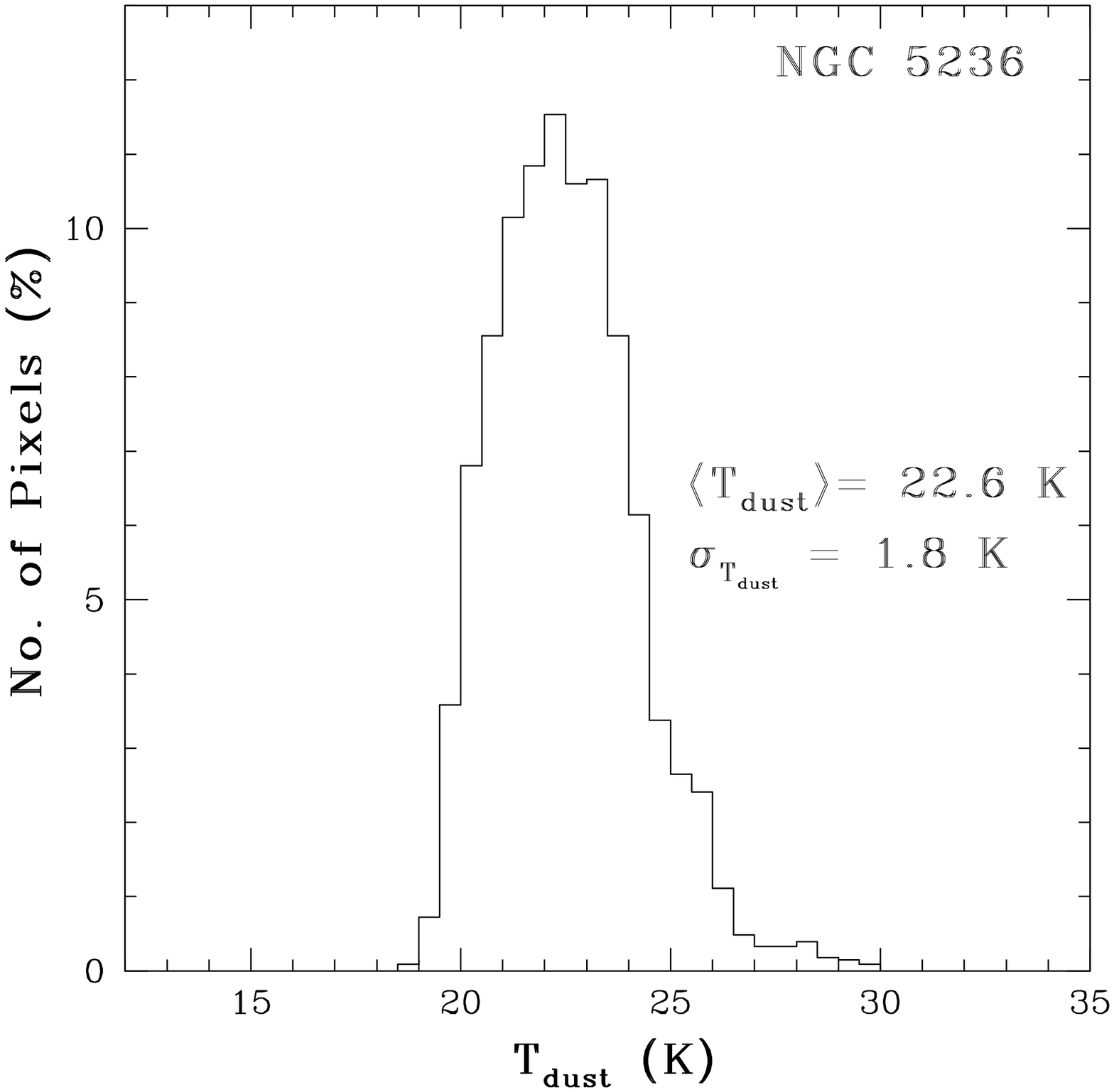}
 \includegraphics[width=5.5cm,height=5.5cm,angle=0]{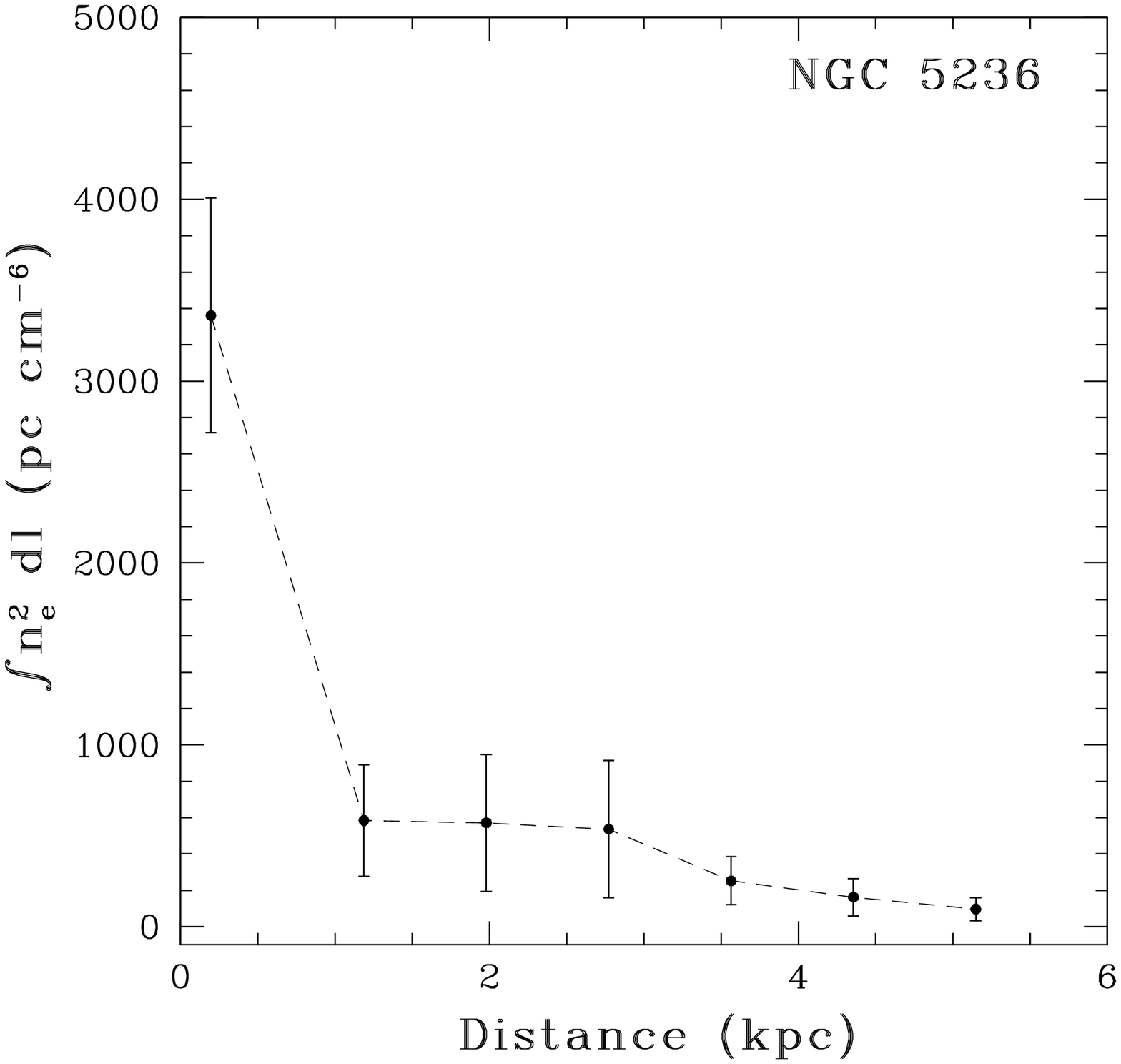}
 \includegraphics[width=5.5cm,height=5.5cm,angle=0]{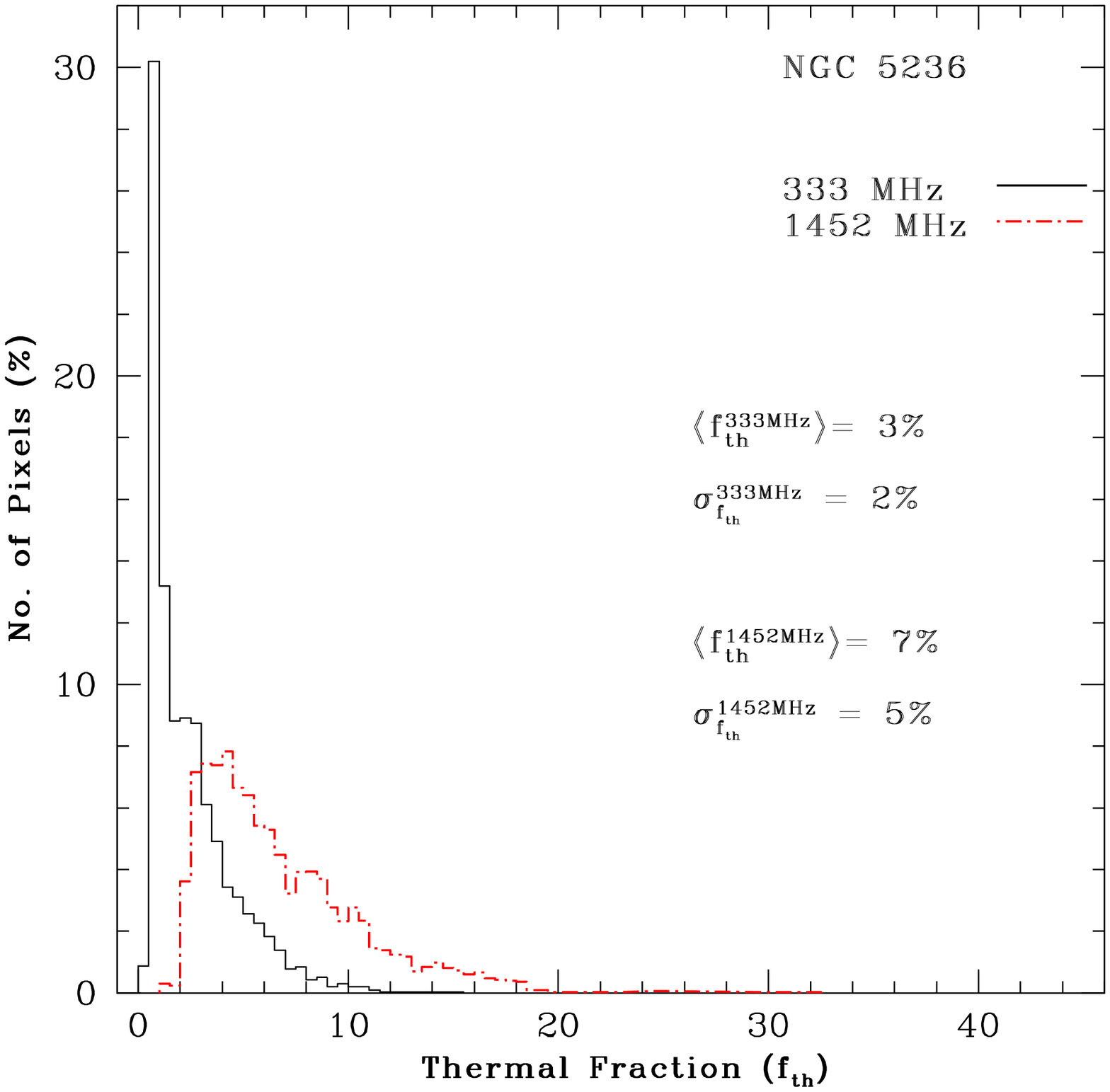}
 \includegraphics[width=5.5cm,height=5.5cm,angle=0]{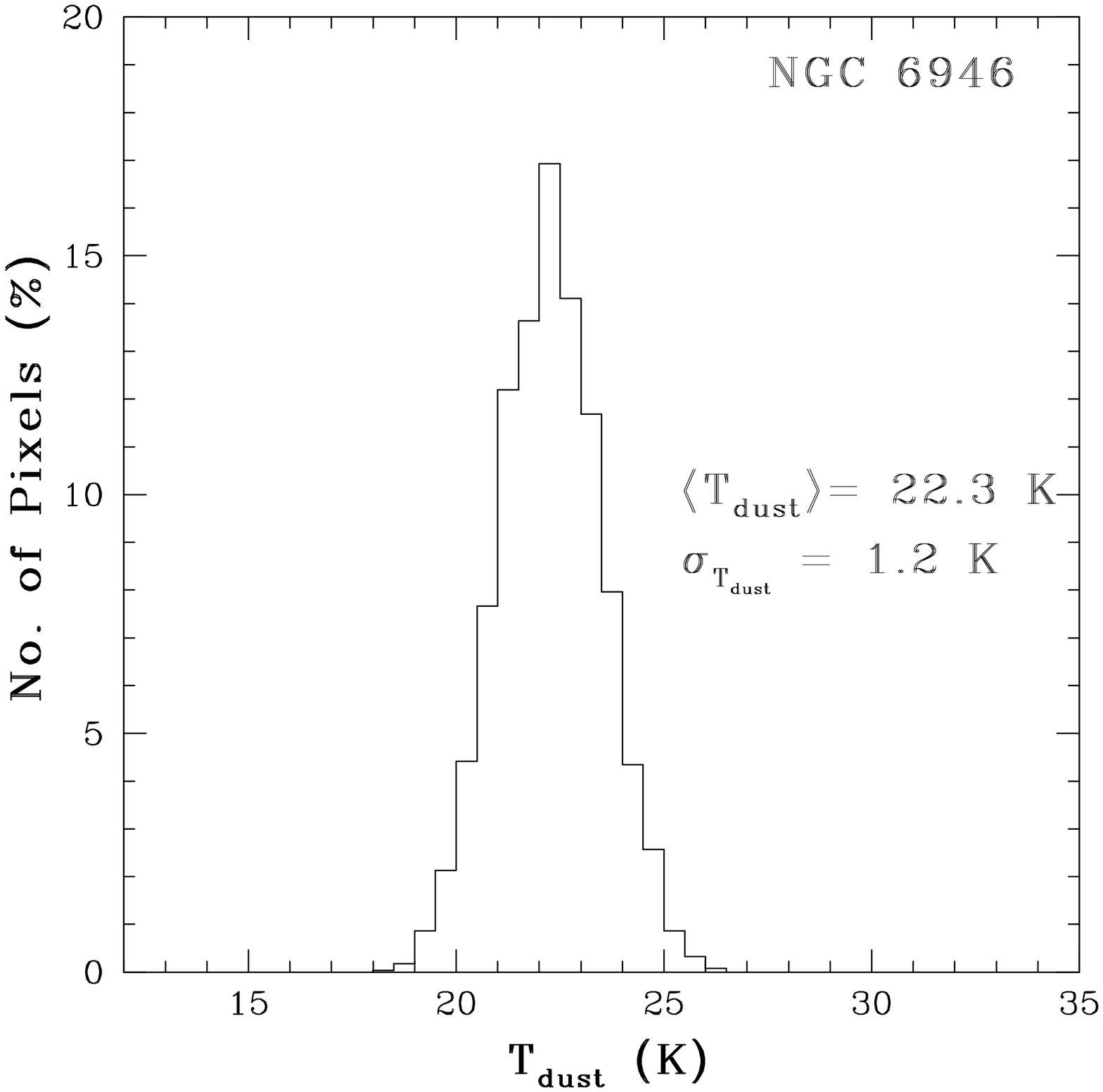}
 \includegraphics[width=5.5cm,height=5.5cm,angle=0]{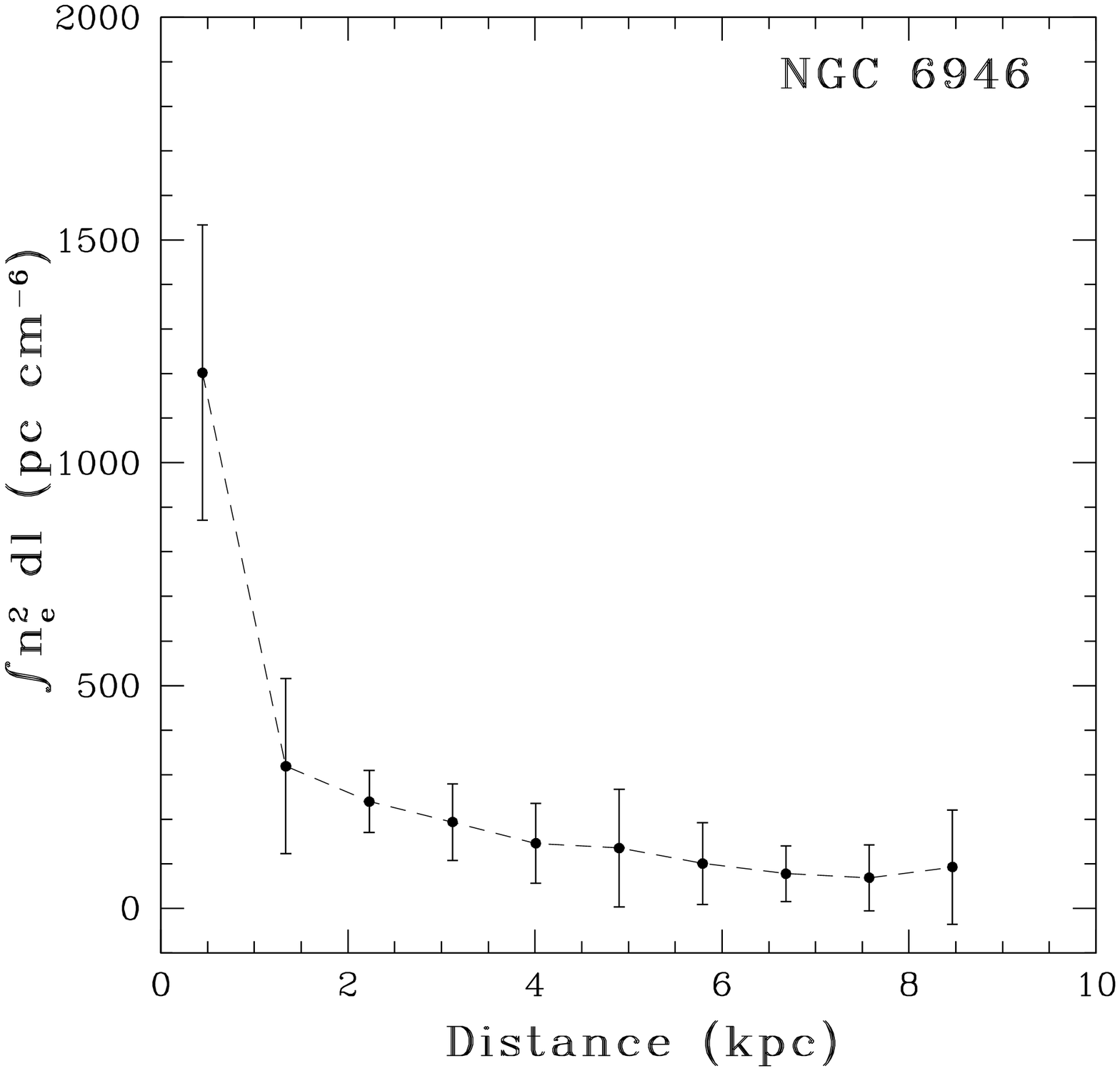}
 \includegraphics[width=5.5cm,height=5.5cm,angle=0]{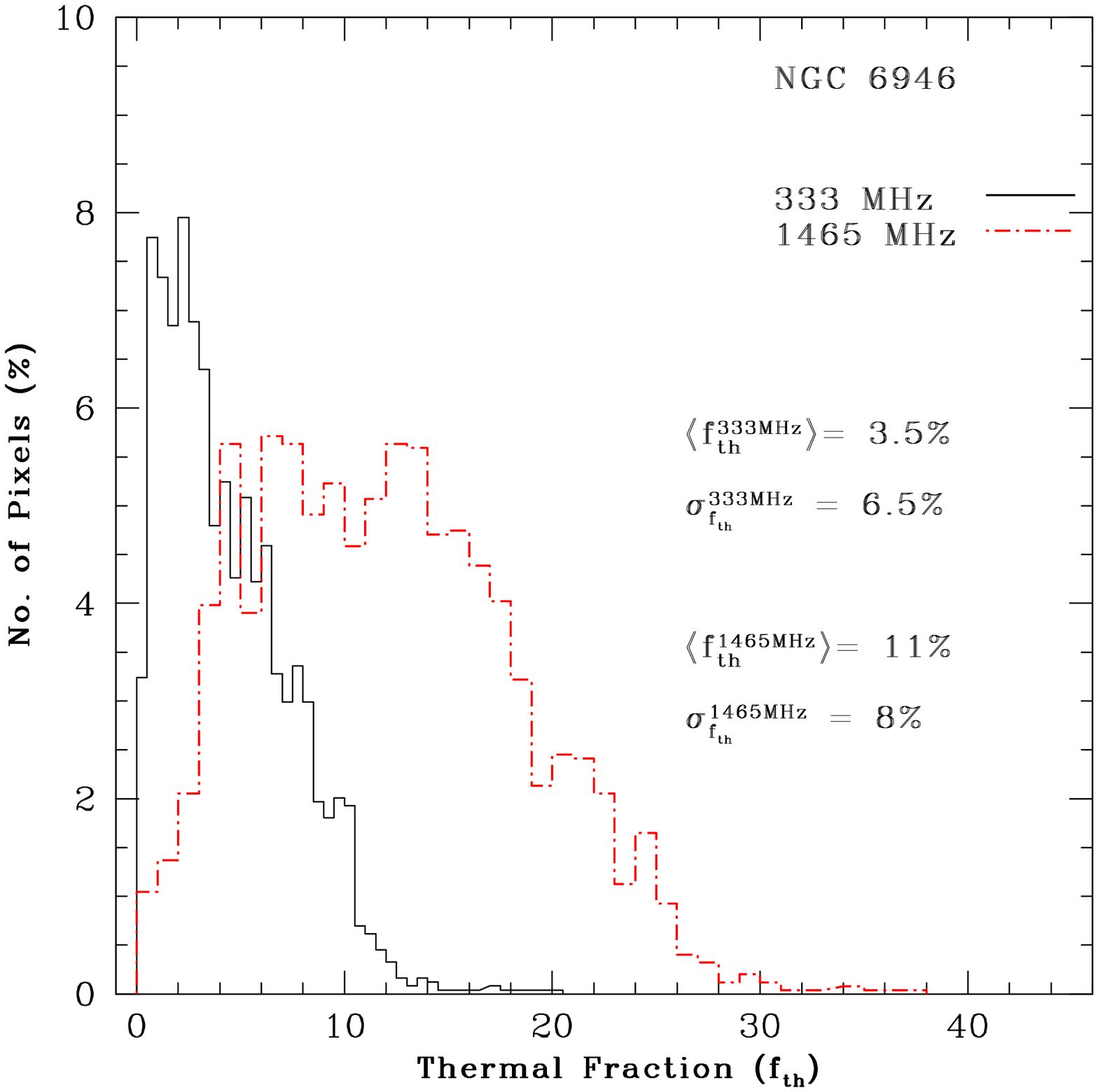}
 \caption{{\it contd...} {\it Left panel:} The pixel-wise distribution of the estimated dust 
temperature ($\rm T_{dust}$). The bins are at intervals of 0.5 K.
{\it Middle panel:} The radial profile of the estimated emission measure ($EM = \int n_{\rm e}^2~{\rm d}l$ $~~\rm pc~cm^{-6}$).
The {\it EM} is azimuthally averaged over annuli of one beamwidth, i.e, $40\arcsec$.
{\it Right panel:} The pixel-wise distribution of the thermal fraction ($\fth$) with a bin size of
0.5\%. The solid black histograms show the distribution at 333 MHz, 
while the dashed red histograms show the distribution near 1.4 GHz. The {\it top} and 
{\it bottom} rows are for NGC 5236 and NGC 6946 respectively.
Each pixel is of size 10$\arcsec$, corresponding to a physical 
scale of about 220 pc and 330 pc respectively.}
\label{figurea1c}
\end{figure*}

\section*{Acknowledgments}

We thank the referee, Elly Berkhuijsen and the editor, for insightful comments that
have improved both the content and presentation of this paper. We
thank the GMRT staff for technical support during the observations.
GMRT is run by the National Centre for Radio Astrophysics of the Tata
Institute of Fundamental Research.  This research has made use of the
NASA/IPAC Extragalactic Database (NED) which is operated by the Jet
Propulsion Laboratory, California Institute of Technology, under
contract with the National Aeronautics and Space Administration.  The
Digitized Sky Surveys were produced at the Space Telescope Science
Institute under U.S. Government grant NAG W-2166. The images of these
surveys are based on photographic data obtained using the Oschin
Schmidt Telescope on Palomar Mountain and the UK Schmidt Telescope.
This work is based [in part] on observations made with the Spitzer
Space Telescope, which is operated by the Jet Propulsion Laboratory,
California Institute of Technology under a contract with NASA.

\label{lastpage}


\begin{thebibliography}{99}
\bibitem[\protect\citeauthoryear{Arp}{1973}]{arp73} Arp, H., 1973, ApJ, 183, 791 
\bibitem[\protect\citeauthoryear{Baars et al.}{1977}]{baars77} Baars J. W. M., Genzel R., Pauliny-Toth I. I. K., et al., 1977, A\&A, 61, 99
\bibitem[\protect\citeauthoryear{Beck \& Grave}{1982}]{beck82} Beck, R., Grave, R., 1982, A\&A, 105, 192
\bibitem[\protect\citeauthoryear{Beck}{1991}]{beck91} Beck, R., 1991, A\&A, 251, 15
\bibitem[\protect\citeauthoryear{Beck et al.}{2002}]{beck02} Beck, R., Shoutenkov, V., Ehle, E., Harnett, J. I., Haynes, R. F., Shukurov, A., Sokoloff, D. D., Thierbach, M., 2002, A\&A, 391, 83
\bibitem[\protect\citeauthoryear{Beck et al.}{2005}]{beck05} Beck, R., Fletcher, A., Shukurov, A., et al., 2005, A\&A, 444, 739
\bibitem[\protect\citeauthoryear{Beck}{2007}]{beck07} Beck, R.\ 2007, A\&A, 470, 539 
\bibitem[\protect\citeauthoryear{Bell}{1978}]{bell78} Bell, R., 1978, MNRAS, 182, 443 
\bibitem[\protect\citeauthoryear{Berkhuijsen, Beck \& Hoernes}{2003}]{berkh03} Berkhuijsen, E. M., Beck, R., Hoernes, P., 2003, A\&A, 398, 937
\bibitem[\protect\citeauthoryear{Berkhuijsen, Mitra \& Mueller}{2006}]{berkh06} Berkhuijsen, E.~M., Mitra, D., \& Mueller, P.\ 2006, Astronomische Nachrichten, 327, 82
\bibitem[\protect\citeauthoryear{Biermann \& Strom}{1993}]{bierm93} Biermann, P. L., Strom, R. G., 1993, A\&A, 275, 659 
\bibitem[\protect\citeauthoryear{Bogdan \& V$\ddot{\rm o}$lk}{1983}]{bogda83} Bogdan,~T.~J., V$\ddot{\rm o}$lk,~H.~J., 1983, A\&A, 122, 129
\bibitem[\protect\citeauthoryear{Braun et al.}{2007}]{braun07} Braun, R., Oosterloo, T. A., Morganti, R., Klein, U., Beck, R., 2007, A\&A, 461, 455
\bibitem[\protect\citeauthoryear{Briggs}{1995}]{brigg95} Briggs, D. S. 1995, Ph.D. Thesis, New Mexico Institute of Mining Technology, Socorro, New Mexico, USA
\bibitem[\protect\citeauthoryear{Broadbent, Osborne \& Haslam}{1989}]{broad89} Broadbent, A., Osborne, J.~L., Haslam, C.~G.~T.\ 1989, MNRAS, 237, 381
\bibitem[\protect\citeauthoryear{Brown \& Hazard}{1961}]{brown61} Brown, R. H., Hazard, C., 1961, MNRAS, 122, 479
\bibitem[\protect\citeauthoryear{de Bruyn}{1977}]{bruyn77} de Bruyn, A. G., 1977, A\&A, 54, 491
\bibitem[\protect\citeauthoryear{Carilli et al.}{1992}]{caril92} Carilli, C. L., Holdaway, M. A., Ho, P. T. P., de Pree, C. G., 1992, ApJ, L59, 399
\bibitem[\protect\citeauthoryear{Caswell \& Wills}{1967}]{caswe67} Caswell, J. L., Wills, D., 1967, MNRAS, 135, 231
\bibitem[\protect\citeauthoryear{Chy$\ddot{\rm z}$y \& Buta}{2008}]{chyzy08} Chy$\ddot{\rm z}$y,~K.~T., Buta,~R.~J., 2008, ApJ, 677, L17
\bibitem[\protect\citeauthoryear{Condon}{1987}]{condo87} Condon, J. J., 1987, ApJ Suppl., 65, 485
\bibitem[\protect\citeauthoryear{Condon}{1992}]{condo92} Condon, J. J., 1992, ARA\&A, 30, 575
\bibitem[\protect\citeauthoryear{Condon et al.}{1998}]{condo98} Condon, et al. 1998, AJ, 115, 1693
\bibitem[\protect\citeauthoryear{Cornwell \& Perley}{1992}]{cornw92} Cornwell, T. J., Perley, R. A., 1992, A\&A, 261, 353
\bibitem[\protect\citeauthoryear{Dickinson et al.}{2003}]{dicki03} Dickinson, C., Davies, R.~D., Davis, R.~J., 2003, MNRAS, 341, 369
\bibitem[\protect\citeauthoryear{Duric \& Dittmar}{1988}]{duric88a} Duric, N., Dittmar, M. R., 1988, ApJ, 332, L67
\bibitem[\protect\citeauthoryear{Gioia \& Gregorini}{1980}]{gioia80} Gioia, I. M., Gregorini, L., 1980, A\&A Suppl., 41, 329
\bibitem[\protect\citeauthoryear{Gioia, Gregorini \& Klein}{1982}]{gioia82} Gioia, I. M., Gregorini, L., Klein, U., 1982, A\&A, 116, 164
\bibitem[\protect\citeauthoryear{Green}{1998}]{green98} Green D.A., 1998, A Catalogue of 
Galactic Supernova Remnants (1998
September version). Mullard Radio Astronomy Observatory, Cambridge, United Kingdom (available on the World-Wide-Web at
http://www.mrao.cam.ac.uk/surveys/snrs/)
\bibitem[\protect\citeauthoryear{Heesen et al.}{2009}]{heese09} Heesen,~V., Beck,~R., Krause,~M., Dettmar,~R.~J., 2009, A\&A, 494, 563
\bibitem[\protect\citeauthoryear{Hummel}{1980}]{humme80} Hummel, E., 1980, A\&AS, 41, 15 
\bibitem[\protect\citeauthoryear{Hummel \& Bosma}{1982}]{humme82} Hummel, E., Bosma, A., 1982, AJ, 87, 242 
\bibitem[\protect\citeauthoryear{de Jong}{1966}]{jong66} de Jong, M. L., 1966, ApJ, 144, 153
\bibitem[\protect\citeauthoryear{de Jong}{1967}]{jong67} de Jong, M. L., 1967, ApJ, 150, 1
\bibitem[\protect\citeauthoryear{Karachentsev, Sharina \& Huchtmeier}{2000}]{karac00} Karachentsev, I. D., Sharina, M. E., Huchtmeier, W. K., 2000, A\&A, 362, 544
\bibitem[\protect\citeauthoryear{Karachentsev et al.}{2002}]{karac02} Karachentsev, I. D., Sharina, M. E., Dolphin, A. E., et al., 2002, A\&A, 385, 21
\bibitem[\protect\citeauthoryear{Karachentsev et al.}{2003}]{karac03} Karachentsev, I. D., Sharina, M. E., Dolphin, A. E., et al., 2003, A\&A, 398, 467
\bibitem[\protect\citeauthoryear{Kellermann et al.}{1969}]{kelle69} Kellermann, K. I., Pauliny-Toth, I. I. K., 
Williams, P. J. S., 1969, ApJ, 157, 1
\bibitem[\protect\citeauthoryear{Kennicutt et al.}{2003}]{kenni03} Kennicutt, Jr., R. C., Armus, L., Bendo, G., et al. 2003, PASP, 115, 928
\bibitem[\protect\citeauthoryear{Klein \& Emerson}{1981}]{klein81} Klein, U., Emerson, D. T., 1981, A\&A, 94, 29
\bibitem[\protect\citeauthoryear{Klein et al.}{1982}]{klein82} Klein, U.,  Beck, R., Buczilowski, U. R., Wielebinski, R., 1982, A\&A, 108, 176
\bibitem[\protect\citeauthoryear{Klein, Grave \& Wielebinski}{1983}]{klein83} Klein, U., Grave, R., Wielebinski, R., 1983, A\&A, 117, 332
\bibitem[\protect\citeauthoryear{Klein, Wielebinski \& Morsi}{1988}]{klein88} Klein, U., Wielebinski, R., Morsi, H. W., 1988, A\&A, 190, 41
\bibitem[\protect\citeauthoryear{Knapen et al.}{2004}]{knape04} Knapen,~J.~H., Stedman,~S., Bramich,~D.-M., Folkes,~S.~L., Bradley,~T.~R., 2004, A\&A, 426, 1135
\bibitem[\protect\citeauthoryear{Knapik et al.}{2000}]{knapi00} Knapik,~J., Soida,~M.,Dettmar,~R.-J., Beck, R., Urbanik,~M., 2000, A\&A, 362, 910
\bibitem[\protect\citeauthoryear{Kothes et al.}{2006}]{kothe06} Kothes, R., Fedotov, K., Foster, T.~J., Uyan{\i}ker, B., 2006, A\&A, 457, 1081
\bibitem[\protect\citeauthoryear{Kr$\ddot{\rm u}$gel}{2003}]{kruge03} Kr$\ddot{\rm u}$gel, E. 2003, The physics of interstellar dust, The physics of interstellar dust,
ed. E. Kruegel (IoP Series in astronomy and astrophysics, Bristol, UK: The
Institute of Physics).
\bibitem[\protect\citeauthoryear{Kuril'chik, Onishchenko \& Turyvskii}{1967}]{kuril67} Kuril'chik, V. N., Onishchenko, L. V., Turyvskii, V. M., 1967, Sov. Astron. J., 11, 528
\bibitem[\protect\citeauthoryear{Kuril'chik et al.}{1970}]{kuril70} Kuril'chik, V. N., Andrievskii, A. E., Ivanov, V. N., Spangenberg, E. E., 1970, Sov. Astron. J., 13, 881
\bibitem[\protect\citeauthoryear{van der Kruit, Allen \& Rots}{1977}]{kruit77} van der Kruit, P. C., Allen, R. J., Rots, A. H., 1977, A\&A, 55, 421
\bibitem[\protect\citeauthoryear{Longair}{2011}]{longa11} Longair, ~M.~S., 2011, High Energy Astrophysics 3-ed, Cambridge University Press.
\bibitem[\protect\citeauthoryear{Maddox et al.}{2006}]{maddo06} Maddox, L. A., Cowan, J.~J., Kilgard, R. E., Lacey, C. K., Prestwich, A. H., Stockdale, C. J., Wolfing, E., 2006, AJ, 132, 310
\bibitem[\protect\citeauthoryear{Mayya, Carrasco \& Luna}{2005}]{mayya05} Mayya, Y.~D., Carrasco, L., Luna, A., 2005, ApJ, 628, L33 
\bibitem[\protect\citeauthoryear{Murgia et al.}{2005}]{murgi05} Murgia, M., Helfer, T. T., Ekers, R., Blitz, L., Moscadelli, L., Wong, T., Paladino, R., 2005, A\&A, 437, 389
\bibitem[\protect\citeauthoryear{Murphy et al.}{2010}]{murph10} Murphy, E. J., Helou, G., Condon, J. J., Schinnerer, E., Turner, J. L., Beck, R., Mason, B. S., Chary, R.-R., Armus, L., 2010, ApJ, 709, L108
\bibitem[\protect\citeauthoryear{Nagar, Falcke \& Wilson}{2005}]{nagar05} Nagar, N.~M., Falcke, H., Wilson, A.~S., 2005, A\&A, 435, 521 
\bibitem[\protect\citeauthoryear{Niklas \& Beck}{1997}]{nikla97a} Niklas, S., Beck, R., 1997, A\&A, 320, 54
\bibitem[\protect\citeauthoryear{Niklas, Klein \& Wielebinski}{1997}]{nikla97} Niklas, S., Klein, U., Wielebinski, R., 1997, A\&A, 322, 19
\bibitem[\protect\citeauthoryear{Ondrechen \& van der Hulst}{1983}]{ondre83} Ondrechen, M. P., van der Hulst, J. M., 1983, ApJ, 269, L47
\bibitem[\protect\citeauthoryear{Ondrechen}{1985}]{ondre85} Ondrechen, M. P., van der Hulst, J. M., Hummel, E., 1985, AJ, 90, 1474
\bibitem[\protect\citeauthoryear{Ondrechen, van der Hulst \& Hummel}{1989}]{ondre89} Ondrechen, M. P., 1989, ApJ, 342, 39
\bibitem[\protect\citeauthoryear{Paladino, Murgia \& Orr$\rm\grave{u}$}{2009}]{palad09} Paladino, R., Murgia, M., Orr$\rm \grave{u}$, E., 2009, A\&A, 503, 747 
\bibitem[\protect\citeauthoryear{Rieke et al.}{2004}]{rieke04} Rieke, G., et al., 2004, ApJS, 154, 25
\bibitem[\protect\citeauthoryear{Roy \& Rao}{2004}]{roy04} Roy, S., Rao P. A., 2004, MNRAS, 349, L25
\bibitem[\protect\citeauthoryear{Schommer \& Sullivan}{1976}]{schom76} Schommer, R.~A., Sullivan, W.~T., 1976, Ap. Letters, 17, 191
\bibitem[\protect\citeauthoryear{Seaquist \& Odegard}{1991}]{seaqu91} Seaquist, E. R., Odegard, N., 1991, ApJ, 369, 320
\bibitem[\protect\citeauthoryear{Segalovitz}{1977a}]{segal77a} Segalovitz, A., 1977a, A\&A, 61, 59 
\bibitem[\protect\citeauthoryear{Segalovitz}{1977b}]{segal77b} Segalovitz, A., 1977b, A\&A, 54, 703 
\bibitem[\protect\citeauthoryear{Smith et al.}{1991}]{smith91} Smith, B.~J., Lester, D.~F., Harvey, P.~M., Pogge, R.~W., 1991, ApJ, 380, 677 
\bibitem[\protect\citeauthoryear{Steer, Dewdney \& Ito}{1984}]{steer84} Steer, D. G., Dewdney, P. E., Ito, M. R., 1984, A\&A, 137, 159
\bibitem[\protect\citeauthoryear{Sukumar, Klein \& Grave}{1987}]{sukum87} Sukumar, S., Klein, U., Grave, R., 1987, A\&A, 184, 71
\bibitem[\protect\citeauthoryear{Swarup et al.}{1991}]{swaru91} Swarup, G., Ananthakrishnan, S., Kapahi, V. K., et al. 1991, Current Science, 60, 95
\bibitem[\protect\citeauthoryear{Tabatabaei et al.}{2007}]{tabat07} Tabatabaei, F., Beck, R., Krugel, E., , Krause, M., Berkhuijsen, E.~M., Gordon, K.~D., Menten, K.~M.\ 2007, A\&A, 475, 133
\bibitem[\protect\citeauthoryear{Valls-Gabaud}{1998}]{valls98} Valls-Gabaud,~D., 1998, PASA, 15, 111 
\bibitem[\protect\citeauthoryear{de Vaucouleurs et al.}{1991}]{vauco91} de Vaucouleurs, G., et al. 1991, Third Reference Catalogue of Bright Galaxies (Berlin: Cambridge University Press)
\bibitem[\protect\citeauthoryear{Wong \& Blitz}{2001}]{wong01} Wong, T., Blitz, L., 2001, ApJ, 540, 771
\bibitem[\protect\citeauthoryear{Wright et al.}{1990}]{wrigh90} Wright, A. E., Wark, R. M., Troup, E., Otrupcek, R., Hunt, A., Cooke, D. J., 1990, PASA, 8, 261.
\end{thebibliography}
\end{document}